\documentclass[twocolumn]{aa}

\usepackage{natbib,twoopt}
\usepackage[breaklinks=true]{hyperref} 

\usepackage{graphicx}
\usepackage{txfonts}
\usepackage{titlesec}
\usepackage{multibib}
\usepackage{epstopdf}
\usepackage{xcolor}
\usepackage{enumitem}
\usepackage{fancyhdr}
\usepackage{wasysym} 

\usepackage{orcidlink}

\bibpunct{(}{)}{;}{a}{}{,} 

\makeatletter
\newcommandtwoopt{\citeads}[3][][]{\href{http://adsabs.harvard.edu/abs/#3}%
{\def\hyper@linkstart##1##2{}%
\let\hyper@linkend\@empty\citealp[#1][#2]{#3}}}
\newcommandtwoopt{\citepads}[3][][]{\href{http://adsabs.harvard.edu/abs/#3}%
{\def\hyper@linkstart##1##2{}%
\let\hyper@S-curve_Fe_ions_AGN1_zoom.pltlinkend\@empty\citep[#1][#2]{#3}}}
\newcommandtwoopt{\citetads}[3][][]{\href{http://adsabs.harvard.edu/abs/#3}%
{\def\hyper@linkstart##1##2{}%
\let\hyper@linkend\@empty\citet[#1][#2]{#3}}}
\newcommandtwoopt{\citeyearads}[3][][]%
{\href{http://adsabs.harvard.edu/abs/#3}
{\def\hyper@linkstart##1##2{}%
\let\hyper@linkend\@empty\citeyear[#1][#2]{#3}}}

\makeatother

\titleformat{\subsubsection}
{\normalfont\fontfamily{phv}\fontsize{8}{17}\selectfont}{\thesubsubsection}{1em}{}
\newcommand{\RomanNumeralCaps}[1]
{\MakeUppercase{\romannumeral #1}}
{
\newcommand\T{\rule{0pt}{2.6ex}}       
\newcommand\B{\rule[-1.2ex]{0pt}{0pt}} 

\begin{document} 
	
	
	\title{New radiative loss curve from updates to collisional excitation in the low-density, optically thin plasmas in SPEX}

	\author{L. {\v S}tofanov{\' a} \inst{1, 2}\orcidlink{0000-0003-0049-6205},
		J. Kaastra \inst{1, 2},
		M. Mehdipour \inst{2, 3},
		\and
		J. de Plaa \inst{2}
	}

	\institute{Leiden Observatory, Leiden University,
		PO Box 9513, 2300 RA Leiden, The Netherlands\\
		\email{stofanova@strw.leidenuniv.nl}
		\and
		SRON Netherlands Institute for Space Research, Sorbonnelaan 2, 3584 CA Utrecht, The Netherlands\\ \vspace*{-2.8mm}
		\and 
		STScI Space Telescope Science Institute, 3700 San Martin Drive, Baltimore, MD 21218, USA\\
	}
	
	\date{}
	
	
	\abstract
    {Understanding and modelling astrophysical plasmas on atomic levels while taking into account various assumptions (for example, collisional ionisation equilibrium or photoionisation equilibrium) became essential with the progress of high-resolution X-ray spectroscopy. In order to prepare for the upcoming X-ray spectroscopy missions such as XRISM or Athena, the plasma codes with their models and the atomic databases need to be up to date and accurate. One such update for the plasma code SPEX is presented in this paper where we focus on the radiative loss due to collisional excitation in the low-density, optically thin regime. We also update the atomic data for neutral hydrogen and include the contribution of the dielectronic recombination. With all these updates being implemented in SPEX we finally present the new cooling curve. We include the comparison to other plasma codes (MEKAL, APEC, Cloudy) and other atomic databases (CHIANTI, ADAS). We show how the updated cooling impacts the stability curve for photoionised plasmas and find a new stable branch.}
	

    \authorrunning{{\v S}tofanov{\' a} et al.}
	\titlerunning{Update of the cooling in PIE model in SPEX}
	\maketitle

\section{Introduction}

Hot astrophysical plasmas heated to temperatures of around $10^4$--$10^8$\,K can be found in a wide range of astrophysical sources at various scales throughout the Universe. Depending on the properties of the plasma (as for example, temperature, density) and its environment, different atomic processes contribute to the overall shape of the spectrum. These processes all need to be taken into account in models for X-ray emission from these hot plasmas.

Over the last couple of decades, X-ray spectroscopy has undergone significant improvements that have allowed us to obtain X-ray spectra with a high level of detail. The observations that have been made so far, however, also reveal the importance of accurate atomic data as well as the improvement of the plasma codes that are needed in order to properly understand the observed X-ray spectra. Such models in the plasma codes help us to understand the radiation coming from the X-ray sources and allow us to study their physical properties.

In general, plasma codes can be either collisional, photoionised or include both regimes at the same time. A few examples are APEC \citep{2001ApJ...556L..91S}, CHIANTI \citep{1997A&AS..125..149D}, MEKAL \citep{mewe1995update}, SPEX \citep{kaastra1996_spex}, Cloudy \citep{1998PASP..110..761F}, Mocassin \citep{2003Mocassin}, Titan \citep{2000A&A...357..823D}, and XSTAR \citep{2001ApJS..133..221K}. The discrepancies between different plasma codes can be relatively high, especially for lower temperatures (for instance, see \citealp{2018PASJ...70...12H}). These discrepancies can be caused by different approaches and approximations used for solving the Schr$\ddot{\rm o}$dinger (Dirac) equation or different ionisation balance and calculation of level populations (for more about atomic data in X-rays, see reviews by \citealt{2005AIPC..774.....S,2007RvMP...79...79K} and references therein).  

One of many consequences of such discrepancies are the estimates of the cooling rates. Plasma can cool down via multiple processes; for example, bremsstrahlung or more complex cooling via radiative recombination, collisional excitation, ionisation, etc. The process of cooling has been extensively studied in many works such as, for example \citet{1976ApJ...204..290R}, \citet{1989A&A...215..147B}, \citet{1993ApJS...88..253S}, \citet{1999A&A...347..401L}, \citet{2009A&A...508..751S}, and \citet{2013MNRAS.429.3133L}. In this paper, we use the term cooling for the cooling of the population of free electrons only, which we express in units of keV m$^3$ s$^{-1}$. For more details about the terminology of cooling, we refer the reader to \citet{2012ApJS..199...20G}, for example.

The radiative loss function and the cooling function should give the same results if the collisional ionisation equilibrium (CIE) is considered since the plasma in an enclosed and non-expanding box can lose the energy only by radiating photons if for example, magnetic fields, heat conduction, radioactive processes, etc. are not taken into account. For the collisional excitation process in a low-density plasma, as discussed in the present paper, the radiative loss is the same as the cooling. This is different from radiative recombination, which leads to radiative losses equal to the cooling by that process plus the ionisation potential of the recombined ion and different from collisional ionisation of the valence electrons, which does not produce radiation at all.

This work serves as a report of the updates made in the plasma code SPEX. More specifically, we focus on the updates of electron collisional excitation rates and the impact of these updates on the cooling for the plasma in the low-density regime. Collisional excitation occurs when a free electron collides with an atom or ion and brings the bound electron into the excited state. Because these excited states are less energetically favourable, the excited electron falls down to a lower level by radiative transition.

In previous versions of SPEX, the collisional excitation rates were calculated with the codes by \citet{1985A&AS...62..197M}, \citet{1986A&AS...65..511M} and  MEKAL \citep{mewe1995update} which originally operated in the regimes $\lambda < 300 \, \AA$, $\lambda < 1000 \, \AA$, and $\lambda < 2000 \, \AA$, respectively. Over the years, the X-ray line database of SPEX has been updated and has grown, also extending to the ultraviolet (UV), optical, and infrared (IR) bands. Since the first version of SPEX, many updates have been made and the most recent works are the updated radiative recombination rates \citep{2016A&A...587A..84M, 2017A&A...599A..10M}, the updated collisional ionisation rates \citep{2017A&A...601A..85U}, a model for the charge exchange transfer \citep{2016A&A...588A..52G}, updates of the collisional excitation rates \citep{2017JInst..12C8008K}, and the updated data for the Fe-L complex (focusing mainly on \ion{Fe}{XVII} to \ion{Fe}{XXIV} ions) by \citet{2019A&A...627A..51G}. The latest version, $3.06.01$, also includes the updated radiative loss function due to collisional excitation that we present in this paper and that is used in the photoionisation model pion (for a plasma in photoionisation equilibrium -- PIE) and updates of the collision strengths of \ion{H}{I} (see Section \ref{Sec:total_cooling_curve} for more details).

The paper is structured as follows. In Section \ref{Sec:cooling_curves} we describe how to calculate the radiative loss rates from the excitation rates and how the radiative loss curves for specific ions are obtained. We show our main results in Section \ref{Sec:cooling_curves_results} where we focus on the plasma in collisional ionisation equilibrium and address the accuracy as well as the role of collisional excitation, resonant excitation, metastable levels and updated atomic data for neutral hydrogen. We also discuss the discrepancies in the cooling curves obtained from different plasma codes. In Section \ref{Sec:application-S-curve} we implement tables created for the radiative loss rates due to collisional excitation to the photoionisation model in SPEX and study the effect of the updated cooling on the stability curves. We discuss the results in Section \ref{Sec:discussion} and give the main conclusions in Section \ref{Sec:conclusions}.

\section{Methods}
\label{Sec:cooling_curves}

\subsection{Radiative loss due to collisional excitation in low-density plasmas}
\label{Sec:coll_exc}

The radiative loss rate for collisional excitation quantifies the energy loss of plasma due to collisions of free electrons with atoms or ions (we consider the radiative loss by the population of free electrons only). To calculate this radiative loss rate, we first define the excitation rate $S_{ij}{(T)}$ for the excitation from level $i$ to level $j$ (while assuming the Maxwellian velocity distribution for electrons) as
\begin{equation}
	S_{ij}{(T)} =  \dfrac{S_0 \bar{\Omega}{(y)}}{\sqrt{T} w_i e^y} \,, \qquad \qquad y = \frac{E_{ij}}{kT} \;,
	\label{eq:exc_rate}
\end{equation} 
where $T$ is the temperature of plasma (electron temperature), $w_i$ is the statistical weight of level $i$, $\bar{\Omega}{(y)}$ is the effective collision strength (\citealt{1940ApJ....92..408H}, also called the Upsilon value),  $E_{ij}$ is the excitation energy for the transition $i \rightarrow j$ and $S_0$ is the constant defined as
\begin{equation}
	S_0 =  h^2 (2 \pi m_e)^{-3/2} k^{-1/2}  \;,
\end{equation}
where $m_e$ is the electron mass and $h$ and $k$ represent the Planck constant and the Boltzmann constant, respectively. The units of $S_{ij}$ are m$^3$s$^{-1}$, such that the total number of excitations per second in a volume V with electron density $n_{\rm e}$ and ion density $n_{\rm i}$ is given by $n_{\rm e}n_{\rm i}V S_{ij}$. As the temperature approaches lower values, the excitation rate decreases exponentially, while for higher temperatures the excitation rate decreases as $1/\sqrt{T}$. The total radiative loss function due to collisional excitation only per specific atom or ion from the lower level \textit{i} is then obtained by summation of the product of the excitation rate and the excitation energy over all transitions in the atom or ion:
\begin{equation}
	\Lambda_{i, \rm exc}{(T)} =  \sum_{j}^{}{S_{ij}{(T)} E_{ij}}  \;.
	\label{eq:cool_rate}
\end{equation} 
We note that the Equation \eqref{eq:cool_rate} is valid only in the low-density plasmas. That allows us to ignore the effect of the collisional de-excitation when the collision of an ion with a free electron or another ion results in the bound electron being brought into the lower energy orbit.

\subsection{Radiative loss curves in SPEX, MEKAL and CHIANTI}
\label{Sec:cool_databases}

Throughout the years, SPEX \citep{kaastra1996_spex} has been extended up to the UV/IR regime, and its atomic database SPEXACT is still under development. The number of elements present in the SPEX atomic database increased from $15$ (H, He, C, N, O, Ne, Na, Mg, Al, Si, S, Ar, Ca, Fe, Ni) to $30$, and now SPEX includes all elements up to and including zinc. It contains all ions of the hydrogen and helium iso-electronic sequences and most of the ions of lithium, beryllium, and boron iso-electronic sequences up to Na-like ions. For all ions where we do not yet have data in our SPEXACT database, we used the MEKAL code to calculate line fluxes when available. Originally, the MEKAL code took into account approximately $5500$ lines, while the new updated SPEXACT database contains around $4.2 \times 10^6$ lines. One of the biggest differences between these two codes lies in calculating the line fluxes, for which MEKAL uses temperature-dependent parametrisation (see \citealt{1985A&AS...62..197M} and references therein, mainly \citet{1972SoPh...22..459M} and Eq.\,(10) and Eq.\,(11)), whereas SPEX calculates the line fluxes by obtaining the level populations from the transition rate balance equations. Then, the total line power in SPEX is calculated using the radiative transition probabilities.



To calculate the radiative loss rates in SPEX we used version $3.05.00$ \citep{kaastra1996_spex, kaastra2018_spex} and made these rates public as a part of SPEX version $3.06.00$ and higher. The most updated version $3.06.01$ includes, among others, updates of the radiative loss rates in the SPEXACT database as well as the cooling rates in the photoionisation model pion \footnote{\url{https://spex-xray.github.io/spex-help/}} (which we describe in this paper). Unless stated otherwise, we used the \citet{2017A&A...601A..85U} ionisation balance and the protosolar abundances by \citet{2009LanB...4B..712L}. 


For comparison, we also calculated the radiative loss rates in version 9.0 of the CHIANTI database \citep{1997A&AS..125..149D, Dere_2019} using version $0.9.1$ of the python package ChiantiPy. The excitation energies of individual transitions are considered to be theoretical values (due to more complete dataset of line energies for different ions), and we used the observed values only if the theoretical values were not available. The choice of using theoretical values does not have a significant impact on the final radiative loss rates for the vast majority of available ions. We comment on these differences in more detail in Section \ref{Sec:Discussion_theoretical_calc}.

To estimate the uncertainties in different databases, we also used the upsilon values from ADAS \citep{summers2004}, for which we used the free electron excitation data from the ion data collections of OPEN-ADAS\footnote{\url{https://open.adas.ac.uk/}}. We calculated the radiative loss rates for a set of oxygen ions and prioritized the most complete datasets with the highest number of levels (if multiple data files contain the same number of levels, we prioritized the data file with the widest temperature range that can be compared to our calculations). The ADAS datasets together with references, number of levels, and temperature range for oxygen ions are shown in Table \ref{table:ADAS_files}. We then calculated the total radiative loss rate on the temperature grid available for each ion (significantly narrower than the temperature grid we choose for SPEX, MEKAL, and CHIANTI). 

\begin{table*}[!t]
	\centering       
	\caption{Summary of \ion{O}{I}-\ion{O}{VIII} data files used for calculation of the radiative loss rates due to collisional excitation in ADAS.}                
		\begin{tabular}{l|c|c|c|c}
			\hline\hline 
			ion & number of levels & file name in OPEN-ADAS & temperature range & reference  	\\
			 &  & & [eV] &  	\\  \hline
			\rule{0pt}{2.5ex}  \ion{O}{I} & $554$ & \href{https://open.adas.ac.uk/detail/adf04/cophps][o/dw/ic][o0.dat}{ic\#o0.dat} & $0.009$ -- $172$  &	[1]	\\ 
			\ion{O}{II} & $668$ & \href{https://open.adas.ac.uk/detail/adf04/cophps][n/dw/ic][o1.dat}{ic\#o1.dat} & $0.034$ -- $689$ &	[1]	 \\ 
			\ion{O}{III} & $590$ & \href{https://open.adas.ac.uk/detail/adf04/copaw][c/clike_jm19][o2.dat}{clike\_jm19\#o2.dat} & $0.155$ -- $1551$ &	[2]		 \\ 
			\ion{O}{IV} & $204$ & \href{https://open.adas.ac.uk/detail/adf04/copaw][b/blike_lgy12][o3.dat}{blike\_lgy12\#o3.dat} & $0.276$ -- $2757$ &	[3]	\\ 
			\ion{O}{V} & $238$ & \href{https://open.adas.ac.uk/detail/adf04/copaw][be/belike_lfm14][o4.dat}{belike\_lfm14\#o4.dat} & $0.431$ -- $4308$ & [4]		\\ 
			\ion{O}{VI} & $204$ & \href{https://open.adas.ac.uk/detail/adf04/copaw][li/lilike_lgy10][o5.dat}{lilike\_lgy10\#o5.dat} & $0.620$ -- $6204$ & [5]		\\ 
			\ion{O}{VII} & $49$ & \href{https://open.adas.ac.uk/detail/adf04/cophps][he/dw/ic][o6.dat}{ic\#o6.dat} & $0.422$ -- $8444$ &	[1]			\\ 
			\ion{O}{VIII} & $25$ & \href{https://open.adas.ac.uk/detail/adf04/cophps][h/dw/ic][o7.dat}{ic\#o7.dat} & $0.551$ -- $11030$ & [1]			\\ \hline
		\end{tabular}
		\tablebib{[1]~Giunta 2012 (available online); [2] \citet{2020A&A...634A...7M}; [3] \citet{2012A&A...547A..87L}; [4] \citet{2014A&A...566A.104F}; [5] \citet{2011A&A...528A..69L}.}
	\label{table:ADAS_files}      
\end{table*}



\section{Results}
\label{Sec:cooling_curves_results}

The excitation rate $S_{ij}$ and the excitation energy $E_{ij}$ for all the transitions from the ground level to upper levels allows us calculate the total radiative loss rate due to collisional excitation for individual ions using Eq.\,\eqref{eq:cool_rate}. In the following sections, we calculate the radiative loss rates in SPEX and show the comparison to different databases or codes. We also address other aspects that need to be considered while obtaining these rates.

\subsection{Effect of the maximum principal quantum number}
\label{Sec:principal_quantum_number}

\begin{figure}[!t]
	\centering
	\includegraphics[angle=-90, width=0.5\textwidth]{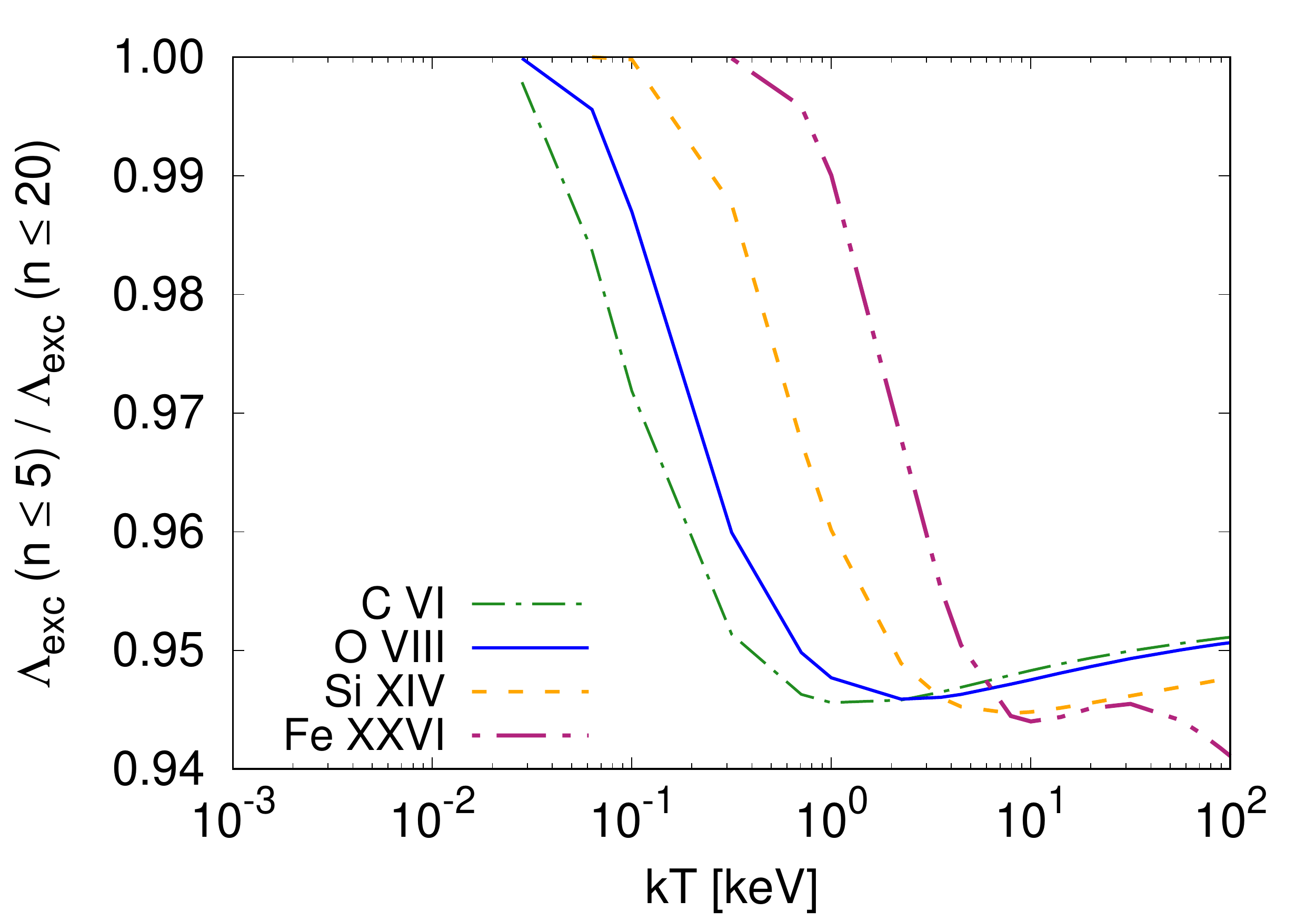} \\
	\includegraphics[angle=-90, width=0.5\textwidth]{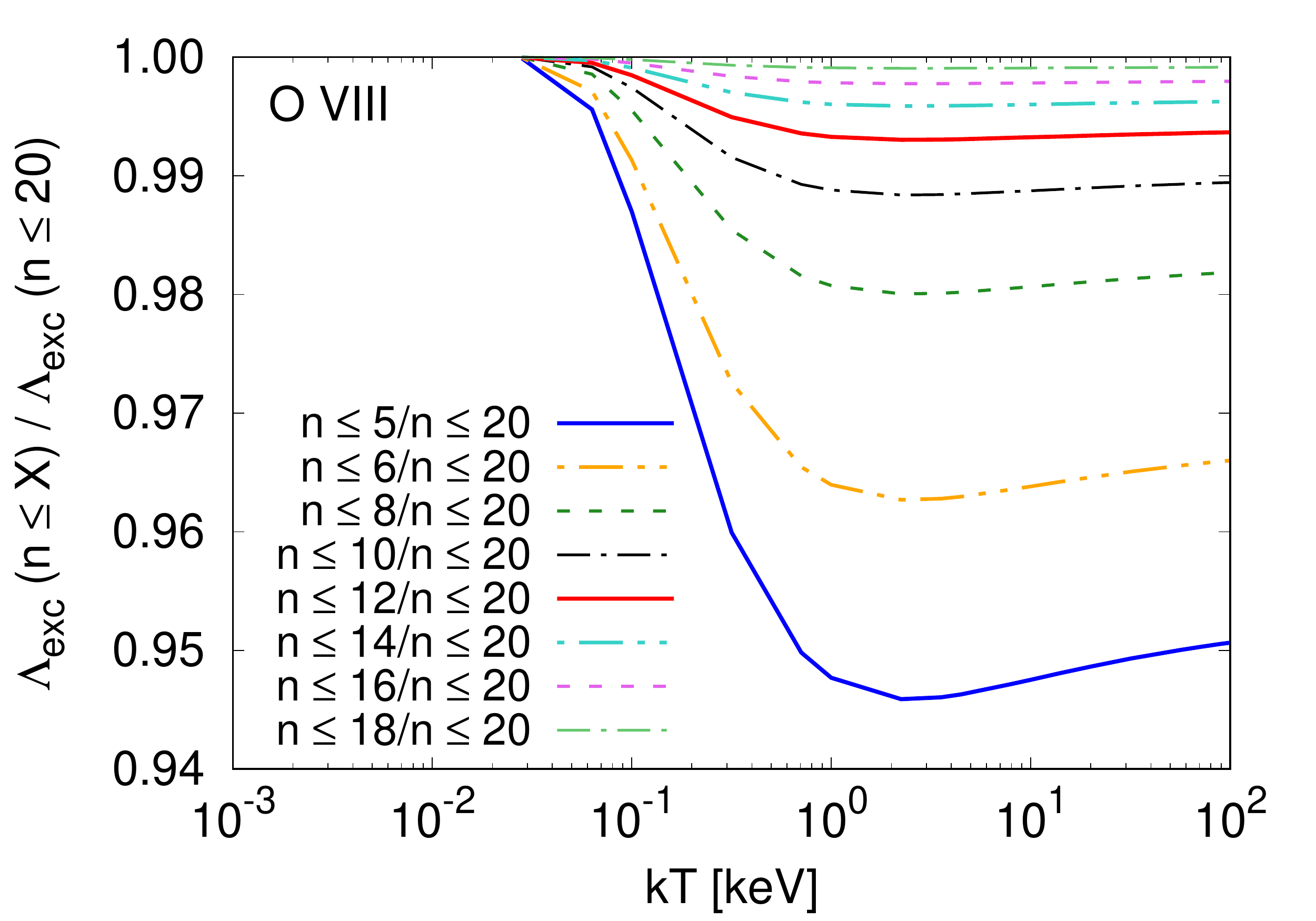} \\
	\caption{\textit{Top panel}: Cumulative radiative loss rate for all transitions up to the principal quantum number $5$ $\Lambda_{\rm exc}(n \leq 5)$ for H-like ions plotted relatively to the cumulative radiative loss rate for all transitions up to the principal quantum number $20$ $\Lambda_{\rm exc}(n \leq 20)$ as a function of temperature. \textit{Bottom panel}: Cumulative radiative loss rates for \ion{O}{VIII} up to various values of $n$ relative to $\Lambda_{\rm exc}(n \leq 20)$. }
	\label{Fig:cool_rate_vs_temp}
\end{figure}

The highest energy loss and therefore the strongest emission lines in the case of the collisional excitation come from the lowest levels of the atom or ion. But with the improvement of the X-ray spectroscopy, the emission lines originating from the higher levels become visible in measured spectra as well. To quantify the effect of the maximum principal quantum number $n$ used in the calculations we studied the dependency of the cumulative radiative loss rate (due to the collisional excitation only) on various $n$ for a set of ions (as well as temperatures) for which the SPEX database includes the transitions up to the principal quantum number $20$. In particular, these ions are \ion{C}{VI}, \ion{O}{VIII}, \ion{Si}{XIV} and \ion{Fe}{XXVI}. We calculated the difference between the cumulative radiative loss rate for transitions $n\leq5$ and compared it to the radiative loss rate when all transitions with $n\leq20$ are taken into account. For all mentioned ions, the difference between $n\leq5$ and $n\leq20$ cumulative radiative loss rates for low temperatures is less than $1$\%, while for higher temperatures this difference increases up to $5$--$6$\,\% (see Fig.\,\ref{Fig:cool_rate_vs_temp}). Therefore, not including the transitions for $n > 5$ introduces uncertainties in the radiative loss rates that are smaller than $6$\%. A similar conclusion was obtained in \cite{2018PASJ...70...12H}. 

Fig.\,\ref{Fig:cool_rate_vs_temp} only considers H-like (one-electron) systems. If more complex systems are taken into account, the number of levels that are included in the calculations can affect the resulting radiative loss rate. This is discussed in \citet{2013MNRAS.429.3133L} for the case of collisional plasma as well as photoionised plasma. Authors show that, in general, more levels are needed in the collisional case in comparison with the photoionised case. The final optimal number of levels that is considered is $100$ ($25$) for iron ions and $50$ ($25$) for other ions in the collisional (photoionised) case. If such a number of levels is included in the calculations, the cooling rates can be reproduced within 1\% ($0.1$\%) up to maximum of $5$\% for the collisional (photoionised) case. As we show in Table \ref{table:levels_SPEX} for a selection of ions, SPEX includes a sufficient number of levels and fulfills this requirement.

\begin{table}[!t]
	\centering       
	\caption{Number of levels included in the SPEX database for C, O, Si, and Fe from H-like to C-like iso-electronic sequences. }                
	\begin{tabular}{l | c c c c}      
		\hline\hline              
		 & \multicolumn{4}{c}{number of levels in SPEX} \\
		iso-electronic sequence & C & O & Si & Fe \\
		\hline                    
		H-like	& 284 & 284 & 284 & 284	  \\
		He-like	& 678 & 678	& 678 & 678	  \\
		Li-like	& 885 & 885	& 885 & 889	  \\
		Be-like	& 753 & 727	& 753 & 808	  \\
		B-like	& 719 & 973	& 972 & 1100  \\
		C-like	& 1199& 1218& 1218& 1525  \\

		\hline                  
	\end{tabular}
	\label{table:levels_SPEX}      
\end{table}

\subsection{Contribution of resonant excitation}
\label{Sec:res_excitation}

In the process of a dielectronic recombination, a free electron is captured while exciting a bound electron. This creates a doubly excited state that is not stable and can be followed by a radiation-free transition to a non-ground level and a loss of an electron. This process is called resonant excitation (RE). As a result, the ion gets to the excited state using less energy in comparison with the energy that would have been needed in the case of direct excitation.

The SPEX database for resonant excitation is currently incomplete but it contains a significant amount of resonances for the whole H-like iso-electronic sequence and almost all He-like ions (besides \ion{He}{I}, \ion{Li}{II}, \ion{Be}{III} and \ion{B}{IV}). If lines produced by RE are available in the SPEX database, they are already accounted for in the total excitation rate. To quantify the difference between the radiative loss rates including or excluding RE (for transitions to the ground level), we calculated the total radiative loss curve per ion for both scenarios. By calculating the ratio of the total radiative loss rate with RE and the total radiative loss rate without RE, we found that if RE was taken into account, the radiative loss rates were higher than the radiative loss rates excluding RE, and its effect was typically $\sim 15$\% (for example, for \ion{Si}{XIV} and $kT \sim 0.2$\,keV). 




\subsection{Comparison of radiative loss function for SPEX and MEKAL}
\label{Sec:cool_curve_comp_with_MEKAL}

\begin{figure*}[!t]
	\centering
	\resizebox{0.85\textwidth}{!}{
		\includegraphics[angle=-90]{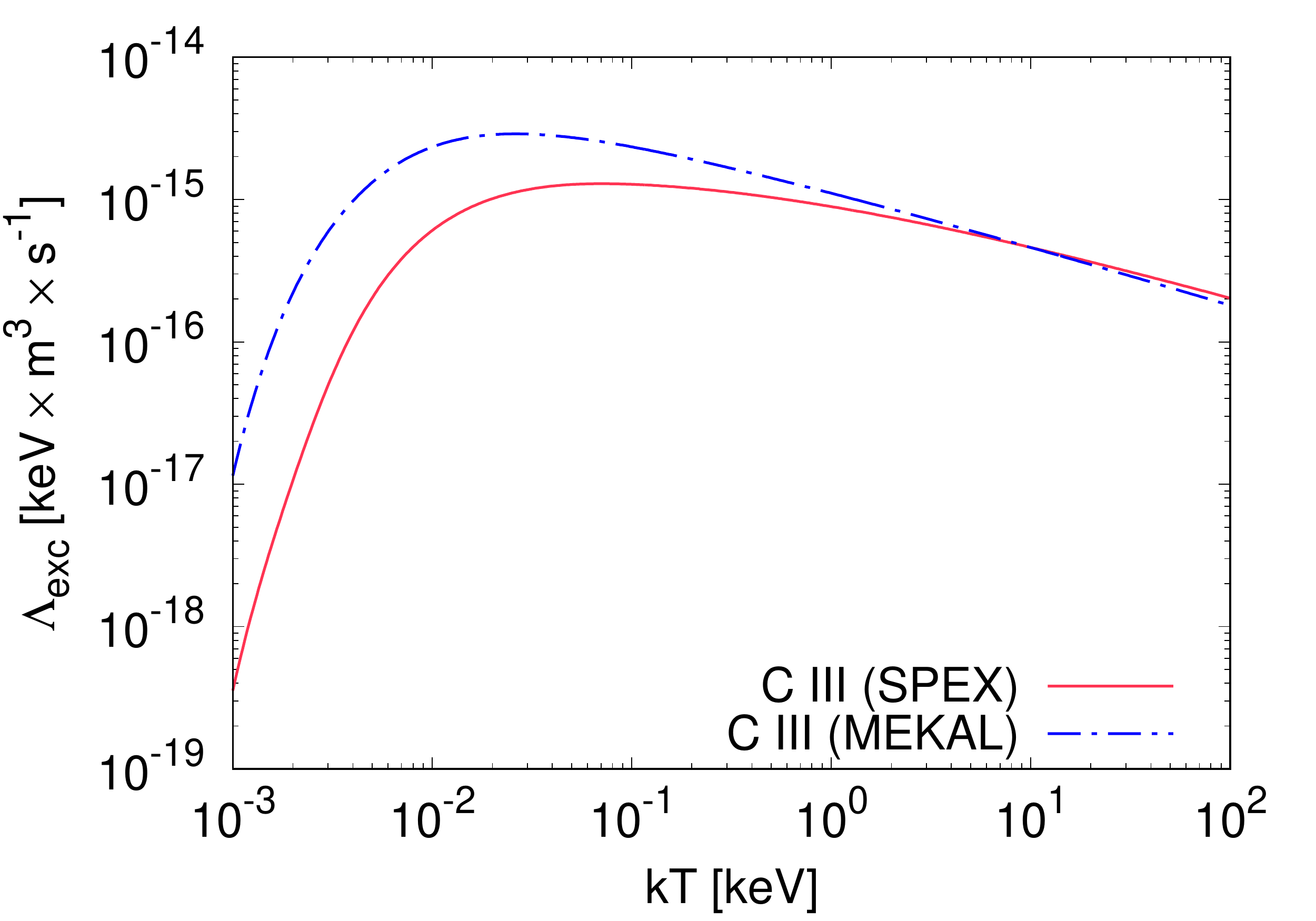}
		\includegraphics[angle=-90]{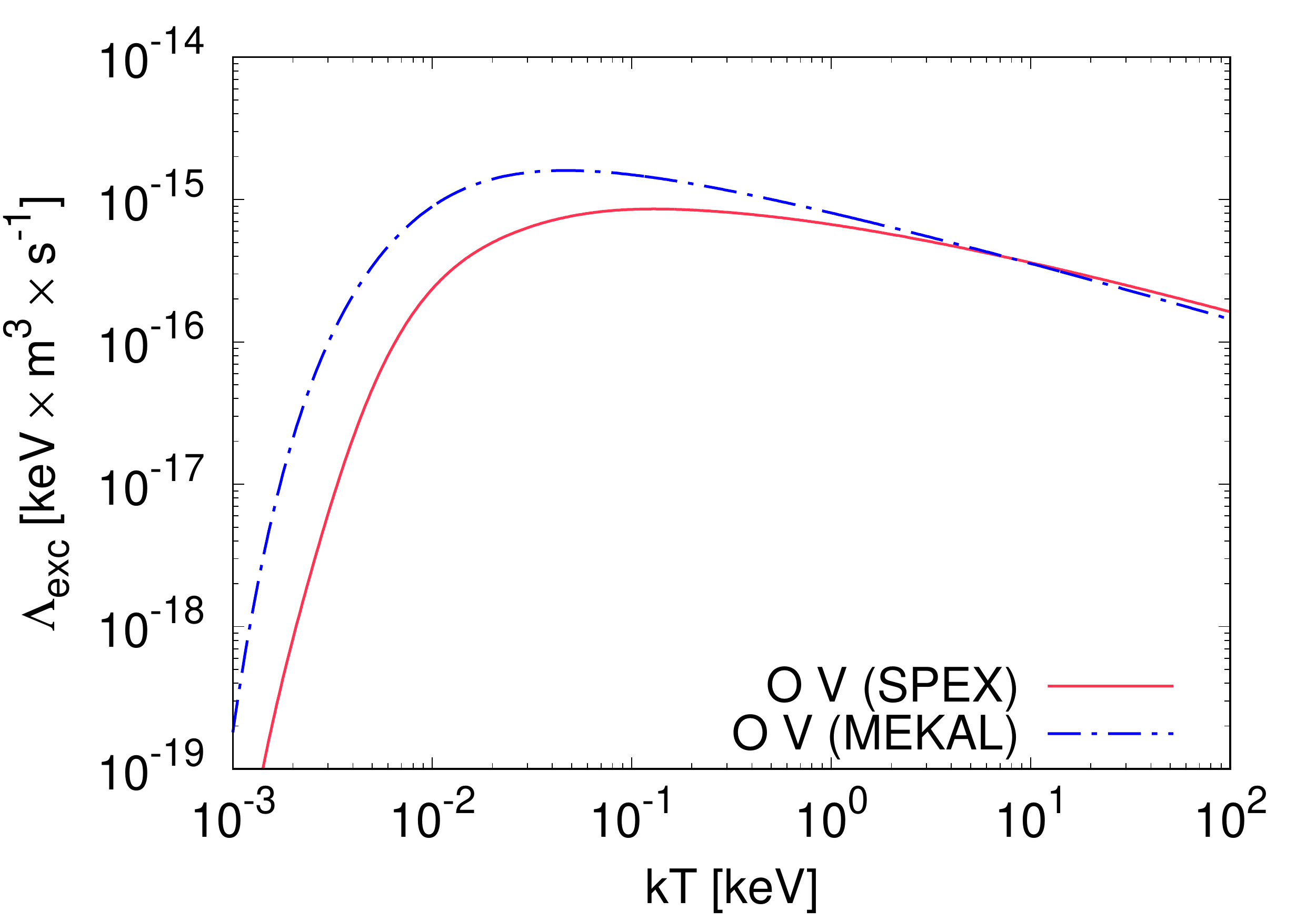}}
	\resizebox{0.85\textwidth}{!}{
		\includegraphics[angle=-90]{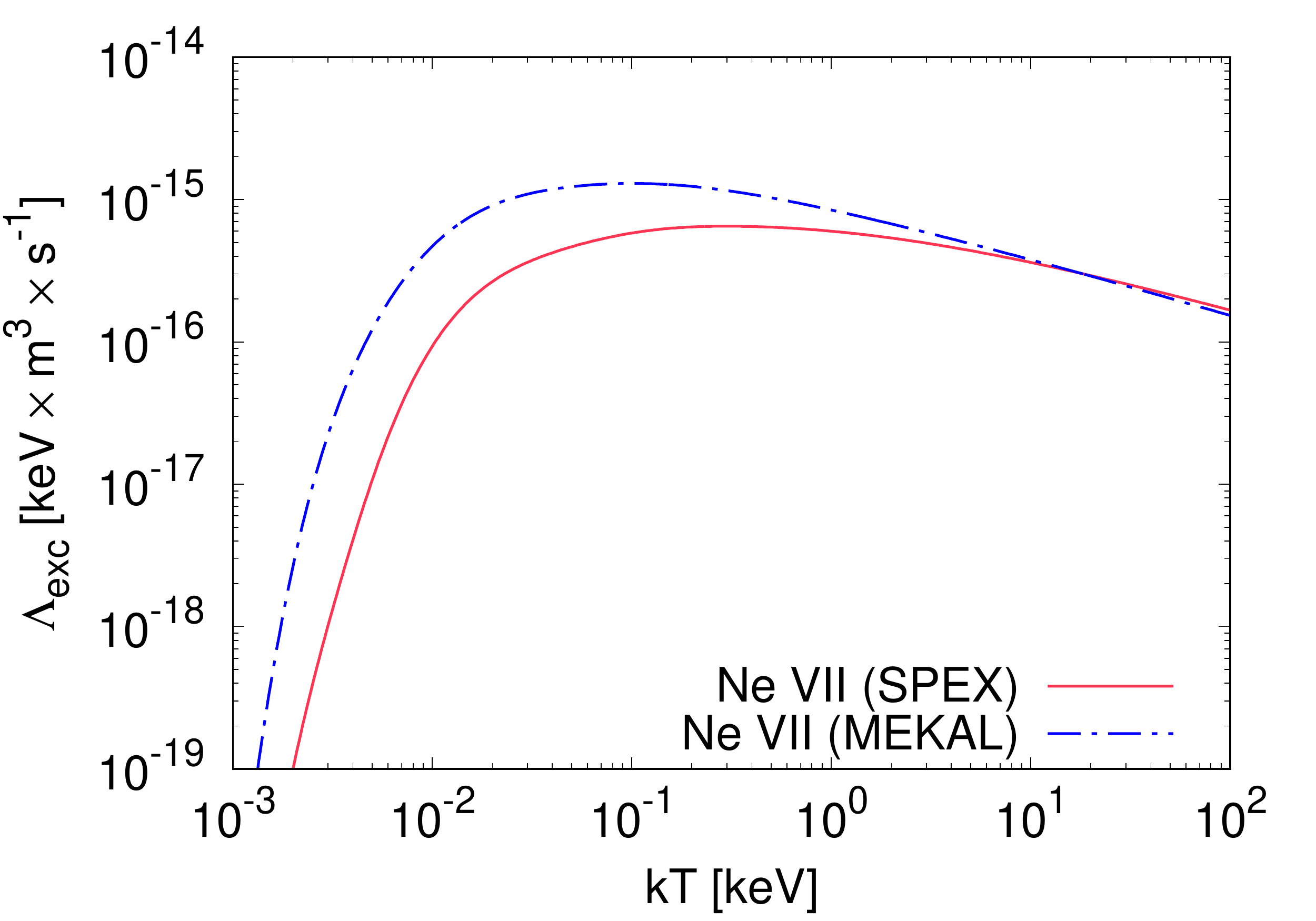} 
		\includegraphics[angle=-90]{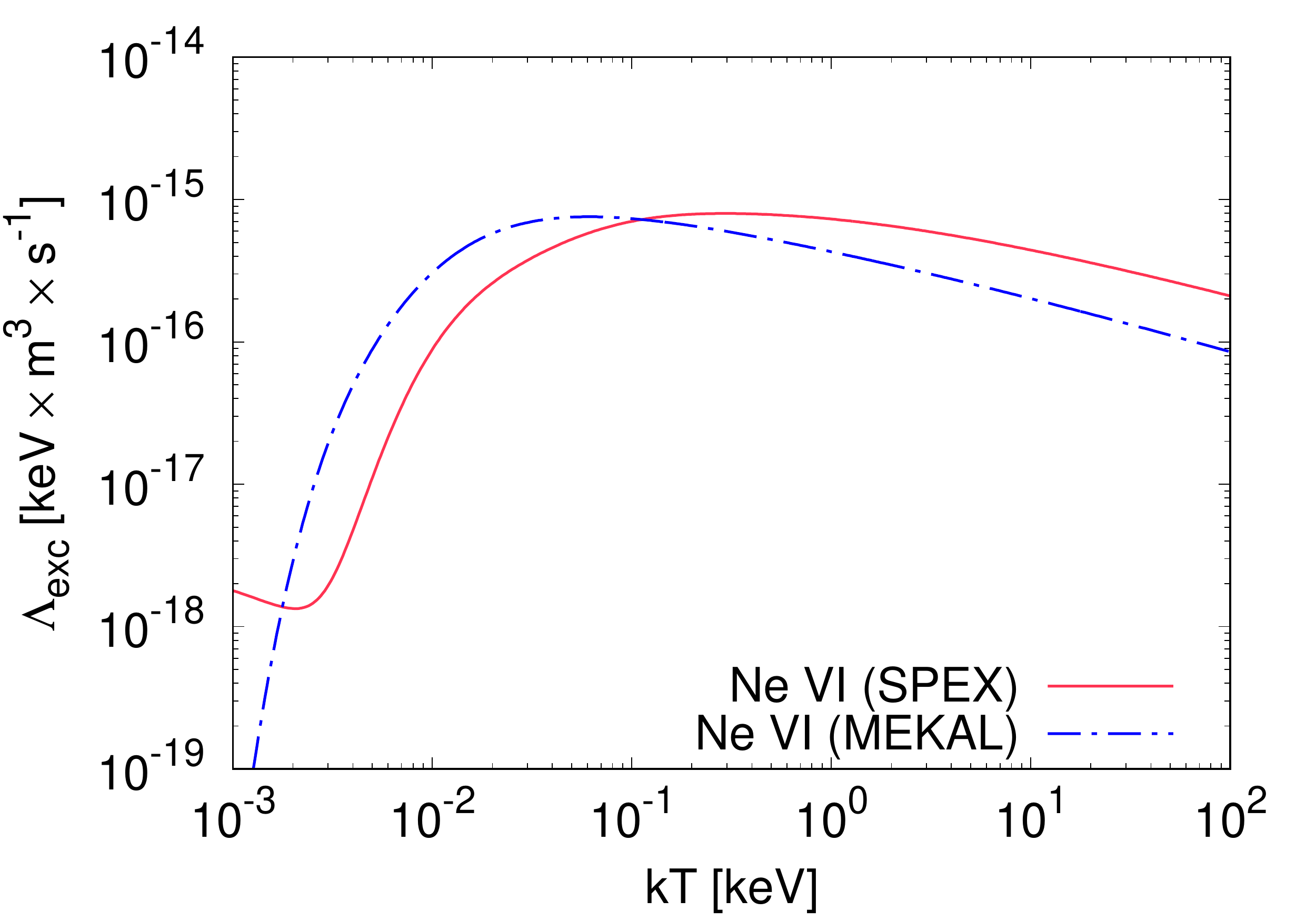} }
	\resizebox{0.85\textwidth}{!}{
		\includegraphics[angle=-90]{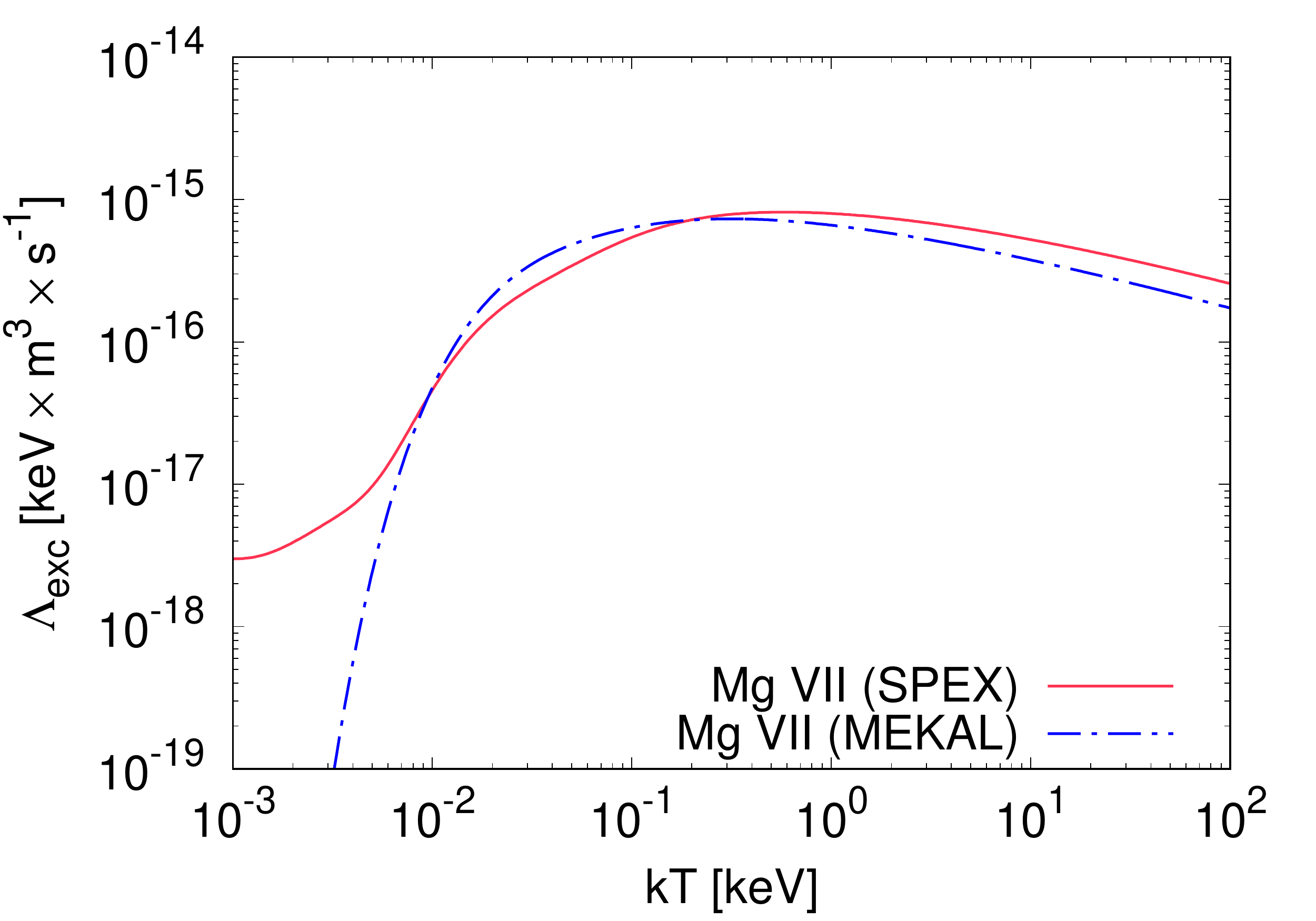}
		\includegraphics[angle=-90]{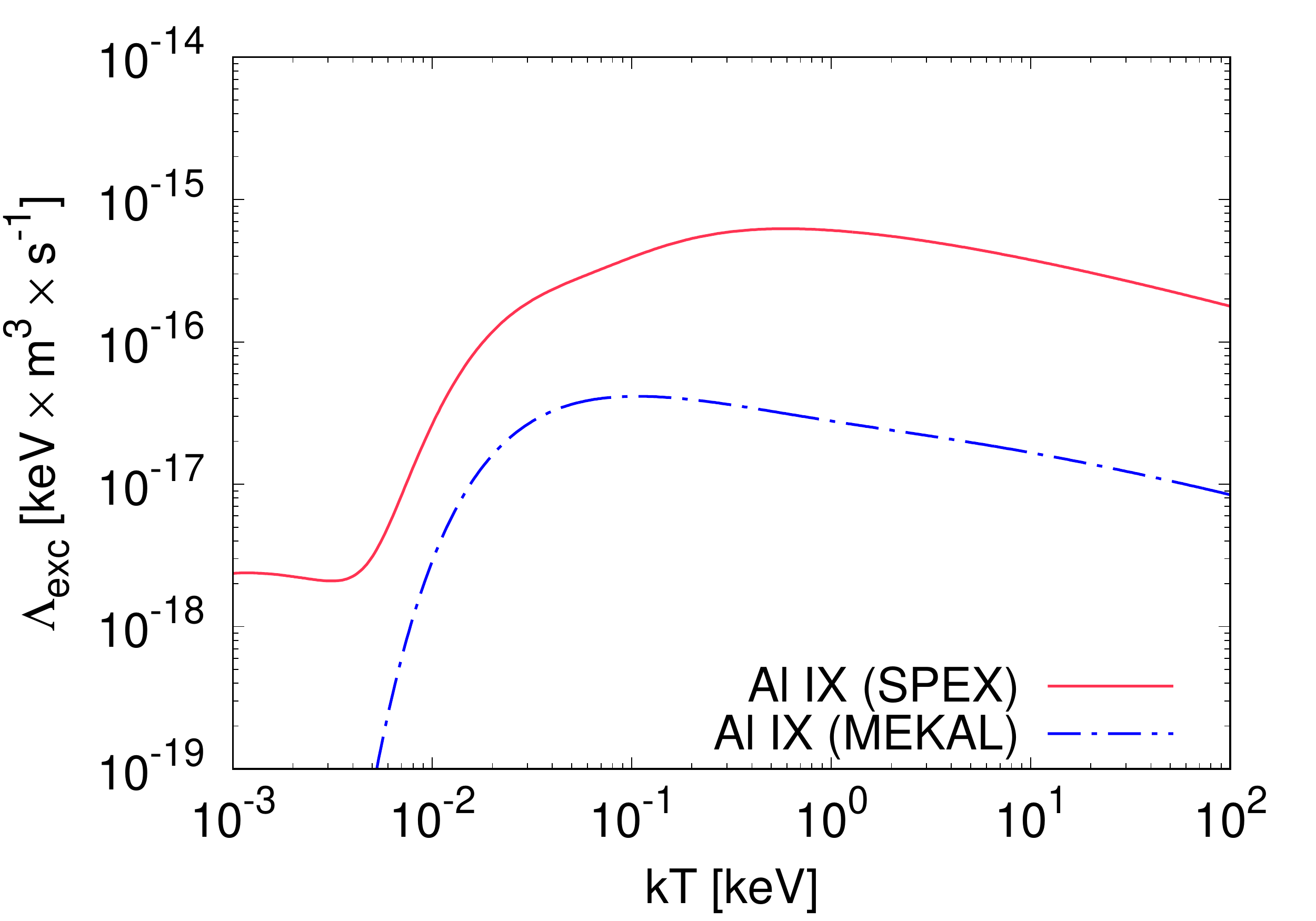}}
	\resizebox{0.85\textwidth}{!}{
		\includegraphics[angle=-90]{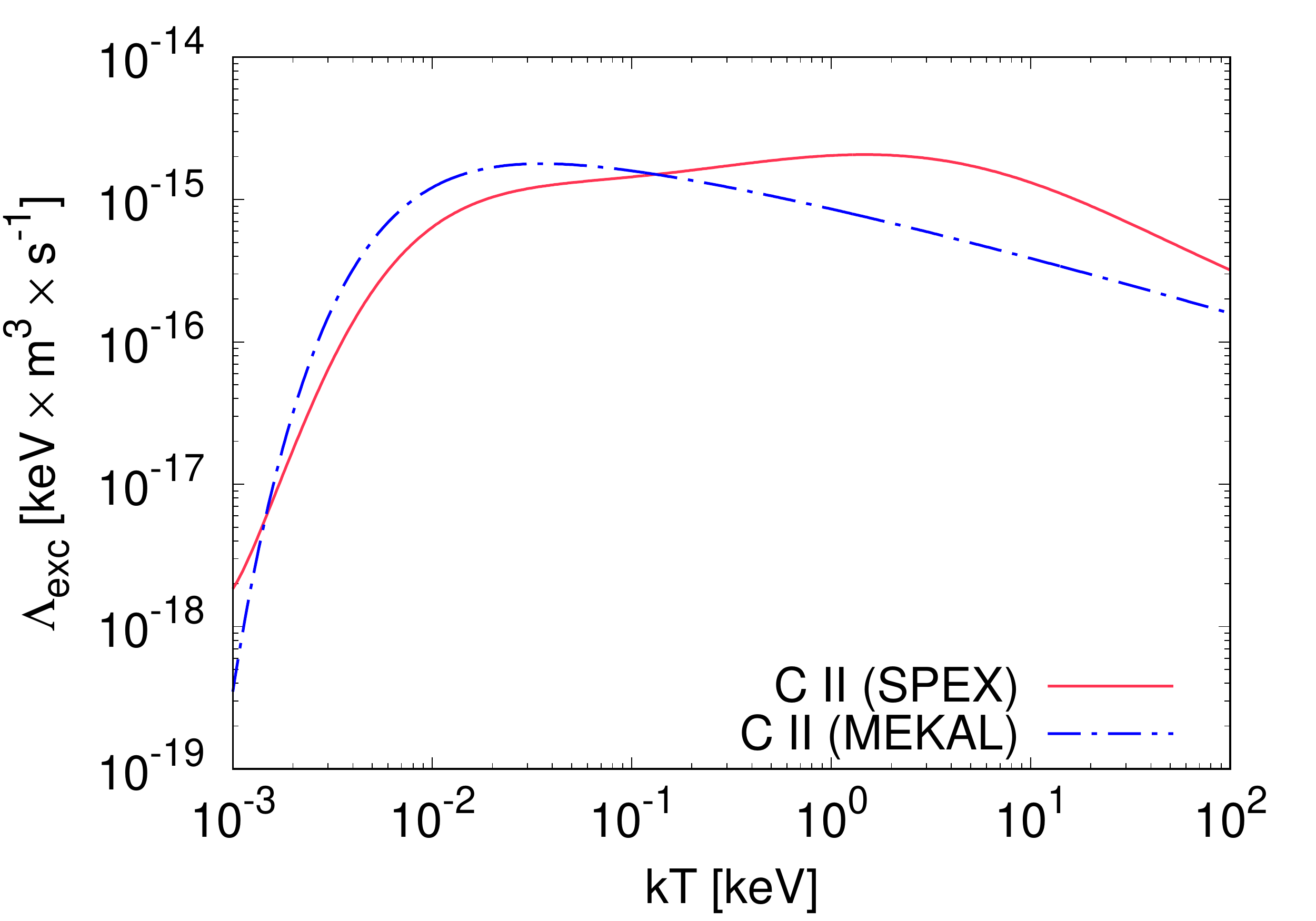}
		\includegraphics[angle=-90]{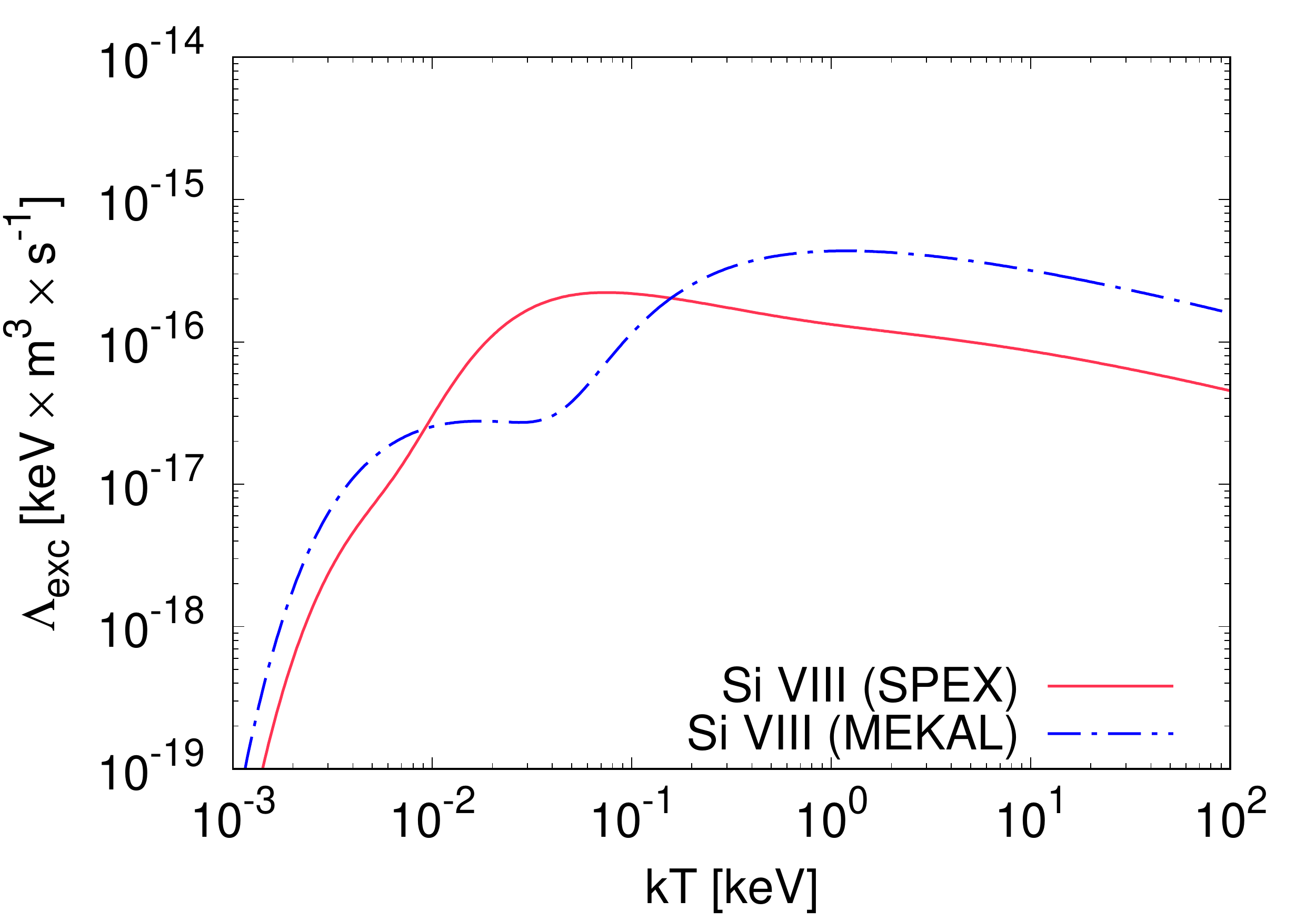}}	
	\caption{Comparison of collisional excitation radiative loss rates from SPEX version $3.06.01$ (red solid line) and MEKAL (blue dash-dotted line) for a representative sample of ions. We refer the reader to Section \ref{Sec:cool_curve_comp_with_MEKAL} for a detailed description of differences between SPEX and MEKAL for these ions.}
	\label{Fig:new_vs_old_B}
\end{figure*}

First, we calculated the radiative loss rates for SPEX (version $3.06.01$) and compared them to the radiative loss rates obtained with MEKAL \citep{1985A&AS...62..197M, 1986A&AS...65..511M}, and \citep{mewe1995update}. In this paper, we showed the comparison to \citet{2009A&A...508..751S}, which is the latest work regarding the updates of the radiative loss function in SPEX. SPEX version $2.00.11$ was based on the MEKAL line-emission model (for all elements including oxygen shown in this work as an example, except for iron, which was calculated using the HULLAC code \citealt{2001JQSRT..71..169B}) with the addition of new lines in the $200$--$2000 \, \AA$ band \citep{mewe1995update}. The main rationale for comparing our results to MEKAL is the fact that the data generated by MEKAL are still widely used while analysing the X-ray spectra. We showed how the update of the collisional excitation affected the radiative loss function. Later in the paper (see Section \ref{Sec:application-S-curve}), we also show the impact on the cooling curve in the photoionisation model pion in which the cooling by collisional excitation used to be calculated based on the MEKAL code (for SPEX version $3.05.00$ and earlier).

In SPEX, we obtained the excitation rates and the excitation energies from the SPEXACT database, and we calculated total radiative loss curves using the Eq.\,\eqref{eq:cool_rate} (we remind the reader that we only consider excitations from the ground level for now). In the case of MEKAL, we obtained the excitation energy and the emission rate (in photons/s) for each transition by excluding the contributions from other processes (such as radiative and dielectronic recombination, inner shell ionisation, proton excitations and dielectronic satellite lines), and we only took into account the contribution from the electron excitations. The total radiative loss rate for each ion was then calculated as the sum of the product of the excitation energy and the emission rate over all transitions.


%

A comparison of the total radiative loss rates per ion can be seen in Fig.\,\ref{Fig:new_vs_old_B}. We see that the updated radiative loss curves either have similar shapes but different normalisations (see Be-like ions \ion{C}{III}, \ion{O}{V}, \ion{Ne}{VII} for which the MEKAL code overestimated the cooling rates at temperatures below $10$\,keV but underestimated the cooling rates for temperatures above $10$\,keV) or the shapes of the curves vary, as can be seen for \ion{C}{II} and \ion{Si}{VIII}. Updated  radiative loss curves for ions such as \ion{Ne}{VI}, \ion{Mg}{VII}, and \ion{Al}{IX} differ from MEKAL by a few orders in the range $10^{-3}$--$10^{-2}$\,keV, which arises from the extension of the energy band to the UV and IR regimes (in comparison with the MEKAL code, which is focused on purely X-ray regime).

More specifically, for Be-like ions \ion{C}{III}, \ion{O}{V}, and \ion{Ne}{VII} the main difference between the SPEX and MEKAL radiative loss rates at low temperatures is caused by a single line for the transition from 2s2p $^1$P$_1$ to the ground state 2s$^2$ $^1$S$_0$. We illustrate this with the example of \ion{C}{III}, where the line occurs at $977$\,$\AA$. In Table\,\ref{table:CIII_line}, we show the values for the Maxwellian-averaged collision strengths $\bar{\Omega}{(T)}$ of \ion{C}{III} at $0.01$\,keV for some spectral codes.

\begin{table*}[!t]
	\centering  
		\caption{Comparison of the Maxwellian-averaged collision strengths $\bar{\Omega}{(T)}$ at $T = 0.01$\,keV of the \ion{C}{III} line at $977$\,$\AA$ (transition from 2s2p $^1$P$_1$ to the ground state 2s$^2$ $^1$S$_0$) between different plasma codes and atomic databases. }                
		\begin{tabular}{c|c|l }     
			\hline\hline 
			\rule{0pt}{2.5ex}  
			code &   $\bar{\Omega}{(T)}$ & remarks       \\  
			\hline                    
			MEKAL  &  $10.48$	&	\begin{tabular}{@{}l@{}} Based on [1] which uses a scaled value from the Fe results given by [2] using $A=0.5$, $B=0.3$ and \\  $E=0.276$ in Eq.\,A7 of that paper. This paper cites the distorted wave calculations of [3]. \end{tabular}  \\ 
			SPEX	 &  $5.93$	&	Based on older FAC\tablefootmark{a} calculations made by A.J.J. Raassen.   \\ 
			CHIANTI v. 9.0   &  $4.93$	&	R-matrix calculations [4].  \\ 
			OPEN ADAS	 &  $4.51$	& R-matrix calculations in combination with an intermediate coupling frame
				transformation [5]. \\
			AtomDB v. $3.0.9$ &  $2.71$	&	Refers to CHIANTI 6 [6], [7].  \\
			\hline                  
		\end{tabular}
	\tablebib{[1]~\citet{1990A&AS...82..229L}; [2] \citet{1985A&AS...62..197M}; [3] \citet{1981A&A...103..324B}; [4] \citet{1985ADNDT..33..195B}; [5] \citet{2014A&A...566A.104F}; [6] \citet{1997A&AS..125..149D}; [7] \citet{2009A&A...498..915D}.}
	\tablefoot{\tablefoottext{a}{Flexible Atomic Code described by \text{\citet{2002APS..APRB17075G, 2008CaJPh..86..675G}} \url{https://github.com/flexible-atomic-code/fac}.}}
	\label{table:CIII_line}      
\end{table*}

From Table \ref{table:CIII_line}, we can see that different codes arrive at values up to two times larger or smaller than the adopted SPEX value. A more careful analysis of the best available data for this transition for a broad range of temperatures is clearly needed, but it is beyond the scope of this paper.

For \ion{Ne}{VI} at low temperatures, most cooling in the SPEX code is caused by the $2$p $^2$P$_{1/2}$ - $2$p $^2$P$_{3/2}$ transition at $75622$\,$\AA$ (magnetic dipole transition); this transition is not present in the MEKAL code. At high temperatures, the main difference comes from the $2$p $^2$P$_{1/2}$ - $3$d $^2$D$_{3/2}$ transition; the excitation energy of this transition is only $101$\,eV, so at high temperatures the excitation cross-section is dominated by the high-energy limit proportional to the logarithm of the energy; in the SPEX code, which is based on calculations with the FAC code, the constant of proportionality is about two times larger than the default value of $0.276$ adopted for MEKAL.

For \ion{Mg}{VII} and \ion{Al}{IX}, similarly to the case of \ion{Ne}{VI} at low temperatures, infrared transitions between the $2$p$^2$ $^3$P$_1$ and $2$p$^2$ $^3$P$_0$ ground state dominate the cooling at low temperatures for SPEX, but these transitions are absent from the MEKAL code. In addition, the MEKAL code only contains two transitions of \ion{Al}{IX} so the lack of flux at higher temperatures is not surprising.

In \ion{C}{II}, the MEKAL code only contains six spectral lines, while the SPEX code has thousands of lines, making it more complete and rendering more flux in most cases. Finally, for \ion{Si}{VIII}, the MEKAL code concentrates a lot of flux in the so-called N1 line at $61.05$\,$\AA$, which is a conglomerate of all $2$p to $3$d transitions in this ion, including non-ground state transitions.


\subsection{Comparison with the CHIANTI and ADAS databases}
\label{Sec:comp_CHIANTI_ADAS}

\begin{figure*}[!t]
	\centering
	\resizebox{0.8\textwidth}{!}{
		\includegraphics{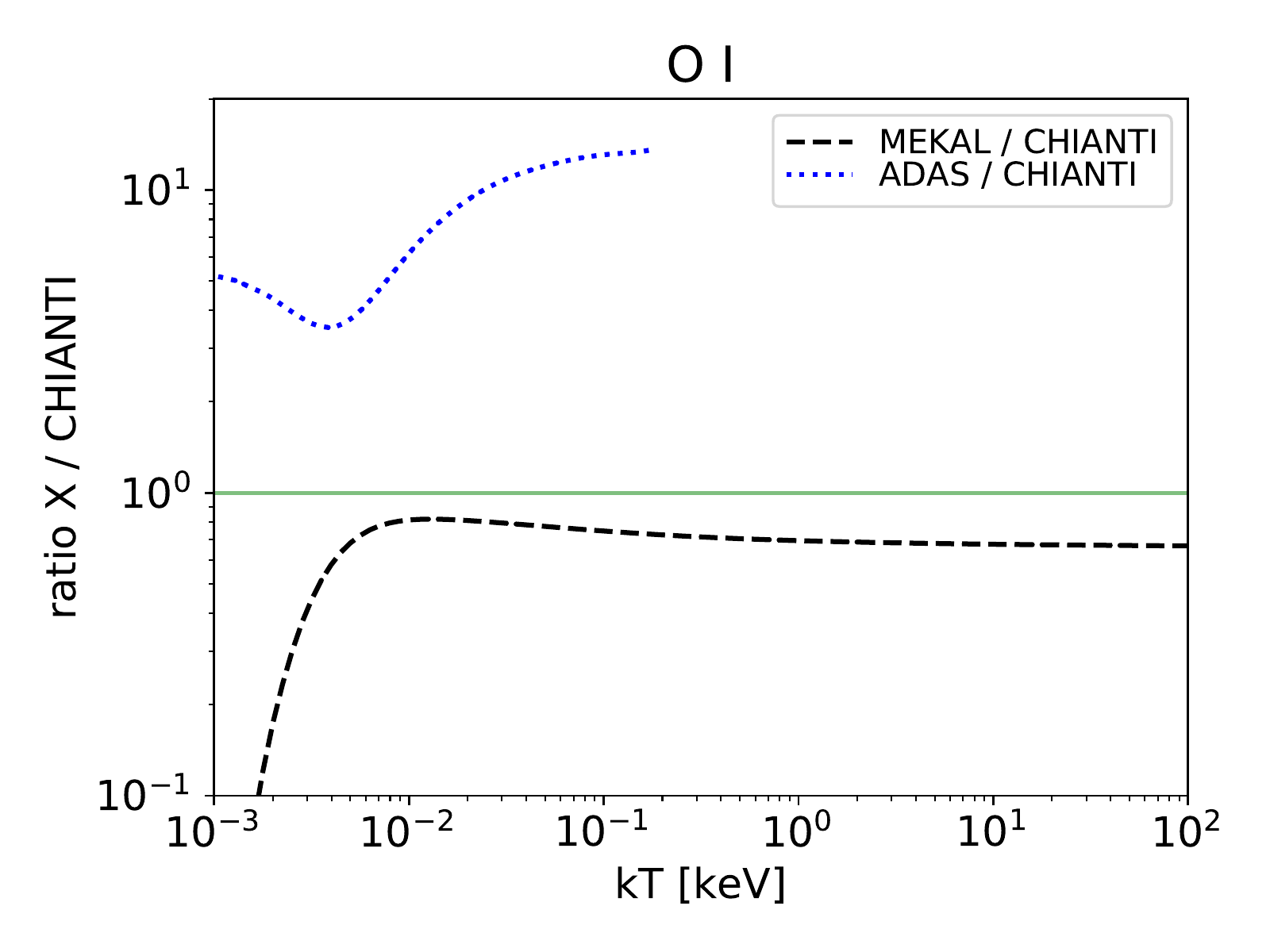}
		\includegraphics{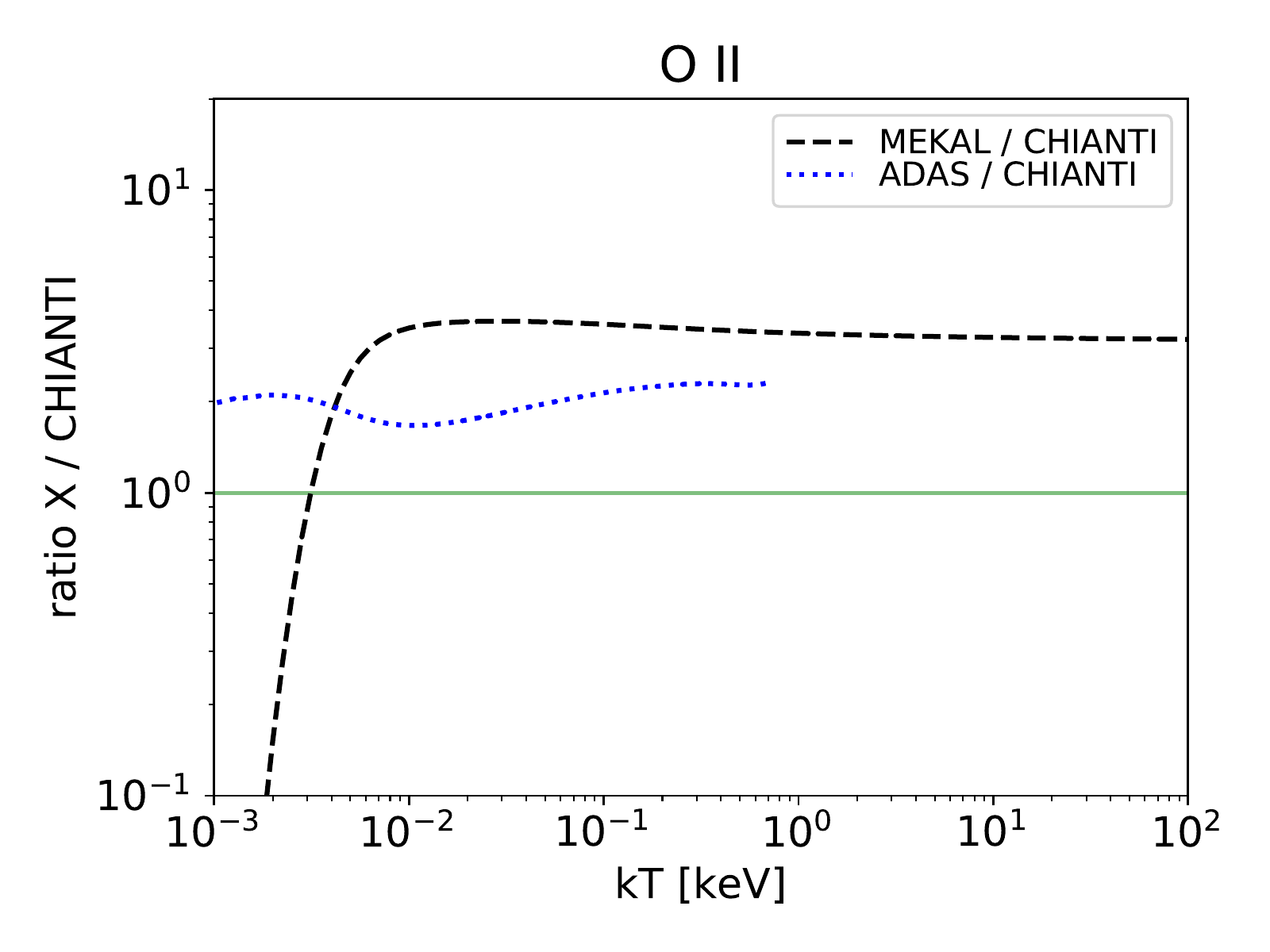} }
	\resizebox{0.8\textwidth}{!}{
		\includegraphics{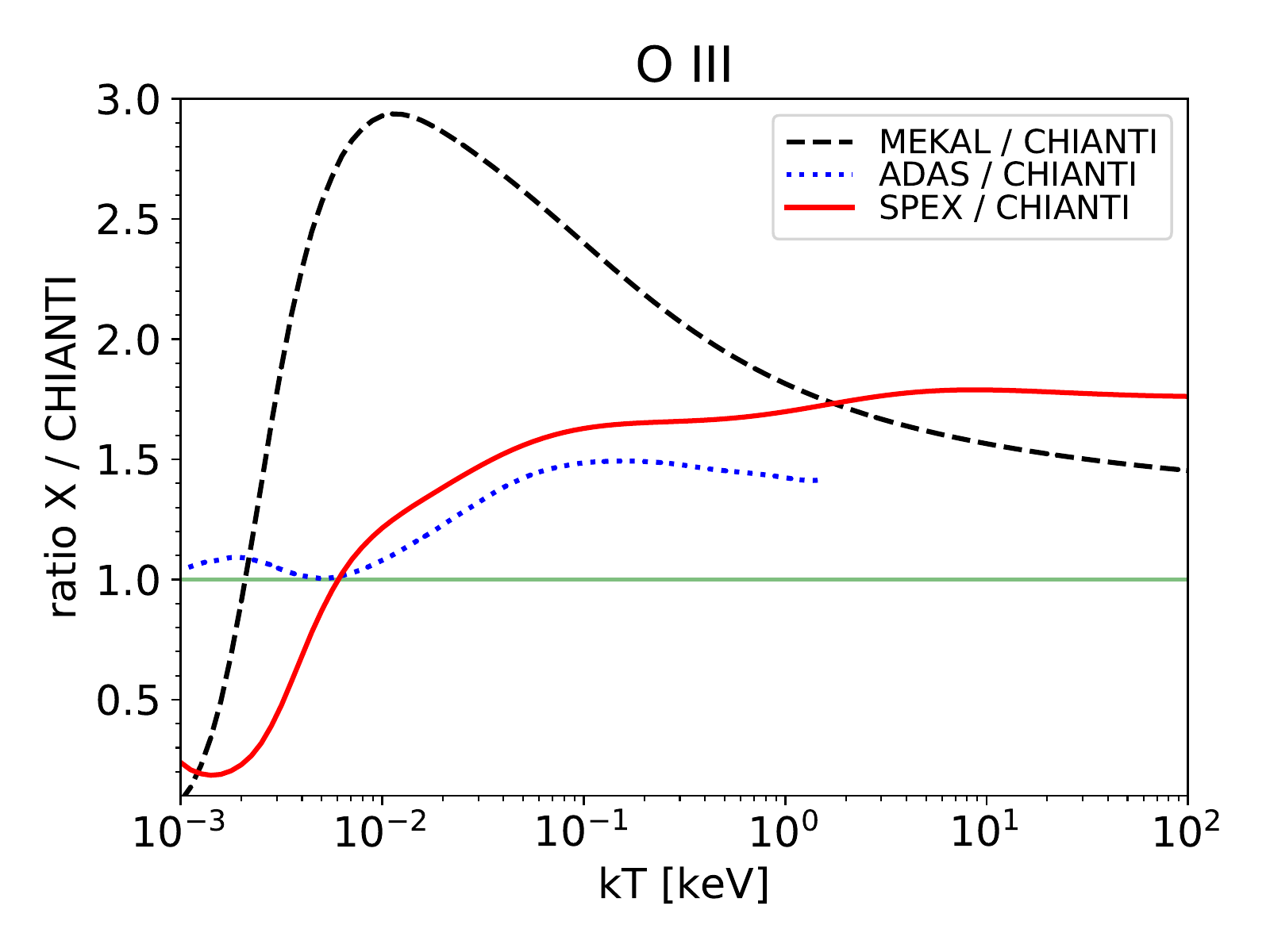}
		\includegraphics{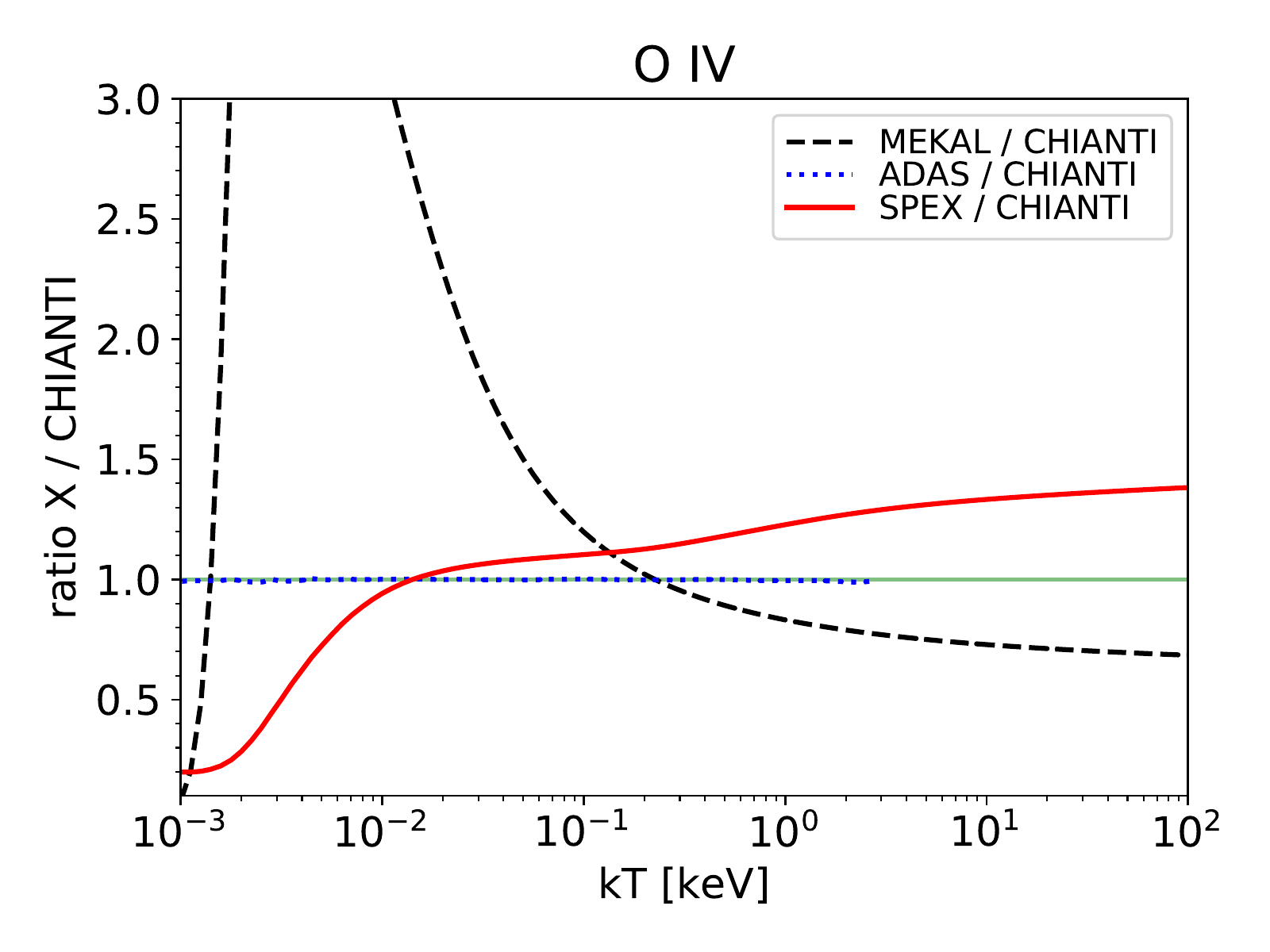} }
	\resizebox{0.8\textwidth}{!}{
		\includegraphics{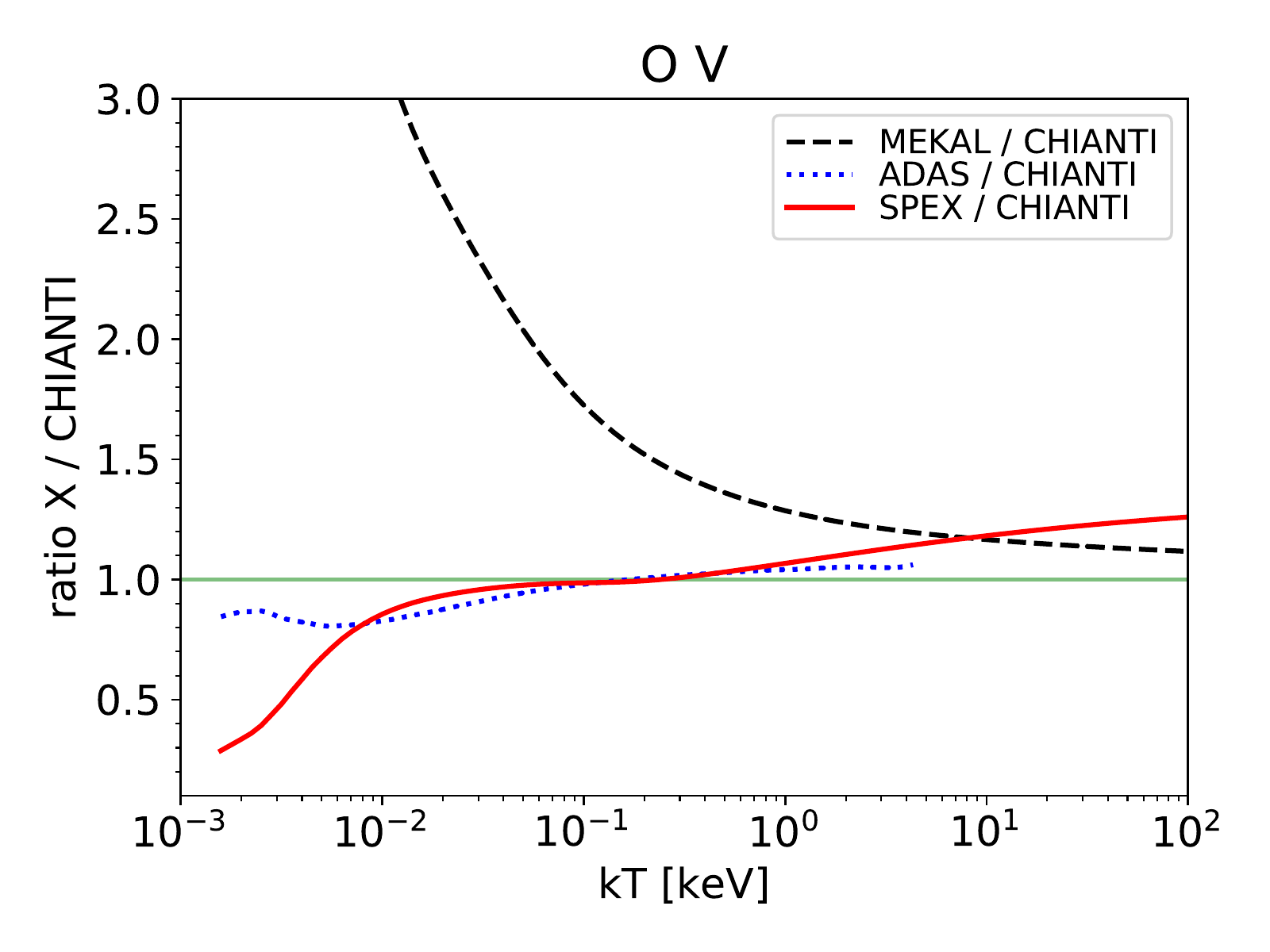}
		\includegraphics{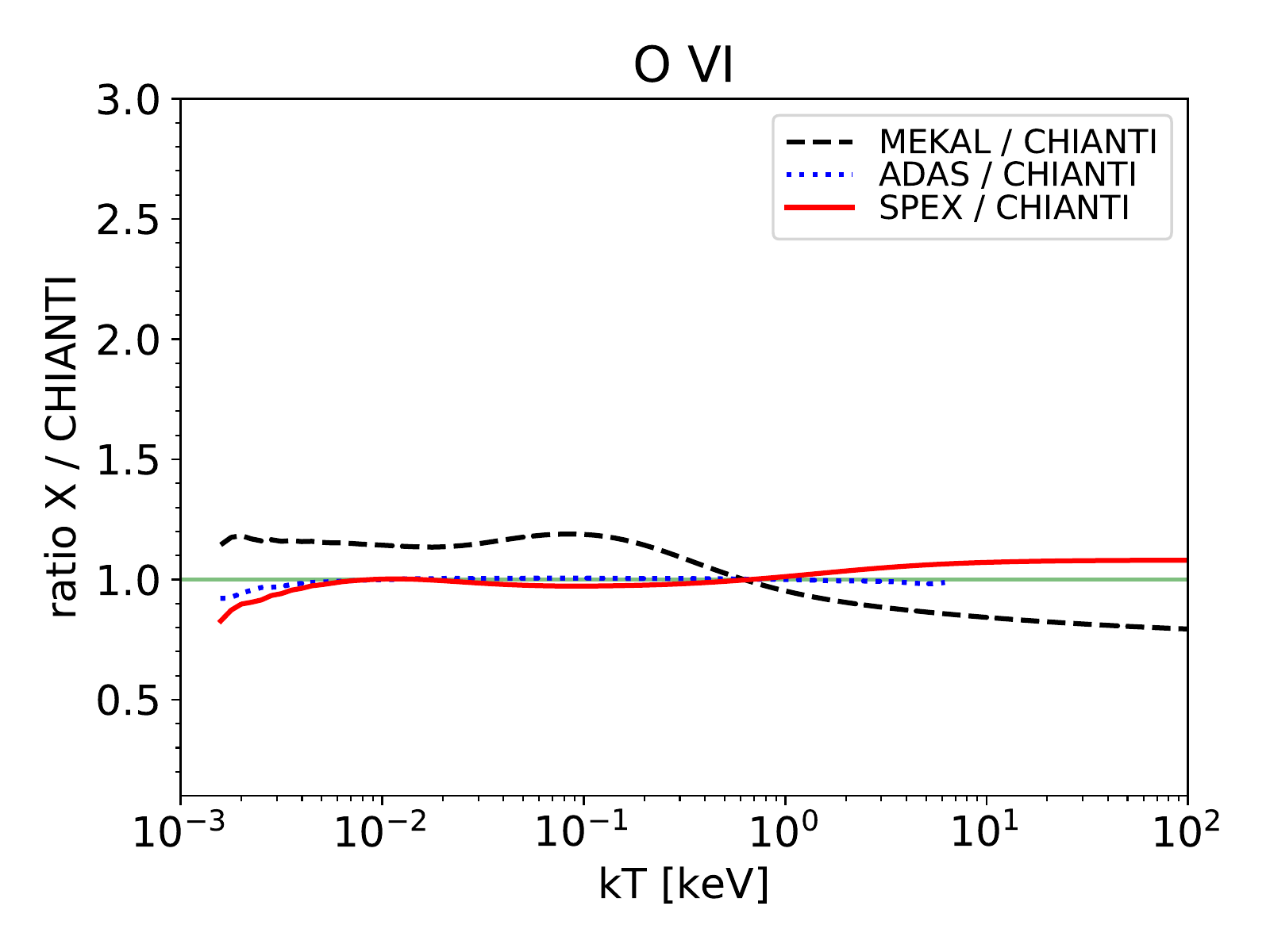} }
	\resizebox{0.8\textwidth}{!}{
		\includegraphics{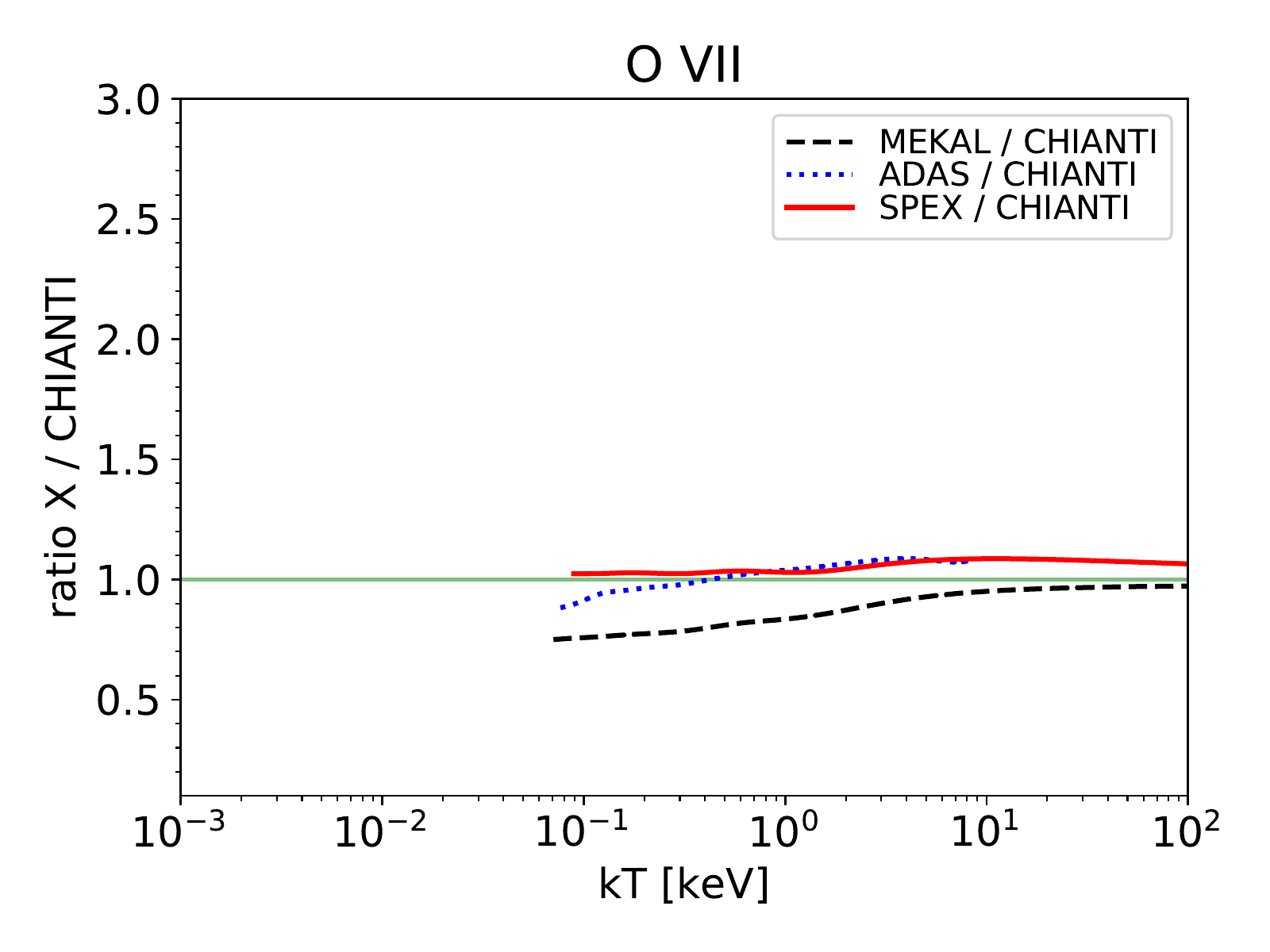}
		\includegraphics{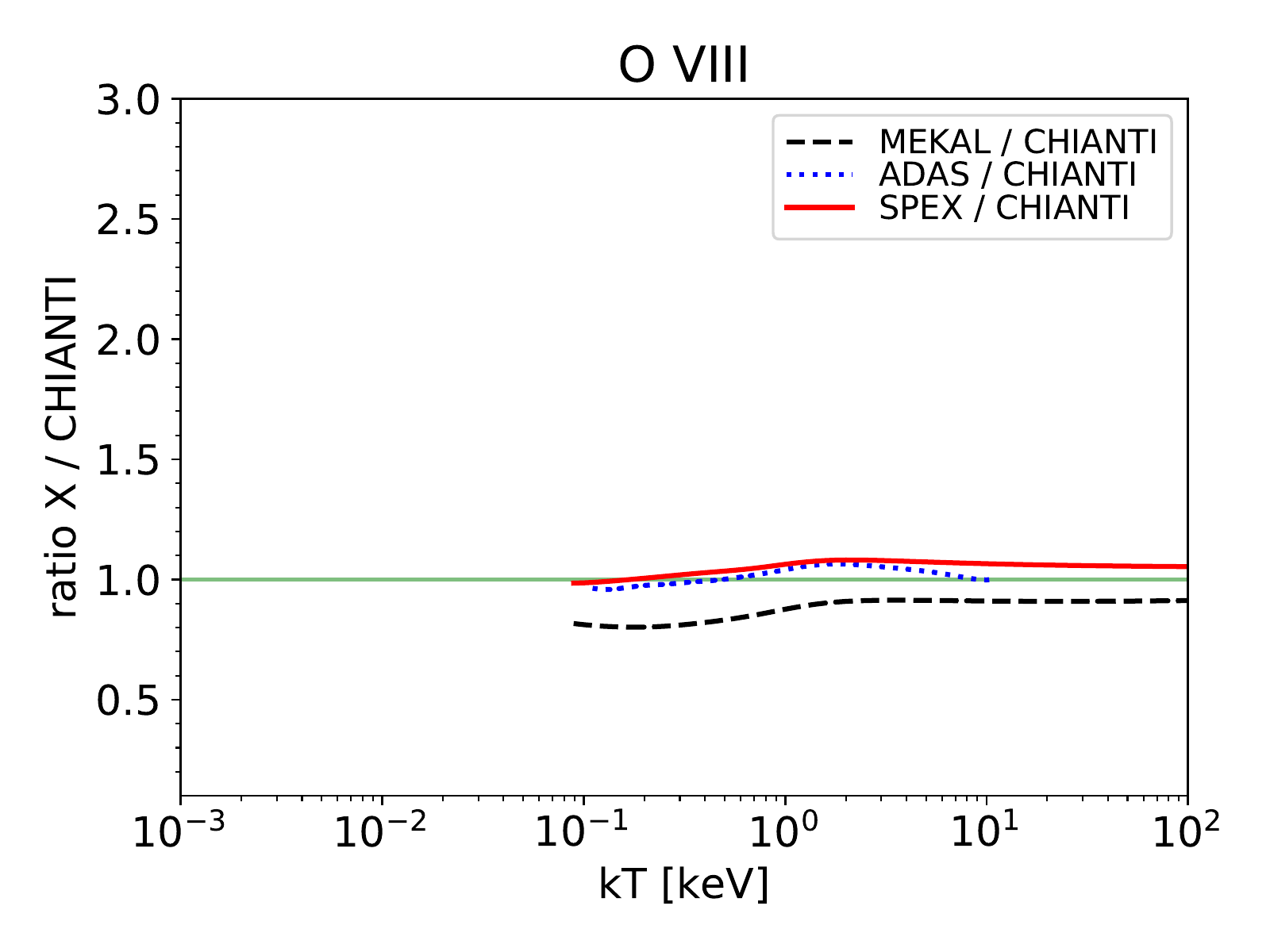} }
	\caption{Radiative loss rates due to collisional excitation for SPEX (red solid line), MEKAL (black dashed line), and ADAS (blue dotted line) plotted relatively to CHIANTI for \ion{O}{I} -- \ion{O}{VIII} ions. We refer the reader to Section \ref{Sec:comp_CHIANTI_ADAS} and Tables \ref{table:ADAS_files} (ADAS) and \ref{table:SPEX_CHIANTI_files} (SPEX and CHIANTI) for more detailed descriptions of the differences seen in these databases.}
	\label{Fig:comp_CHIANTI_ADAS}
\end{figure*}

For the series of oxygen ions (\ion{O}{I} -- \ion{O}{VIII}) plotted in Fig.\,\ref{Fig:comp_CHIANTI_ADAS}, we showed how the radiative loss rates obtained from different atomic databases and codes (SPEX/MEKAL/ADAS) differed from CHIANTI (except for \ion{O}{I} and \ion{O}{II} ions, which are not in the SPEX database). We summarised the references for oxygen ions in Tables \ref{table:ADAS_files} (ADAS) and \ref{table:SPEX_CHIANTI_files} (SPEX and CHIANTI).

For cases of highly ionised oxygen (\ion{O}{VI} to \ion{O}{VIII}) and higher temperatures, SPEX and CHIANTI agree within $10$\%, but for \ion{O}{V} the discrepancies rise to $\sim 30$\%. For lower temperatures around $10^{-3}$\,keV, radiative loss rates differ by $75$\% for \ion{O}{V} and $26$\% for \ion{O}{VI} \footnote{Data for radiative loss rates at these temperatures for \ion{O}{VII} and \ion{O}{VIII} are below the threshold of $10^{-20}$\,keV\,m$^3$\,s$^{-1}$ that we introduced due to numerical uncertainties.}. The discrepancies between SPEX and CHIANTI at high temperatures are slightly bigger for \ion{O}{III} and \ion{O}{IV}. To understand where these differences between SPEX and CHIANTI are coming from, we summarised the collision strengths data and origins in Table \ref{table:SPEX_CHIANTI_files}. We see that the data comes from different types of calculations (for example, FAC versus R-matrix calculations), and different approaches were used to solve the Schr$\ddot{\rm o}$dinger equation (see for example, Section \RomanNumeralCaps{3} of \citealt{2007RvMP...79...79K} for more details about these approximations). The databases of SPEX and CHIANTI also include different numbers of lines and differ in the maximum principal quantum number \textit{n} that is available for the calculations.

When it comes to \ion{O}{I}, MEKAL radiative loss rates are lower than CHIANTI radiative loss rates for the whole temperature range, while for \ion{O}{II} and temperatures above $0.003$\,keV we see that the MEKAL radiative loss rates are higher than in the case of CHIANTI. For \ion{O}{I} MEKAL agrees with CHIANTI within $30$\% for all temperatures above $0.01$\,keV, whereas for \ion{O}{II} it overestimates the radiative loss rates by factor of $3$ above $0.01$\,keV. MEKAL is also an outlier for the case of higher ionised oxygen as the dataset used in MEKAL is less complete and is outdated. For \ion{O}{IV}, the MEKAL values differ substantially from the CHIANTI values, because, as for other ions, several multiplets at long wavelengths are treated as a single line and the MEKAL code has only $11$ lines for this ion.


For all ions besides \ion{O}{I} and \ion{O}{II}, the ADAS radiative loss rates agree with CHIANTI within $10$\% (for \ion{O}{VI} and temperatures around $10^{-3}$\,keV the agreement is within $20$\%). In the case of lower ionisation states of oxygen at low temperatures, we see that ADAS overestimates the radiative loss rates in comparison with CHIANTI by a factor of $2$ and $5$ for \ion{O}{II} and \ion{O}{I}, respectively.  

Overall, we see many discrepancies between SPEX, MEKAL, CHIANTI, and ADAS databases, which might be caused by several factors, for instance different approaches for calculation of the collision strengths, not accurate or even missing atomic data, different values of line energies or the amount of transitions used in the atomic database. This shows that the constant updates of the atomic physics codes and the atomic databases is needed in order to prepare for unprecedented level of details provided by the upcoming X-ray missions, namely XRISM and Athena.


\begin{table*}[!t]
	\centering       
		\caption{References, number of levels included in the database and the maximum principal quantum number \textit{n} for \ion{O}{I}-\ion{O}{VIII} data files used for calculation of the radiative loss rates in SPEX and CHIANTI.}    
        	\begin{tabular}{l|c|c|c||c|c|c}
        	\hline\hline 
        	ion & levels$_{\rm SPEX}$ & $n_{\rm SPEX}$ & reference$_{\rm SPEX}$ & levels$_{\rm CHIANTI}$ & $n_{\rm CHIANTI}$ & reference$_{\rm CHIANTI}$ 	\\ \hline
        	\rule{0pt}{2.5ex} \ion{O}{I} & -- & -- & -- & $7$ & $3$ & [1], [2], [3]	\\ 
        	\ion{O}{II} & -- & -- & -- & $35$ & $3$ & [4], [5]		\\
        	\ion{O}{III} & $1003$ & $5$ & [6] & $46$ & $3$ & [7], [8], [9], [10]	\\ 
        	\ion{O}{IV} & $627$ & $5$ & [6] & $204$ & $4$ & [11]	\\ 
        	\ion{O}{V} & $219$ & $6$ &  [6] & $166$ & $5$ &  [12], [13]		\\
        	\ion{O}{VI} & $282$ & $5$ & [6], [14] & $923$ & $8$ & [15]	\\
        	\ion{O}{VII} & $247$ & $10$ & [6], [16] & $49$ & $5$ & [17]	\\
        	\ion{O}{VIII} & $164$ & $20$ & [6], [18] & $25$ & $5$ &	[19]	\\ \hline
            \end{tabular}
   	\tablebib{[1]~\citet{1998MNRAS.293L..83B}; [2] \citet{2003ApJS..148..575Z}; [3] \citet{2004ADNDT..87....1F}; [4] \citet{2007ApJS..171..331T}; [5] \citet{ 2009MNRAS.397..903K}; [6] T. E. Raassen \tablefootmark{a} using FAC code; [7] \citet{1983ApJS...52..387A}; [8] \citet{1985A&A...146..149A}; [9] \citet{1993ADNDT..54..133B}; [10] \citet{1994A&AS..103..273L}; [11] \citet{2012A&A...547A..87L}; [12] \citet{1990ADNDT..44..133K}\tablefootmark{b}; [13] \citet{2013ApJ...763...86L}; [14] \citet{1990ADNDT..44...31Z}; [15] \citet{2011A&A...528A..69L}; [16] \citet{1983ADNDT..29..467S}; [17] \citet{2001JPhB...34.3179W}; [18] \citet{1991JPhB...24.4583A}; [19] \citet{2003JPhB...36.3707B}.  }
   \tablefoot{\tablefoottext{a}{Data are public as a part of the SPEX plasma code and can be downloaded here \url{https://spex-xray.github.io/spex-help/ }.} \tablefoottext{b}{These data were updated by Berrington, K.A. More information can be found in \citet{2006ApJS..162..261L}.}}
         	\label{table:SPEX_CHIANTI_files}      
\end{table*}



\subsection{Updated dielectronic recombination and its contribution to overall cooling}
\label{Sec:dielectronic_recombination}

In this paper, we put emphasis on the role of cooling due to radiative losses caused by collisional excitation. In order to quantify its role in astrophysical situations, we also studied the plasmas in photoionisation equilibrium in Section \ref{Sec:application-S-curve}, where we described the plasma by the photoionisation model pion. However, a fraction of the total cooling may also be carried by dielectronic recombination (DR), which was ignored in the previous versions of pion \citep{2016A&A...596A..65M}. Since then, we have included the radiative losses caused by DR in the pion model. For other processes that lead to cooling and are included in SPEX, we invite the reader to consult Section \ref{Sec:S-curve_SPEX}.

We implemented the radiative loss rates from DR in SPEX as follows: for ions for which the new collisional calculations were available, the radiative loss rate was obtained from the product of the DR rate that was used for the computation of the occupation of the excited levels and the corresponding energy of the auto-ionising level. For all other ions, we used the DR contribution as taken from the MEKAL code. 

In all practical cases presented in this paper, we found that the DR did not contribute by more than a few percent to the total cooling rate.

\subsection{Implementation of updates to PIE model in SPEX}
\label{Sec:implementation_to_SPEX}

\begin{figure}[!t]
	\centering
	\includegraphics[angle=-90, width=0.5\textwidth]{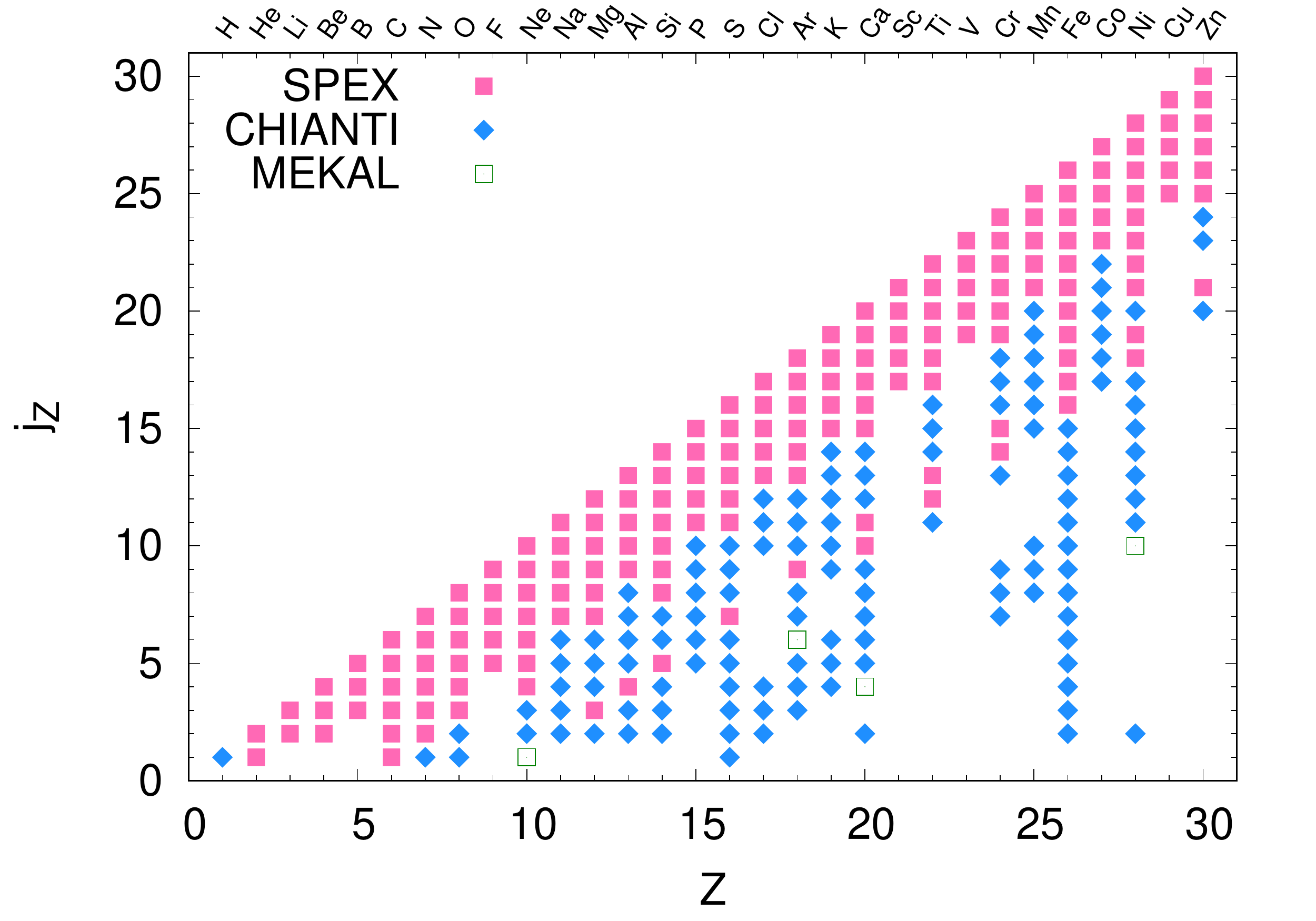}
	\caption{Final set of ions used for the tables implemented to SPEX model pion. Atomic number \textit{Z} is plotted on the x-axis, while the y-axis represents the ionisation stage $j_Z$ (e.g. \ion{H}{I} is neutral hydrogen). }
	\label{Fig:all_used_ions}
\end{figure}

In SPEX versions $3.05.00$ and earlier, in the photoionisation model pion the radiative loss by collisional excitation was calculated using the MEKAL code. Now we update these calculations by applying the results from the previous sections assuming the plasma is in the low-density regime. This new model is then used for calculations in Section \ref{Sec:application-S-curve}, where the sample of ions used for the updated radiative loss by collisional excitation is shown in Fig.\,\ref{Fig:all_used_ions}. We prioritised ions from the SPEX database and extended our sample with ions from the CHIANTI database. The ions such as \ion{Ne}{I}, \ion{Ar}{VI}, \ion{Ca}{IV}, \ion{Ni}{X} were included from the MEKAL database. In order to facilitate the interpolation, we rescaled $\Lambda_{\rm exc}$ according to the following expression:
\begin{equation}
f(T) = \Lambda_{\rm exc} \sqrt{kT} \exp{\left( \dfrac{E_{\rm eff}}{kT} \right)}		\;,
\label{eq:f_T}
\end{equation}
where $E_{\rm eff}$ is the excitation energy from the ground level to the second level. These rescaled radiative loss rates per ion were then used for the final interpolation to get the total radiative loss curve expressed on a chosen temperature grid. We interpolated $\log f$ as a function of $\log T$ using three-point Lagrangian interpolation for a grid with spacing $0.05$ in $\log T$ between $10^{-3}$\,keV and $100$\,keV and linear extrapolation in $\log f - \log T$ space outside this range in order to secure an accurate interpolation.

\subsection{Contribution from metastable levels}
\label{Sec:metastable_lev}
In the previous sections, we analysed the radiative loss rates by the collisional excitation for the transitions from the ground level to upper levels in the low-density plasma regime.

In this section, we discuss the contribution of the metastable levels to the overall radiative loss. This plays an important role in the high-density plasmas. We showed how the radiative loss by collisional excitation from the metastable levels differed in comparison to the radiative loss by collisional excitation from the ground level. If the density of plasma is high enough, the metastable levels can be highly populated and may contribute to the total radiative loss significantly, even more so than the ground level. For example, \citet{2017A&A...607A.100M} found that if the hydrogen number density $n_{\rm H} \gtrsim 10^{14} \textnormal{m}^{-3}$, the first three excited levels of Be-like ions (2s2p $^3$P$_{0-2}$) are significantly populated ($\gtrsim$1\%  of the ground level population). However, for the same density the metastable levels of Be-like ions are less populated than the metastable levels of B-like and C-like ions.
	
SPEX can be used in the high-density regime for the calculation of spectra in collisional ionisation equilibrium, but it has not been updated yet to include the high-density regime in the calculation of the ionisation balance. Therefore, the present calculations of the radiative loss rates from the metastable levels are not yet used in the current version of SPEX. We note here that, for higher densities, the effect of collisional de-excitation in the radiative loss should be taken into account. The radiative loss by collisional excitation (either from the ground level or from the metastable levels) in this paper  was calculated only in the low-density limit.
	

In Table \ref{table:metastable_levels}, we list all the lower levels from which the radiative loss rates by collisional excitation were calculated for the set of Be-like to Cl-like iso-electronic sequences besides Ne and Na (level 1 represents the ground level). For example, if we calculate the radiative loss rates from level 2, this effectively means taking into account transitions from the second level to all upper levels available in the SPEX database.

\begin{table}[!t]
	\centering       
	\caption{Metastable levels (in energy order) for which the radiative loss rates were calculated (level 1 represents the ground level). }                
	\begin{tabular}{c c || c }     
		\hline\hline              
		\multicolumn{2}{c}{iso-electronic sequence} & levels\\
		\hline                    
		\rule{0pt}{2.5ex} Be-like  &  Mg-	&	$1$--$4$   \\
		B-like	 &  Al-	&	$1$--$2$   \\
		C-like   &  Si-	&	$1$--$5$   \\
		N-like	 &  P-	&	$1$--$5$   \\
		O-like	 &  S-	&	$1$--$5$   \\
		F-like	 &  Cl-	&	$1$--$2$   \\
		\hline                  
	\end{tabular}
	\label{table:metastable_levels}      
\end{table}

\begin{figure}[!t]
	\centering
	\includegraphics[angle=-90, width=0.46\textwidth]{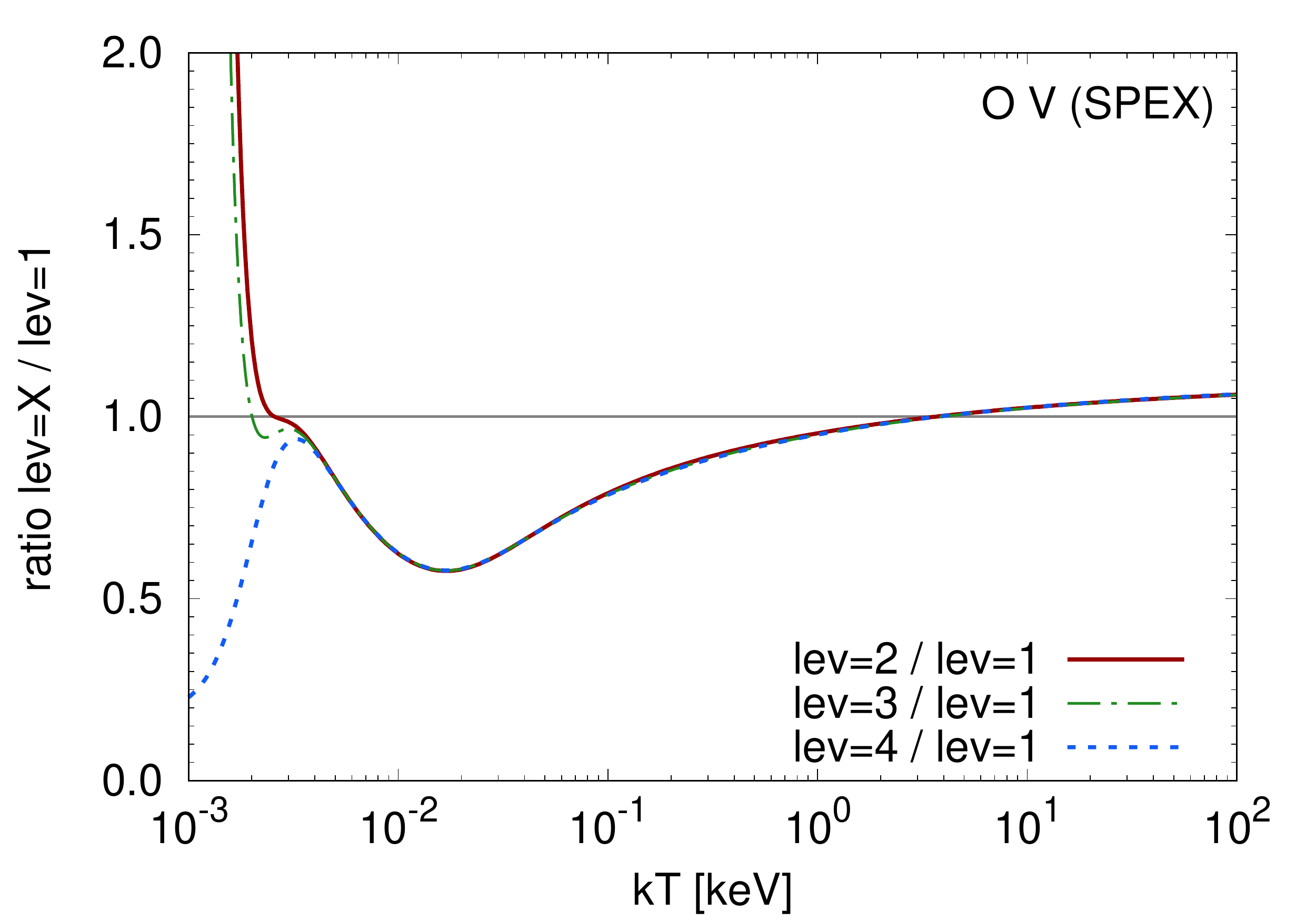} 
	\caption{Contribution of metastable levels to the radiative loss rates shown relatively to the radiative loss rates only from the ground level for Be-like oxygen. The red solid line, green dash-dotted line, and blue dotted line represent the ratio of the total radiative loss rate from the second level (lev=2), the third level (lev=3), and the fourth level (lev=4) relatively to the radiative loss from the ground level (lev=1), respectively. The definition of metastable levels can be found in Table \ref{table:metastable_levels_definition_OV}. }
	\label{Fig:matastable-levels_O_V}
\end{figure}

In Fig.\,\ref{Fig:matastable-levels_O_V}, we show the contributions from metastable levels relatively to the contribution from the ground level using the SPEX database in the example of \ion{O}{V} (the definition of metastable levels for \ion{O}{V} can be found in Table \ref{table:metastable_levels_definition_OV}). For $kT \sim 0.002$\,keV, the contribution from the second and third levels is higher by more than a factor of two in comparison with the contribution from the ground level and on contrary the contribution from the fourth level is lower than the contribution from the ground level. Towards higher temperatures, the ratios decrease with a local minimum around  $kT$ $\sim$ $0.01$--$0.02$\,keV. At very high temperatures above $3$\,keV, the curves indicate a higher contribution to the radiative loss from metastable levels than from the ground level.

\begin{table}[t]
	\centering       
	\caption{Definition of the ground level (level 1) and metastable levels (levels 2--4) and the energies for \ion{O}{V} in SPEX.}                
	\begin{tabular}{c l c || c c c }     
		\hline\hline              
		level & configuration & term & energy  \\
		 & &  & [$\times 10^{-2}$\,keV]  \\
		\hline                    
		\rule{0pt}{2.5ex} $1$ & $1 \textnormal{s}^2$ $2 \textnormal{s}^2$                       & $^{1} \textnormal{S}_{0}$ & $0.000$ \T\B \\  
		$2$ & $1 \textnormal{s}^2$ $2 \textnormal{s}^1$ $2 \textnormal{p}^1$  & $^{3} \textnormal{P}_{0}$ & $1.038$  \T\B \\
		$3$ & $1 \textnormal{s}^2$ $2 \textnormal{s}^1$ $2 \textnormal{p}^1$  & $^{3} \textnormal{P}_{1}$ & $1.039$  \T\B \\
		$4$ & $1 \textnormal{s}^2$ $2 \textnormal{s}^1$ $2 \textnormal{p}^1$  & $^{3} \textnormal{P}_{2}$ & $1.043$  \T\B \\
		\hline                  
	\end{tabular}
	\label{table:metastable_levels_definition_OV}      
\end{table}



\subsection{Calculation of the total cooling curve in SPEX and the comparison to MEKAL, Cloudy, and APEC}
\label{Sec:total_cooling_curve}

\begin{figure}[!t]
	\centering
	\includegraphics[ width=0.54\textwidth]{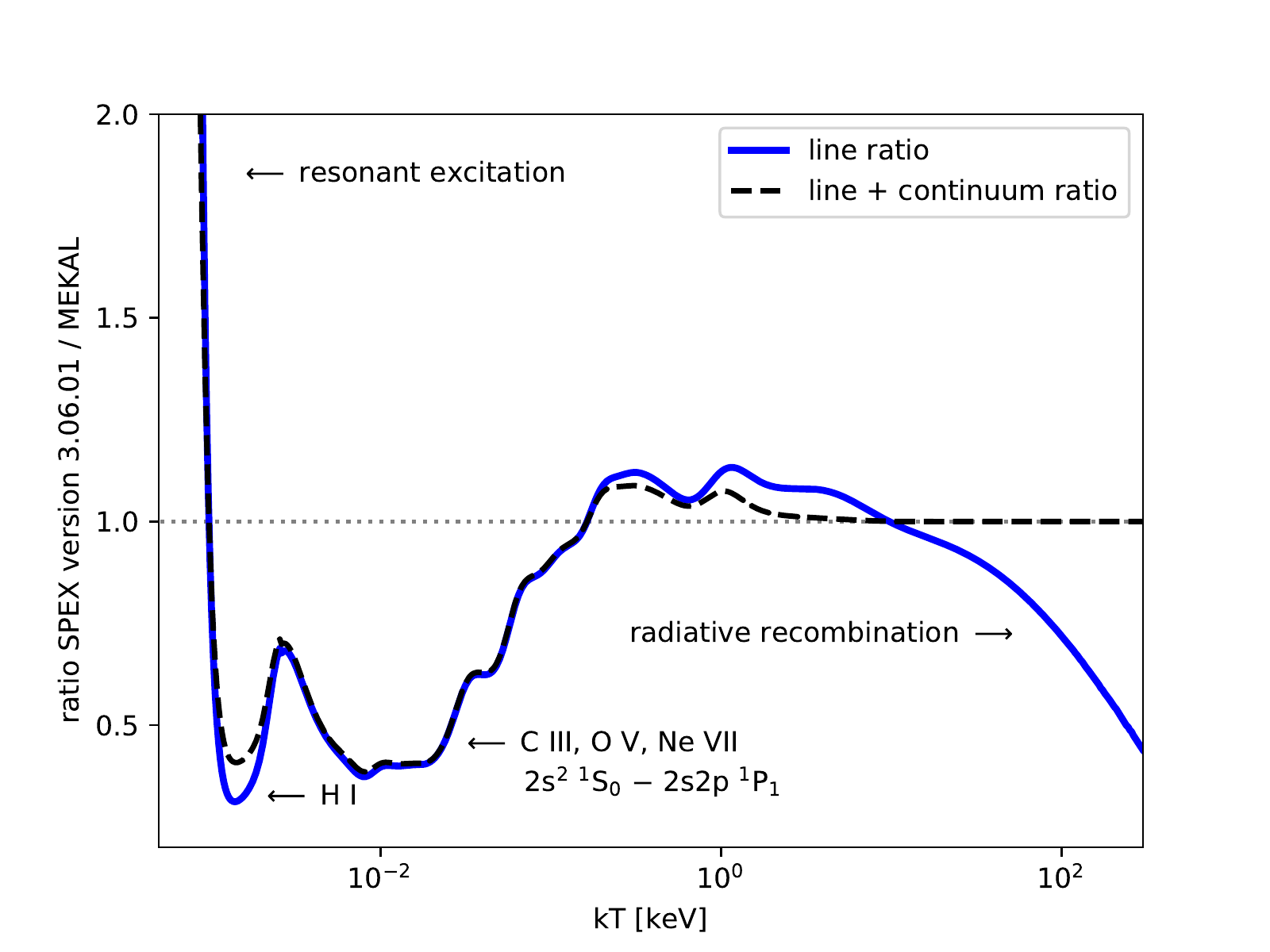}
	\caption{Ratio of radiative loss rates due only to line radiation for the SPEX version $3.06.01$ and MEKAL as a function of temperature (blue solid line) and for both line and continuum radiation (black dashed line). The four biggest differences are pointed out by black arrows and are caused by updates of resonant excitation ($kT < 10^{-3}$\,keV), collisional excitation ($2 \times 10^{-3} < kT < 10^{-1}$\,keV), radiative recombination ($kT > 10$\,keV), and \ion{H}{I} collision strengths ($kT \sim 1.5 \times 10^{-3}$\,keV).}
	\label{Fig:tot_cooling_lines}
\end{figure}

Fig.\,\ref{Fig:tot_cooling_lines} shows the difference between the SPEX version $3.06.01$ and the MEKAL code for the total radiative loss due only to line radiation (solid blue line) and for both line and continuum radiation (black dashed line). We can divide the figure into four sections: (a) $kT > 10$\,keV, where the updates of the radiative recombination were already done by \citet{2017A&A...599A..10M}; (b) $2 \times 10^{-3} < kT < 10^{-1}$\,keV, with the most significant difference caused by the updated database for collisional excitation (this paper); (c) $kT < 10^{-3}$\,keV, where the difference is mainly caused by resonant excitation; and (d) $kT \sim 1.5 \times 10^{-3}$\,keV, where the updates of \ion{H}{I} collision strengths considering data from \citet{2000JPhB...33.1255A} and revised by \citet{2002JPhB...35.1613A} are included in the latest SPEX version. The Be-like ions \ion{C}{III}, \ion{O}{V} and \ion{Ne}{VII} (more specifically transitions from $2\textnormal{s}^2$ to $2\textnormal{s}2\textnormal{p}$) account for the most significant contributions to the dip in the $2 \times 10^{-3}$--$10^{-1}$\,keV range, where for this energy range the MEKAL calculations overestimate the total radiative loss by approximately $60$\% in comparison with SPEX.

\begin{figure}[!t]
	\centering
	\includegraphics[angle=-90, width=0.5\textwidth]{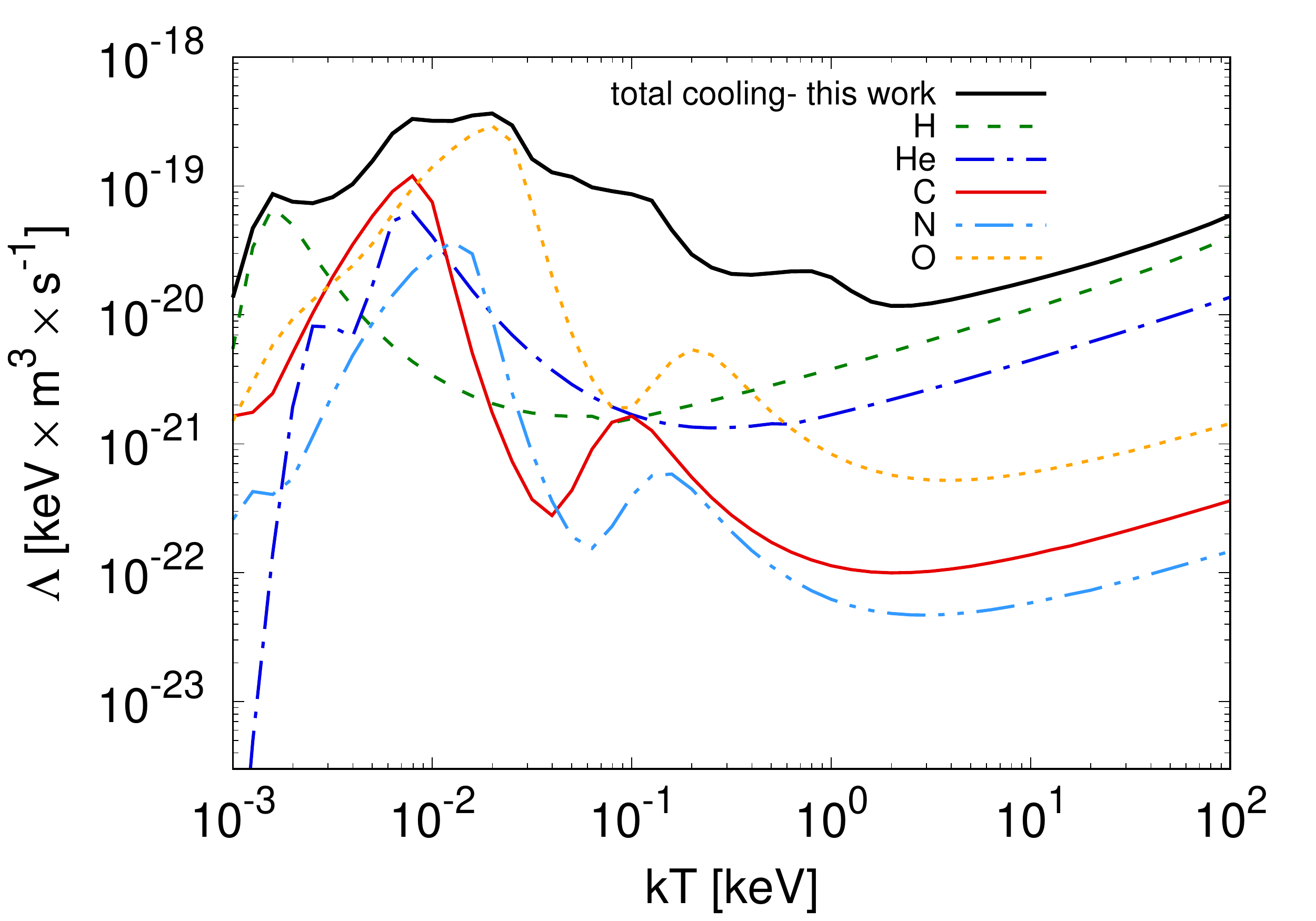} \\
	\includegraphics[angle=-90, width=0.5\textwidth]{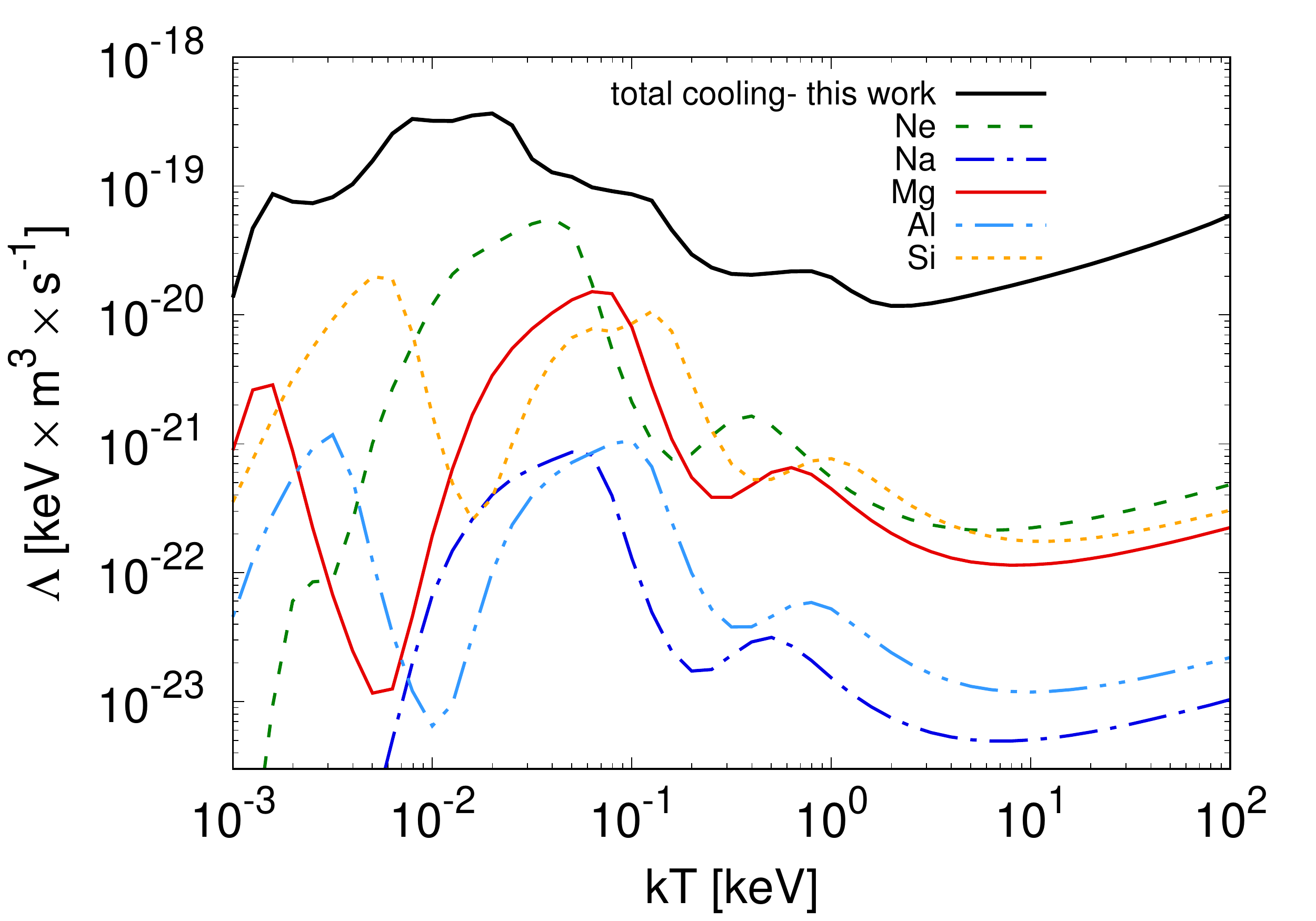} \\
	\includegraphics[angle=-90, width=0.5\textwidth]{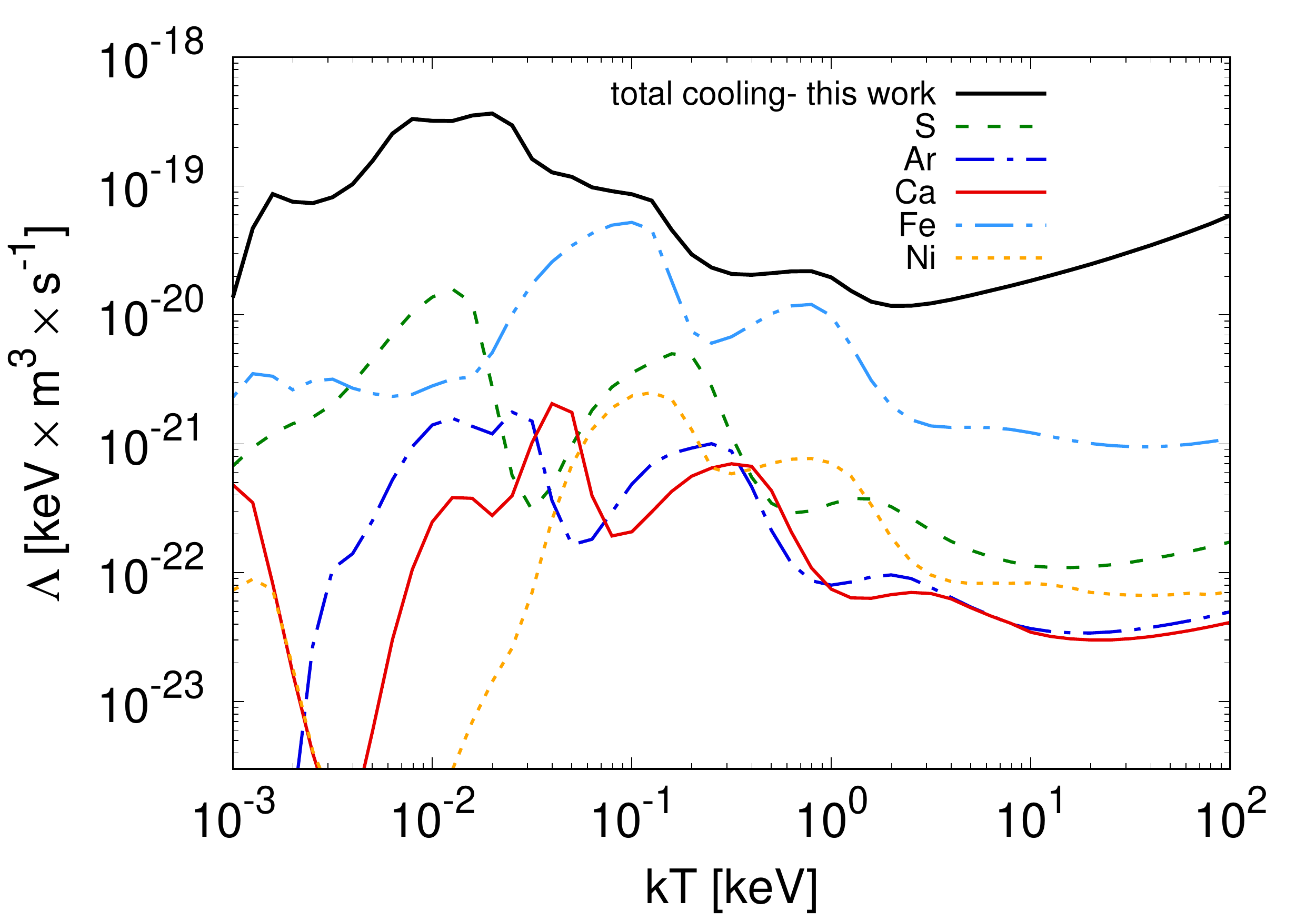}
	\caption{Total cooling curve obtained in this work (black solid line) and individual contributions to the total cooling from different elements.}
	\label{Fig:tot_cool_elem}
\end{figure}

We present the final updated total cooling curve in SPEX as well as individual contributions to this curve from different elements in Fig.\,\ref{Fig:tot_cool_elem}. We note that this is not a radiative loss curve as discussed in the sections before. We also include all transitions from lower to upper levels (as opposed to the previous calculations that considered the transitions only from the ground state to the upper levels) and continuum emission. The calculation assumes \citet{2009LanB...4B..712L} abundances and the \citet{2017A&A...601A..85U} ionisation balance. We plot contributions from the $15$ most abundant elements (H, He, C, N, O, Ne, Na, Mg, Al, Si, S, Ar, Ca, Fe, Ni) that are present in the MEKAL and SPEX database.


\begin{figure}[!t]
	\centering
	\includegraphics[angle=-90, width=0.5\textwidth]{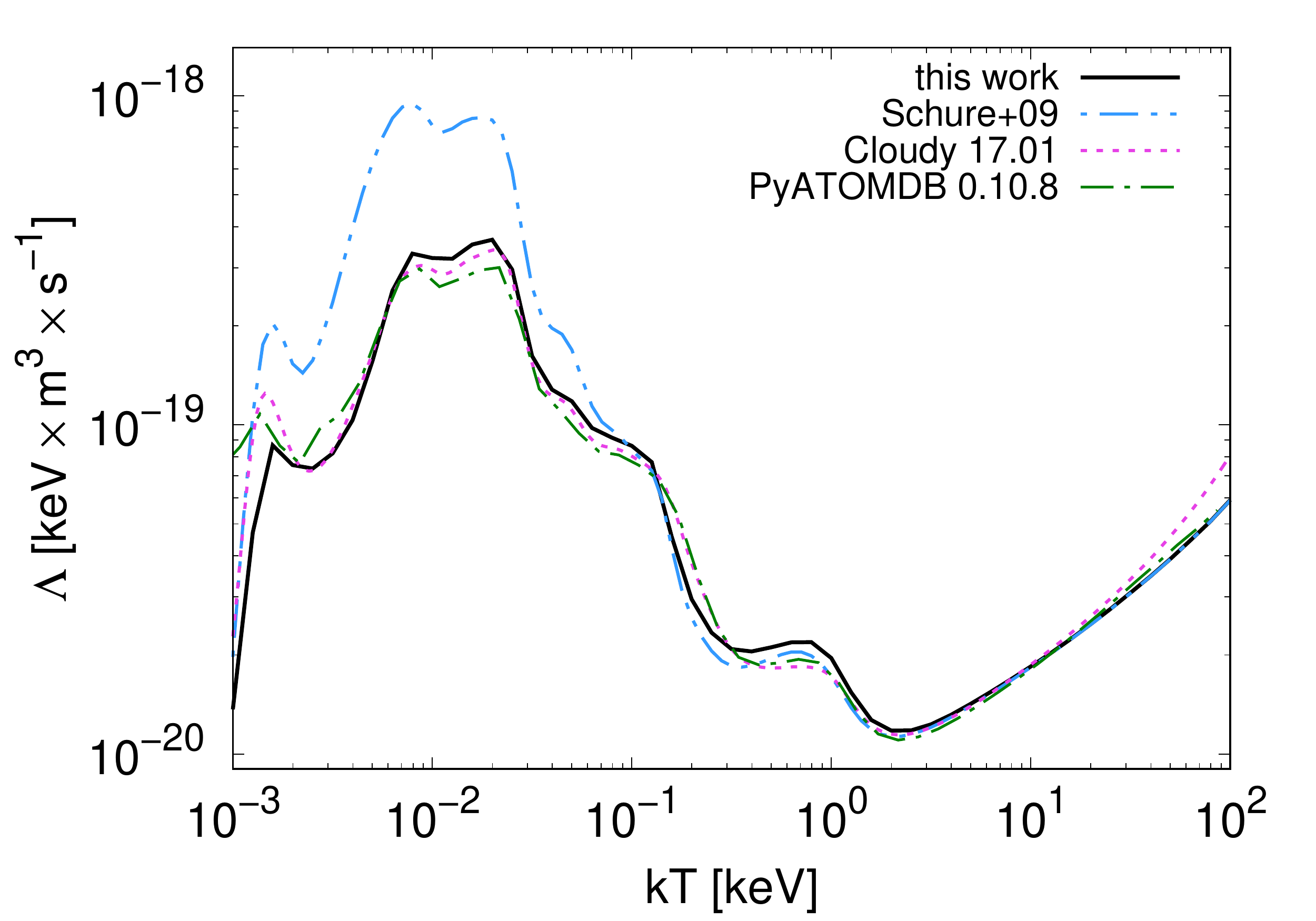}
	\caption{Comparison of the updated total cooling curve from SPEX (black solid line) and Cloudy (purple dashed line) to the total radiative loss curve from \citet{2009A&A...508..751S} (blue dash-double-dotted line) and APEC (green dash-dotted line). }
	\label{Fig:tot_cool_Schure_APEC_Cloudy}
\end{figure}

In Fig.\,\ref{Fig:tot_cool_Schure_APEC_Cloudy} we compare the updated total cooling curve from SPEX to the total radiative loss curve from \citet{2009A&A...508..751S}. We also plot the total cooling curve from Cloudy and the total radiative loss curve from APEC. The total radiative loss curve by \citet{2009A&A...508..751S} uses MEKAL calculations and assumes \citet{1989GeCoA..53..197A} abundances and the ionisation balance from \citet{1985A&AS...60..425A} for all present elements in MEKAL besides iron, for which the \citet{1992ApJ...398..394A} ionisation balance is used. The radiative loss curve from APEC \citep{2001ApJ...556L..91S} is calculated using PyATOMDB\footnote{\url{https://github.com/AtomDB/pyatomdb}} version $0.10.8$, \citet{2009LanB...4B..712L} abundances, and the \citet{2009ApJ...691.1540B} ionisation balance. We also include the total cooling curve from Cloudy for which the calculations were performed with version $17.01$, last described by \citet{2017RMxAA..53..385F}. To obtain the Cloudy curve that can be directly compared to SPEX (therefore we discuss the cooling curve and not the radiative loss curve in the following text), we carefully rescale the Cloudy cooling rates and divide them by the product of the hydrogen and the electron density. For that, we obtain the hydrogen density for the specific temperature from SPEX, use it as an input to Cloudy calculations, and as an output we get the Cloudy cooling rates as well as the electron density that Cloudy assumes for the assumed hydrogen density (for Cloudy it is necessary to give the hydrogen density to run the calculations, whereas SPEX needs the electron density for version $3.06.00$ and lower). These quantities are used for rescaling the Cloudy cooling rates from units of $[\textnormal{erg} \times \textnormal{cm}^{-3} \times \textnormal{s}^{-1}]$ to units of $[\textnormal{keV} \times \textnormal{m}^3 \times \textnormal{s}^{-1}]$ which are used in the graphs throughout this paper.

In the energy range from $2 \times 10^{-2}$ to $30$\,keV, APEC and SPEX are in reasonable agreement. 
For higher energies, the differences are negligible (approximately $2$\% at $86$\,keV), even though SPEX and APEC use different treatments of relativistic corrections to the Bremsstrahlung emission (free-free emission). SPEX uses the correction described in Eq.\,(B4) in \citet{1982ApJS...48..239K} (based on \citet{Gluckstern1953}). On the other side, APEC (to be more precise, its PyAtomDB module) offers the user to use either a non-relativistic \citep{1988ApJ...327..477H}, semi-relativistic \citep{1975ApJ...199..299K}, or relativistic \citep{1998ApJ...507..530N} version of Bremsstrahlung. In Fig.\ref{Fig:tot_cool_Schure_APEC_Cloudy}, we plot the relativistically corrected Bremsstrahlung using \citep{1998ApJ...507..530N}.

The cooling curves from SPEX (this work) and Cloudy agree reasonably well for all temperatures besides very low ones around $2$\,eV (cooling due to neutral hydrogen) and higher temperatures above $10$\,keV, where the main differences come from the relativistic corrections to Bremsstrahlung, where Cloudy uses \citet{2015MNRAS.449.2112V}. Even though both codes use the Born approximation, we still see significant differences in the relativistic corrections to Bremsstrahlung (by approximately $28$\% at $100$\,keV). However, these energies are not the main interest of our paper and therefore we did not investigate the differences in specific treatments of the relativistic corrections to Bremsstrahlung in more detail.

Our new cooling curve is now more in agreement with Cloudy, especially at temperatures around $10^{-2}$\,keV ($\sim 10^5$\,K). This is mainly due to the updates of carbon, oxygen, and neon ions. This agrees well with the findings of \citet{2013MNRAS.429.3133L}, which states that \citet{2009A&A...508..751S} finds significantly more cooling in this regime in comparison with Cloudy and that the most probable reasons for that are the differences in the atomic data for oxygen and carbon, since these are the main coolants for these temperatures.

We also improve the cooling for the temperatures around $2.5$\,eV where we find significant discrepancies between SPEX and other codes. For this temperature, the bulk of the difference between Cloudy and SPEX comes from the cooling by \ion{H}{I} from which 71\% is produced by Lyman $\alpha$ lines and 8\% is produced by Lyman $\beta$ (considering a CIE plasma). The SPEX data for H I in version $3.06.00$ and older were based on the extrapolations of R-matrix calculations of C VI \citep{1991JPhB...24.4583A}, Ne X \citep{1991PhyS...44..517A}, Si XIV \citep{1992PhyS...46..193A}, and Ca XX \citep{1992JPhB...25..751A} with an assumption that the Gaunt factor does not strongly depend on the atomic number Z and is more or less constant along the iso-electronic sequence. This assumption, however, breaks for the neutral atoms \citep{1962ApJ...136..906V} and influences the collision strength calculations. Due to this assumption, there was a significant difference between cooling in SPEX and Cloudy around $2.5$\,eV. 

After a careful comparison of various calculations that are available in the literature, we replaced the collision strengths for neutral hydrogen with data from CHIANTI in SPEX version $3.06.01$ (which is used in this paper). This data is based on \citet{2000JPhB...33.1255A} and revised by \citet{2002JPhB...35.1613A}. The update resulted in a decrease in SPEX cooling due to neutral hydrogen by a factor of three for temperatures around $2.5$\,eV and now shows a better agreement of SPEX and other plasma codes. For our final updated cooling curve based on the calculations and discussions in this paper (including the updates of the radiative loss rates due to collisional excitation and the updates of the collision strengths of neutral hydrogen), we invite the reader to check the black line in Fig.\,\ref{Fig:tot_cool_elem} and Fig.\,\ref{Fig:tot_cool_Schure_APEC_Cloudy}.

\begin{figure}[!t]
	\centering
	\includegraphics[angle=-90, width=0.5\textwidth]{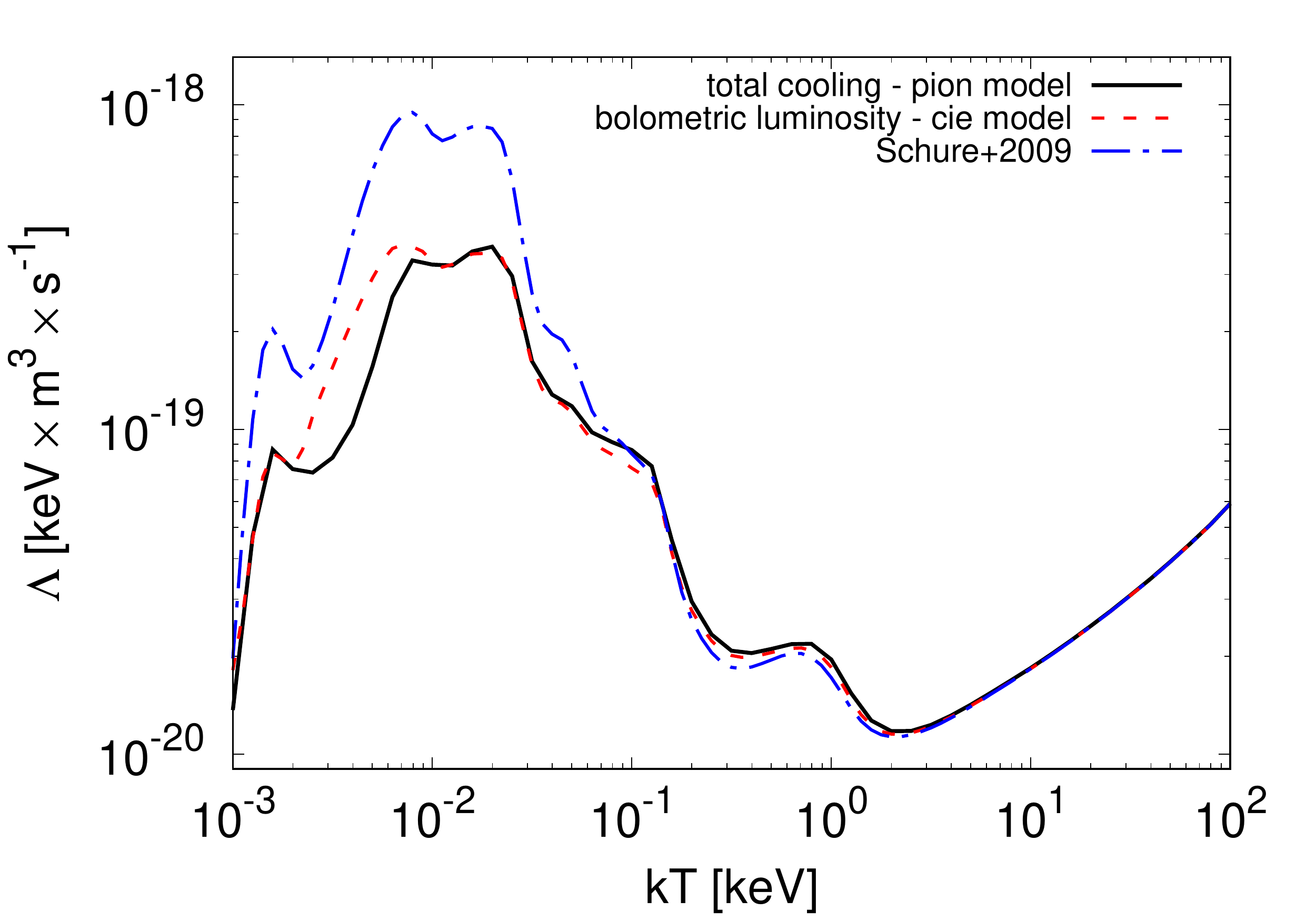}
	\caption{Comparison of the radiative loss curve obtained from \citet{2009A&A...508..751S} (blue dash-dotted line), the radiative loss curve obtained from the cie model (red dashed line) in SPEX version $3.06.01$, and the cooling curve obtained from the pion model (black solid line) in its CIE limit (zero ionisation parameter).}
	\label{Fig:tot_cool_pion_vs_bol_lum_CIE}
\end{figure}

Lastly, Fig.\,\ref{Fig:tot_cool_pion_vs_bol_lum_CIE} shows the comparison of the radiative loss curve obtained with MEKAL (the same as presented in \citet{2009A&A...508..751S}) and the radiative loss curve obtained from the cie model in SPEX (version $3.06.01$), as well as the cooling curve obtained from the pion model used in its CIE limit (ionisation parameter is set to zero). As we already mentioned in the introduction, unless the processes such as heat conduction or magnetic fields are taken into account, for the low-density CIE plasma in an enclosed, non-expanding box, the total cooling is the same as the energy radiated away. This holds true if the cooling is defined as the energy loss of the population of free electrons. However, if we consider individual atomic processes, the cooling and the radiative loss do not have to be the same, as in the case of collisional ionisation or radiative recombination, for example. For collisional ionisation the cooling is greater than the radiative loss (the cooling by the free electron is equal to the ionisation potential of the relevant ion, nothing is radiated away, and the energy loss is zero), whereas for radiative recombination the cooling is less than the radiative loss; the contribution to the cooling comes from the kinetic energy of the free electron that is being captured, while the contribution to the total radiative loss is the kinetic energy of the free electron plus the ionisation potential of the ion into which the free electron is captured. However, since in the balanced plasma the number of ionisations is equal to the number of recombinations, these differences between cooling and radiative loss for the ionisation and the  recombination processes (for instance also dielectronic recombination) cancel out. However, if we pay a close attention to the red and black curves in Fig.\,\ref{Fig:tot_cool_pion_vs_bol_lum_CIE}, we see that the total cooling curve is not the same as the total radiative loss curve for CIE plasma such as that described above. The reason is that the spectral calculation in the CIE model in SPEX still lacks the relevant complete and updated atomic data for several ions, in particular the ions for which we used Chianti data for the cooling contribution (see Fig.\,\ref{Fig:all_used_ions}). Correcting for this discrepancy at low temperatures is out of the scope of this paper and will be addressed in the future releases of SPEX.

\section{Application: Stability curve for photoionised plasmas}
\label{Sec:application-S-curve}

In the previous sections, we focused on a CIE plasma when no additional source of excitation or ionisation other than collisions with free electrons played a role in exciting (ionising) the plasma.  In this section, we examine plasma that is exposed to such external source of ionisation, and we assume the plasma is in photoionisation equilibrium. 

We updated the cooling function of the photoionisation model pion according to Section \ref{Sec:implementation_to_SPEX} (specifically the contribution to the cooling by the collisional excitation described in Sections \ref{Sec:cooling_curves} and \ref{Sec:cool_curve_comp_with_MEKAL}, dielectronic recombination described in Section \ref{Sec:dielectronic_recombination}, and the update of the collision strengths of neutral hydrogen discussed in Section \ref{Sec:total_cooling_curve}) and quantified the changes between the most recently updated SPEX (this work, SPEX version $3.06.01$), and the previous one, SPEX version $3.05.00$. We calculate the stability curve for plasma (the so called S-curve) and compare the results of our new updated cooling to the cooling in SPEX version $3.05.00$, which uses the MEKAL code to calculated the cooling by collisional excitation. We remind the reader that the cooling/heating we refer to in SPEX relates to the cooling/heating of the population of free electrons only.

\subsection{Thermal stability curve}


While determining the ionisation equilibrium in PIE plasma, it is convenient to define the ionisation parameter $\xi$ \citep{1969ApJ...156..943T, 1981ApJ...249..422K}
\begin{equation}
	\xi \equiv \dfrac{L}{n_{\rm H} r^2} \;,
	\label{eq:ionisation_par_xi}
\end{equation}
where the ionising source is described by the luminosity $L$ over the energy band $1$--$1000$ Rydbergs, $n_{\rm H}$ is the hydrogen density, and $r$ is the distance of the plasma from the source of ionisation. To obtain the thermal stability curve, we need to define the ionisation parameter in its dimensionless pressure form $\Xi$: 
\begin{equation}
	\Xi \equiv \dfrac{1}{4 \pi c k} \dfrac{\xi}{T} \;,
	\label{eq:ionisation_par_Xi}
\end{equation}
where $c$ is the speed of light, $k$ is the Boltzmann constant, and $T$ is the electron temperature. On the S-curve, the total cooling rate is equal to the total heating rate. Based on the slope of the S-curve (positive or negative), one can tell whether the plasma is thermally stable ($dT/d\Xi > 0$) or thermally unstable ($dT/d\Xi < 0$).

\subsection{Obtaining the S-curve in SPEX}
\label{Sec:S-curve_SPEX}

After updating the cooling rates for the pion model, we compared new S-curves (following the updates in Section \ref{Sec:cooling_curves} and \ref{Sec:cooling_curves_results}) with the S-curves obtained from SPEX version $3.05.00$ for three different spectral energy distributions (SEDs) based on \citet{2016A&A...596A..65M}: (a) a power-law continuum (labelled PL) in the energy band $10^{-4}$--$10^3$\,keV, with the photon index assumed to be $\Gamma = 2$; (b) an unobscured active galactic nucleus (AGN) that corresponds to NGC\,$5548$ (\citealt{2015A&A...575A..22M}, labelled AGN1); and (c) an obscured version of AGN1 (labelled AGN2). We described plasma photoionised by such SEDs in SPEX with the pion model. The transmitted spectrum was then calculated when the photons emitted from the ionising source ionised the plasma and by using the plasma routines in SPEX; the photoionisation equilibrium (ionisation and energy balance) was calculated self-consistently. In SPEX, the atomic processes that contribute to the total cooling are: inverse Compton scattering, electron ionisation, radiative recombination, free-free emission, collisional excitation, dielectronic recombination, and adiabatic expansion.The atomic processes that contribute to the total heating are: free-free absorption, photo-electrons, Compton ionisation, Auger electrons, collisional de-excitation, and external sources of ionisation.

We selected the grid for the ionisation parameter $\xi$ in the $10^{-7}$--$10^{10}$\,nWm range, and for each $\xi$ we obtained the equilibrium electron temperature $T$. The pressure form of the ionisation parameter $\Xi$ was then calculated using Eq.\,\eqref{eq:ionisation_par_Xi}.

\subsection{A new stable branch}
\label{Sec:S-curve_SPEX_results}

\begin{figure}[!t]
	\centering
	\includegraphics[angle=-90, width=0.46\textwidth]{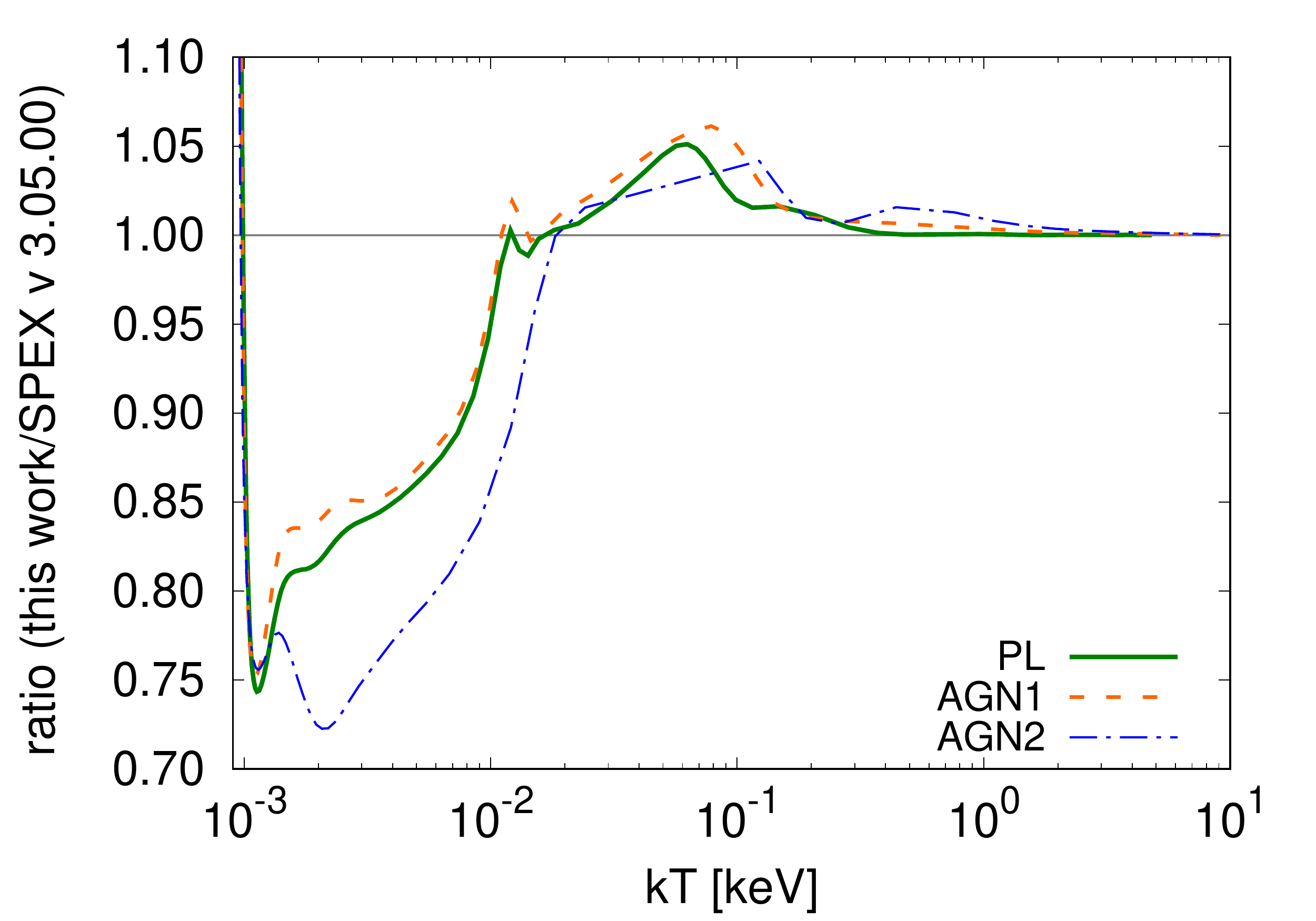}
	\caption{Ratio of the total cooling curve for the updated pion model (this work) and the previous version calculated with SPEX v $3.05.00$ as a function of the equilibrium temperature $kT$.}
	\label{Fig:tot_cooling}
\end{figure}

We show the ratio of the total cooling rate as a function of equilibrium temperature $kT$ for SPEX version $3.06.01$ (this work) and SPEX version $3.05.00$ in Fig.\,\ref{Fig:tot_cooling} for all SEDs mentioned in Section \ref{Sec:S-curve_SPEX}. We show a similar comparison for the radiative loss curve for the CIE case in Fig.\,\ref{Fig:tot_cooling_lines}, where we compared SPEX version $3.06.01$ to its precursor, MEKAL. In Fig.\,\ref{Fig:tot_cooling}, we make a comparison to the latest previously published update by \citet{2016A&A...596A..65M}.

The biggest difference between SPEX $3.06.01$ and $3.05.00$ can be seen for equilibrium temperatures from $10^{-3}$--$2\times10^{-2}$\,keV (apart from the resonant excitation around $10^{-3}$\,keV, which is not the main focus of this paper) where the total cooling decreased by almost $30$\%. This result is expected since the updates of the collisional excitation and the \ion{H}{I} collision strengths caused a decrease in the total radiative loss curve, as we show in Fig.\,\ref{Fig:tot_cooling_lines}, Fig.\,\ref{Fig:tot_cool_Schure_APEC_Cloudy}, and Fig.\,\ref{Fig:tot_cool_pion_vs_bol_lum_CIE}.

\begin{figure}[!t]
	\centering
	\includegraphics[angle=-90, width=0.46\textwidth]{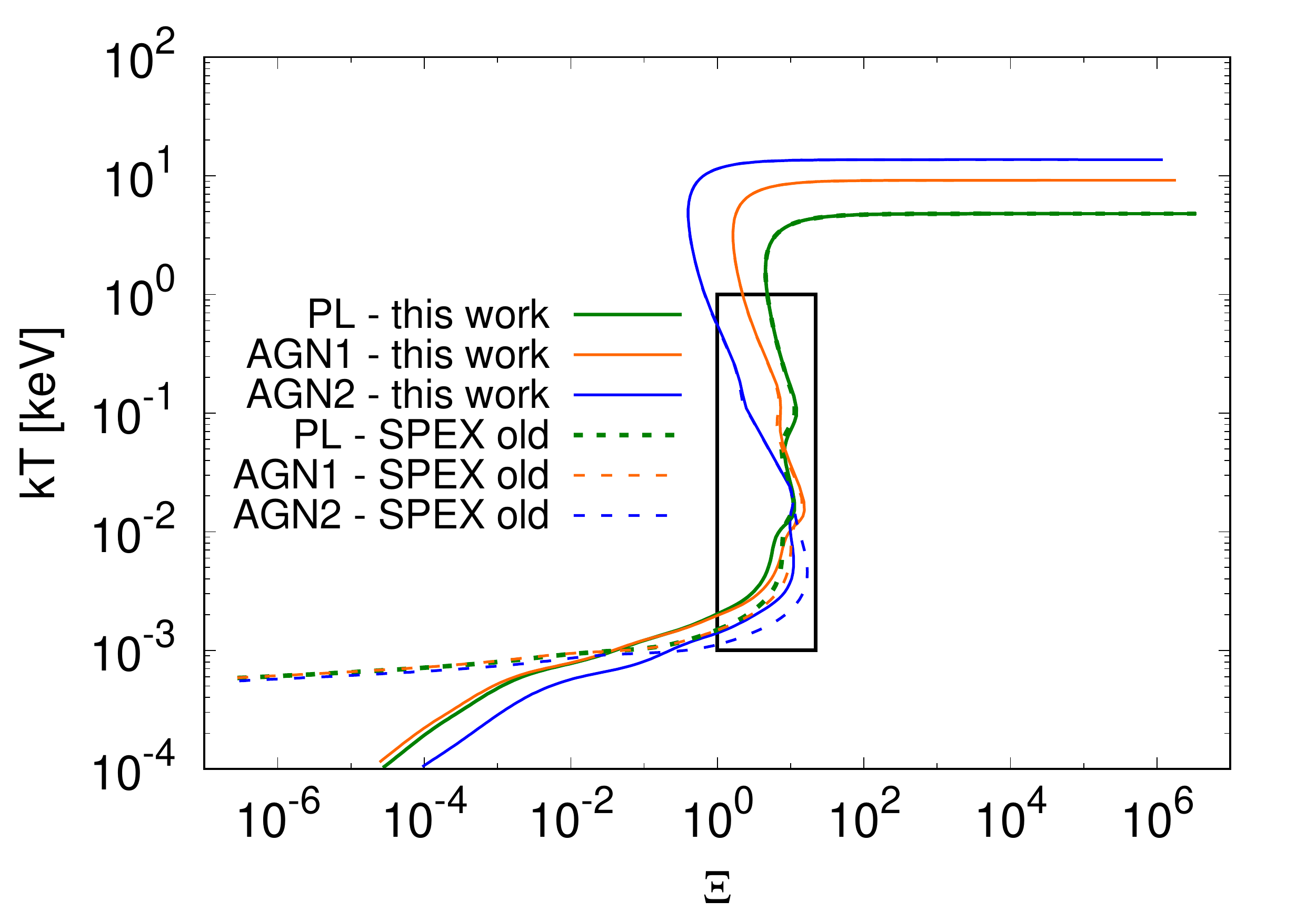} \\
	\includegraphics[angle=-90, width=0.46\textwidth]{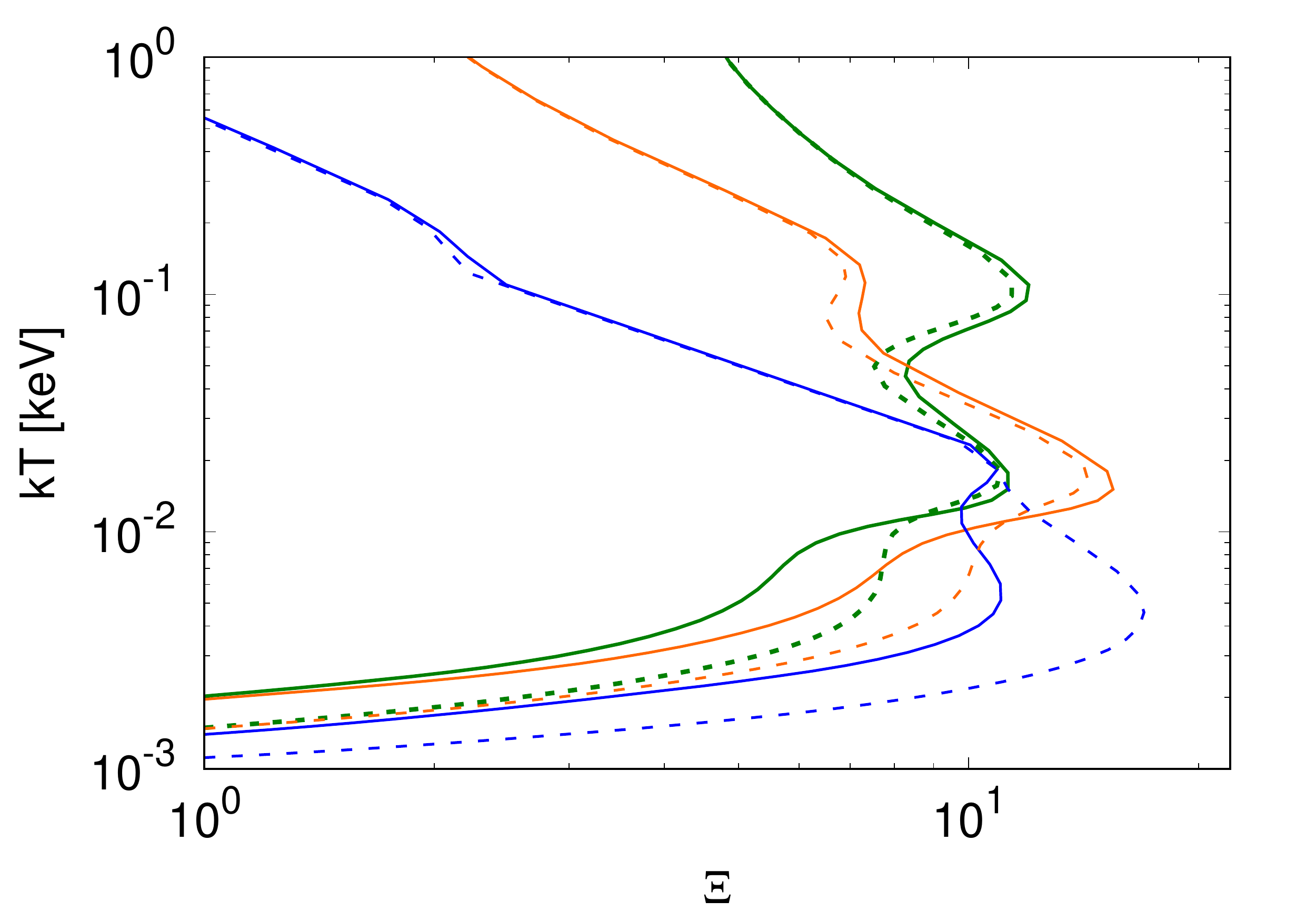}
	\caption{Stability curves for new (this work) and SPEX version $3.05.00$ (labelled as SPEX old) calculations after implementing the updates of the collisional excitation, dielectronic recombination and collision strengths of neutral hydrogen (top panel), and a zoomed-in view for $\Xi \in (1,20)$ and $kT \in (10^{-3},1)$\,keV (bottom panel).} 
	\label{Fig:S-curve}
 \end{figure}

Knowing the equilibrium temperature and the ionisation parameter makes it possible to calculate the dependence of $\Xi = \Xi{(kT)}$ using Eq.\,\eqref{eq:ionisation_par_Xi} and obtain the stability curves. The S-curves for all three SEDs before and after the updates are plotted in the top panel of Fig.\,\ref{Fig:S-curve}, where for values of $\Xi > 20$ the S-curves do not change their shape or normalization, while for $\Xi < 10^{-1}$ the equilibrium temperature decreases significantly towards lower values of $\Xi$. 



If we look at the bottom panel of Fig.\,\ref{Fig:S-curve}, where we see a zoomed-in view of the upper panel for $\Xi \in (1,20)$ and $kT \in$\,$\,(10^{-3},1)$\,keV, we notice a different behaviour and the change of the slope in two specific cases: (a) AGN1, $kT \sim 10^{-1}$\,keV and $\Xi \sim 7$, where the S-curve is almost vertical; and (b) AGN2, $kT \sim $ ($1.43$--$1.82$)\,$\times 10^{-2}$\,keV and $\Xi$ $\in$ ($10.17$--$10.96$), where an additional stable branch is found. We address the ionic column densities that are peaking on these stable branches in the following sub-section. 


\subsection{Ionic column densities}
\label{Sec:ionic_column_densities}

\begin{figure*}[!t]
	\centering
	\resizebox{0.86\textwidth}{!}{
		\includegraphics[angle=-90]{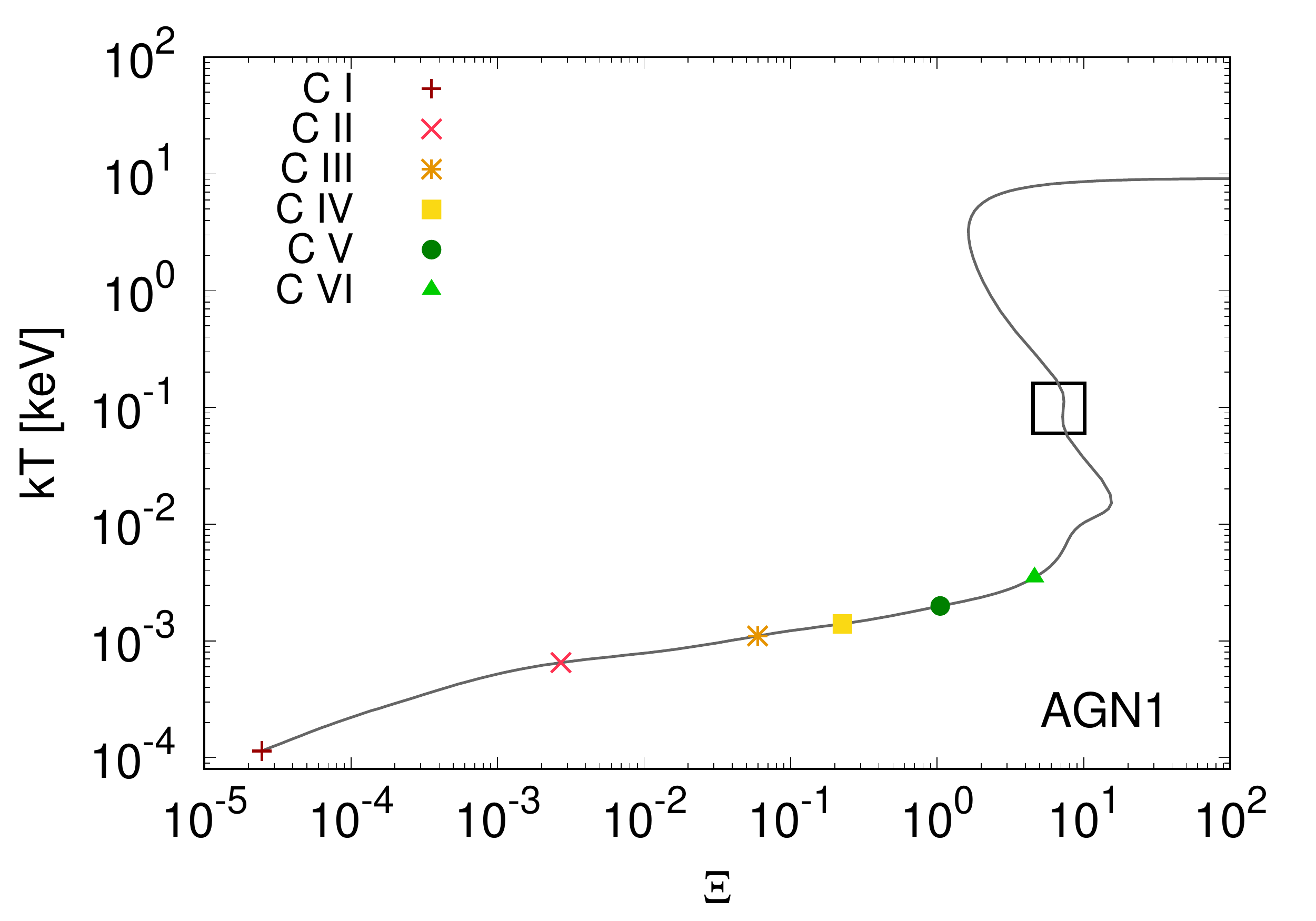}
		\includegraphics[angle=-90]{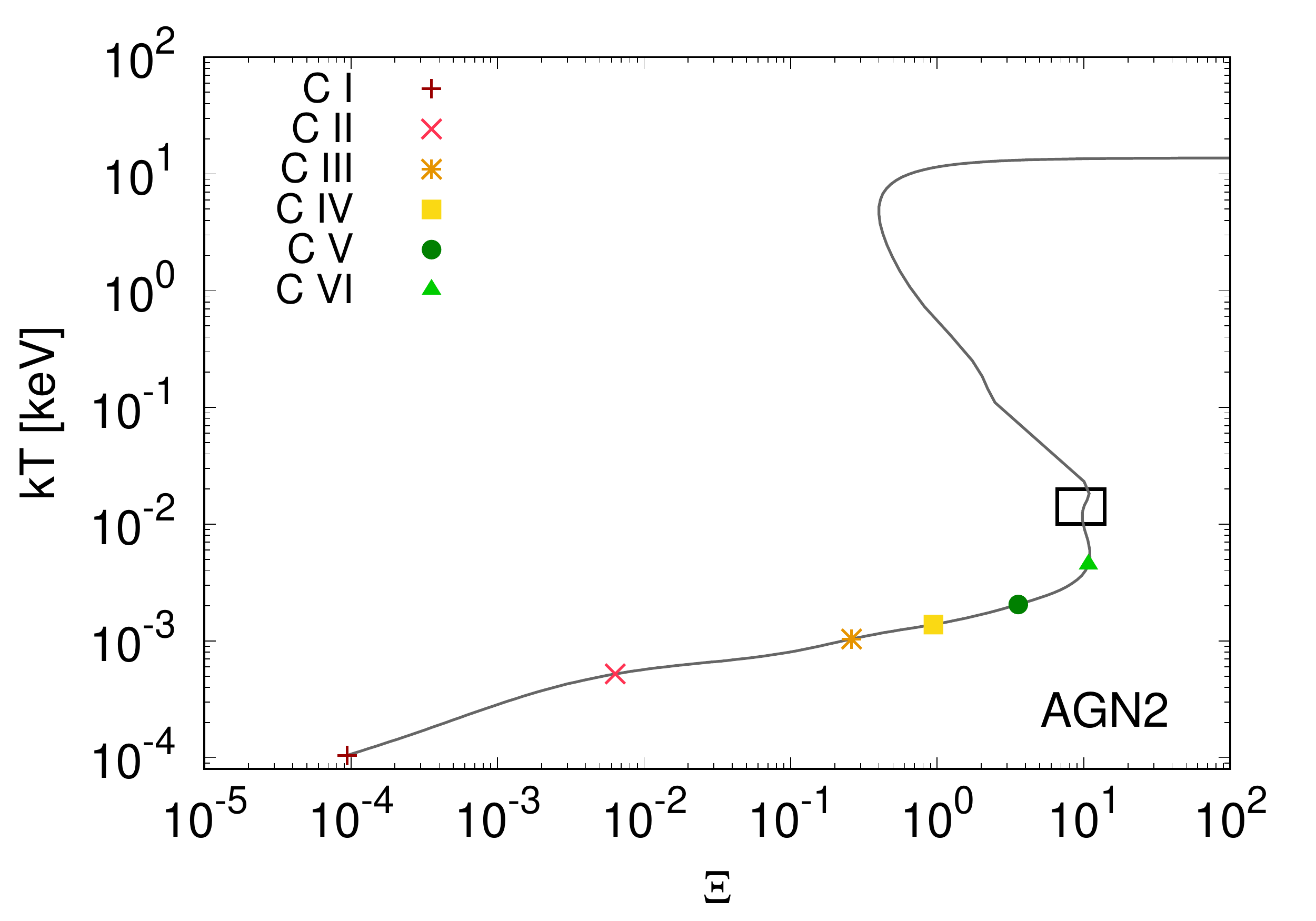} }
	\resizebox{0.86\textwidth}{!}{
		\includegraphics[angle=-90]{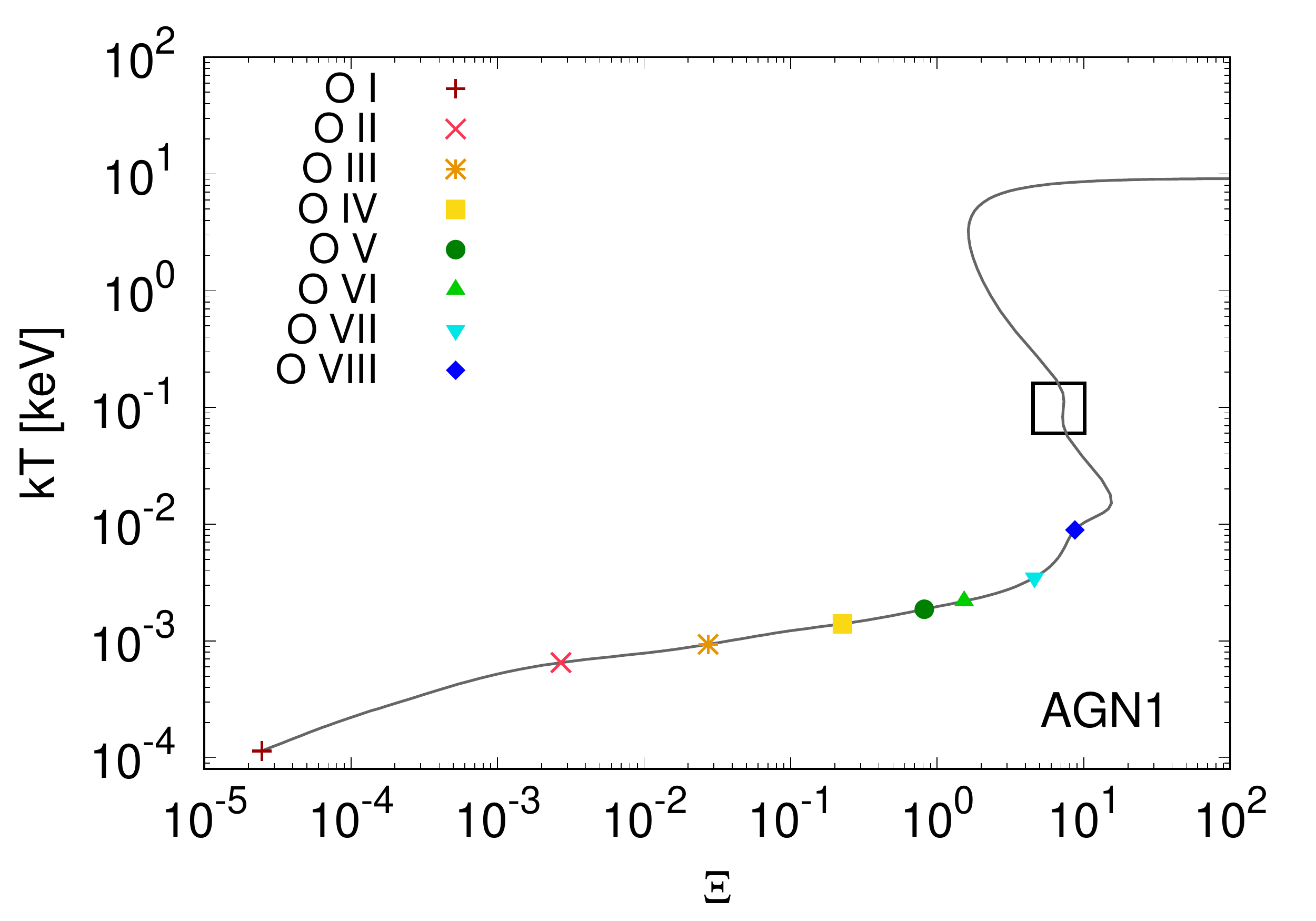}
		\includegraphics[angle=-90]{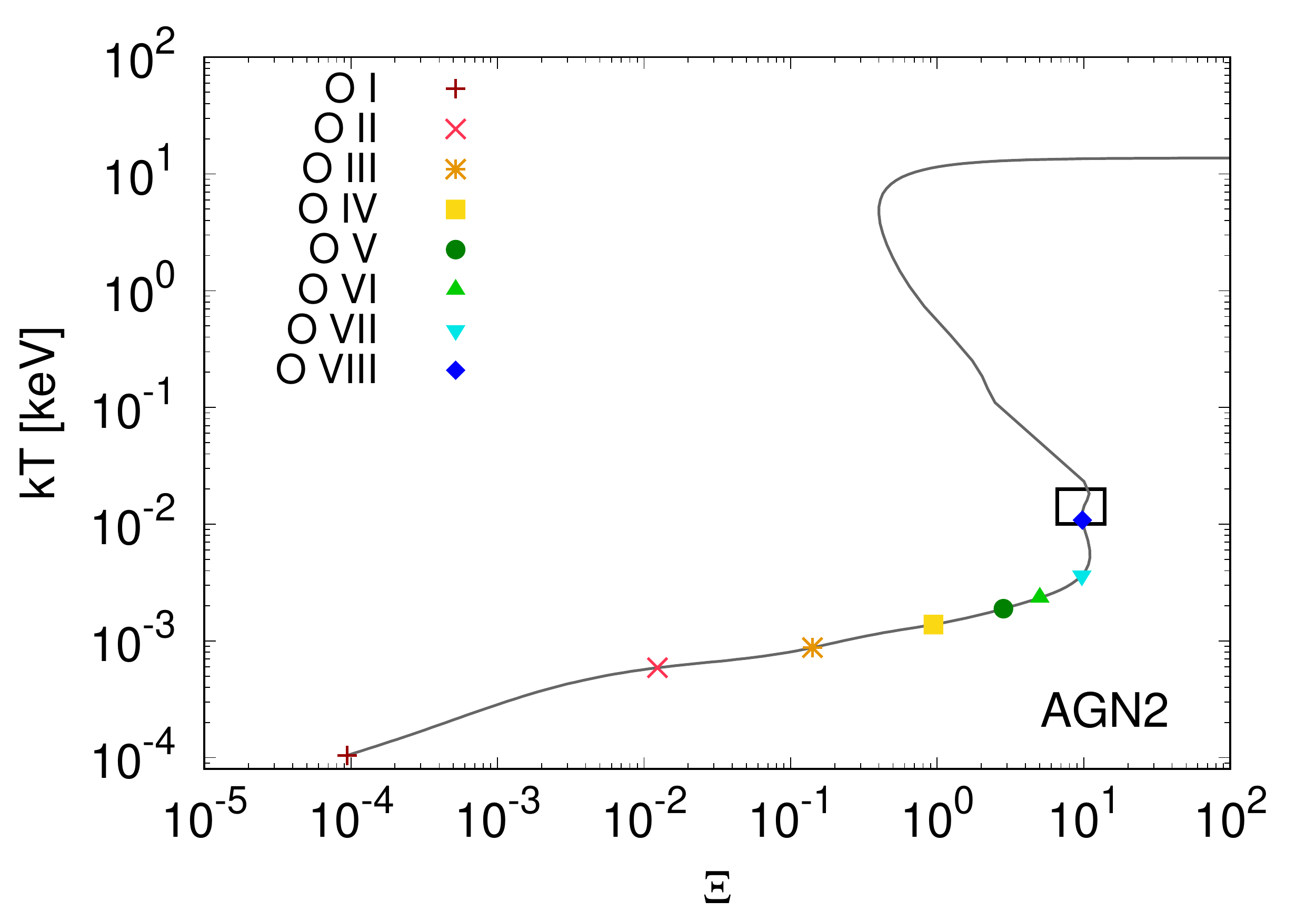} }
	\resizebox{0.86\textwidth}{!}{
		\includegraphics[angle=-90]{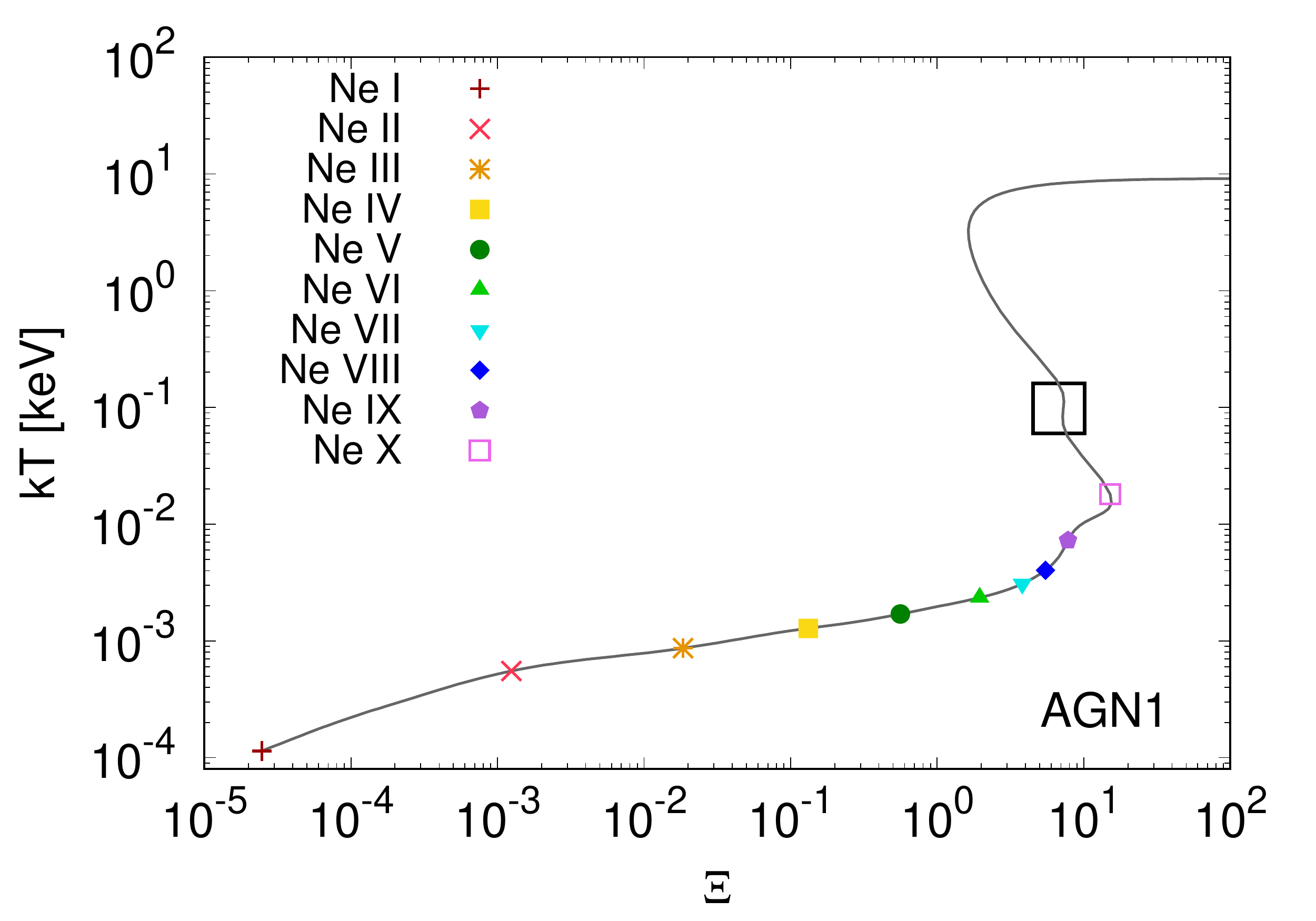}
		\includegraphics[angle=-90]{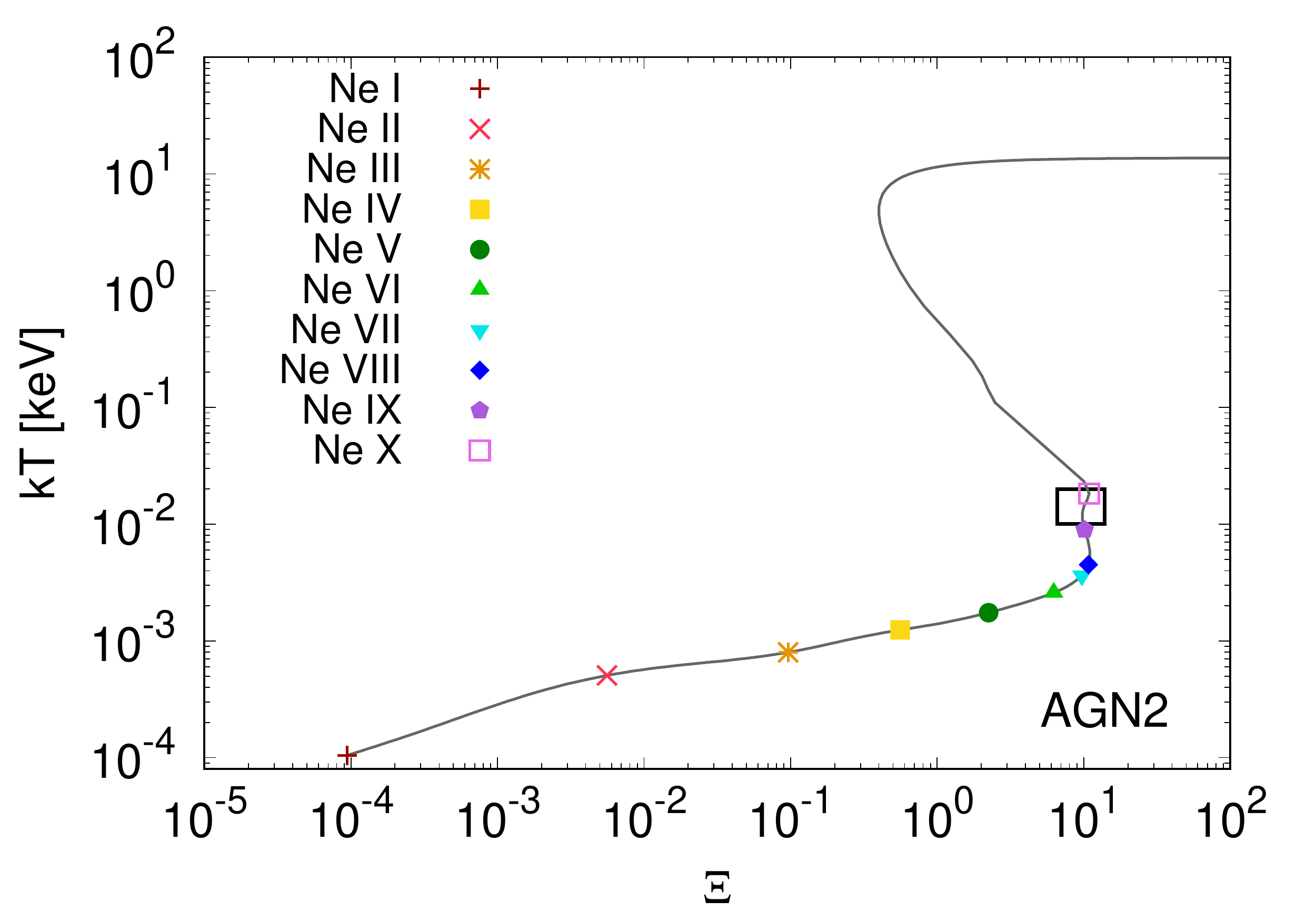} }
	\resizebox{0.86\textwidth}{!}{
		\includegraphics[angle=-90]{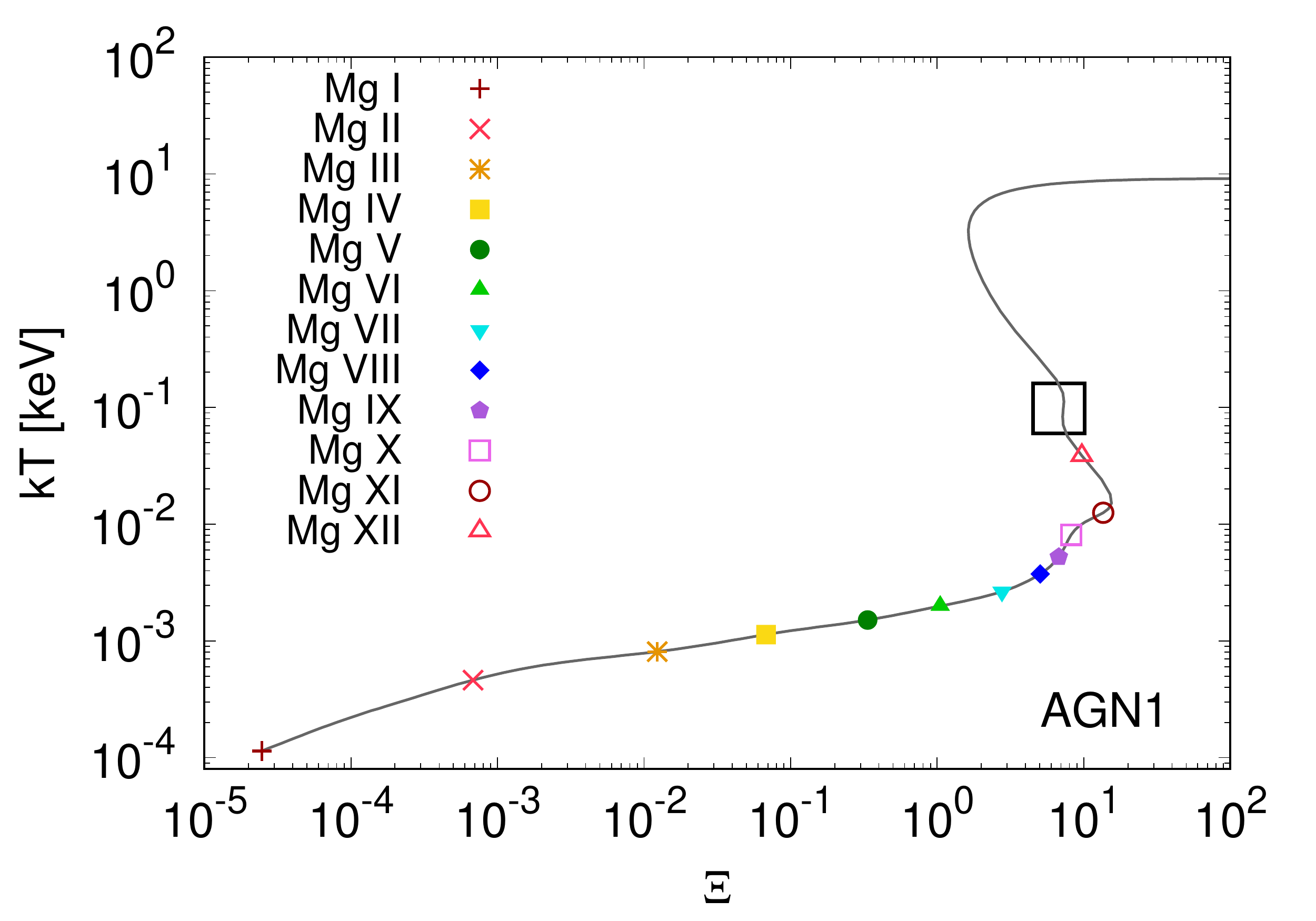}
		\includegraphics[angle=-90]{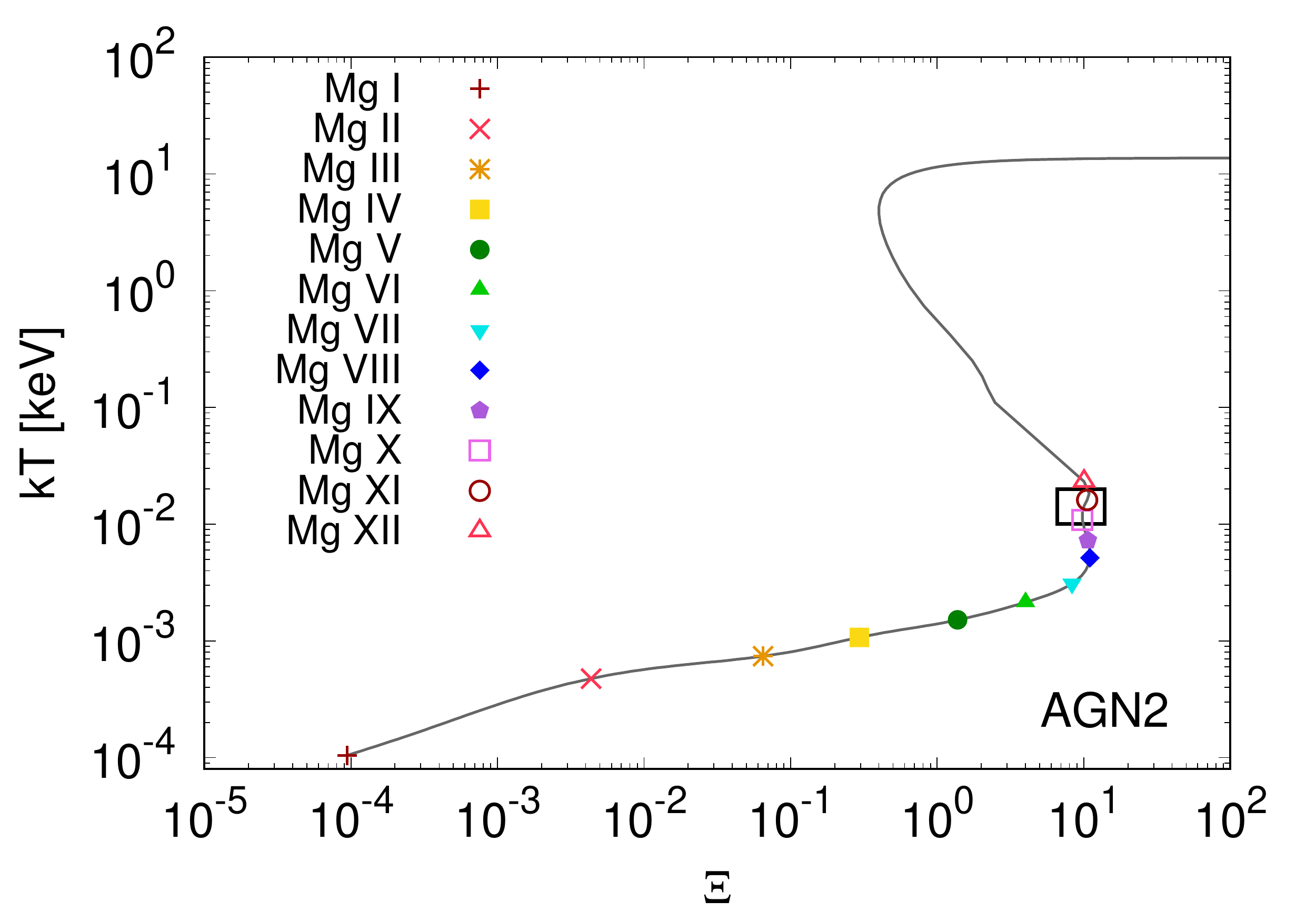} }
	\caption{Stability curves for AGN1 (left panel) and AGN2 (right panel), with coloured points indicating the points in ($kT$, $\Xi$) phase space, where the column densities of C, O, Ne, and Mg peak. The rectangles represent new stable branches found after implementing the updates (described in this paper) to the photoionisation model pion in SPEX. }
	\label{Fig:S-curve_ions_1}
\end{figure*}

\begin{figure*}[!t]
	\centering
	\resizebox{0.85\textwidth}{!}{
		\includegraphics[angle=-90]{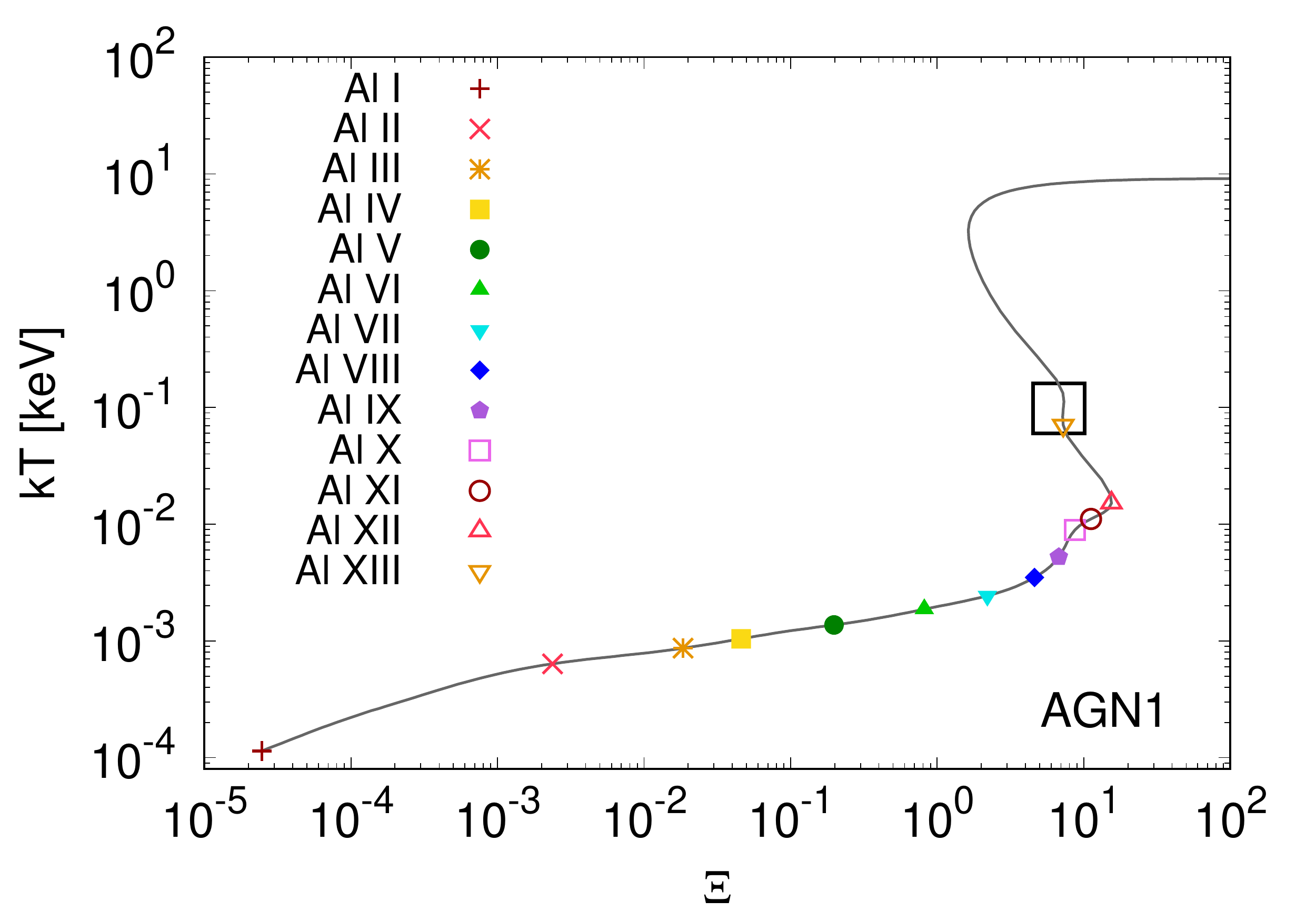}
		\includegraphics[angle=-90]{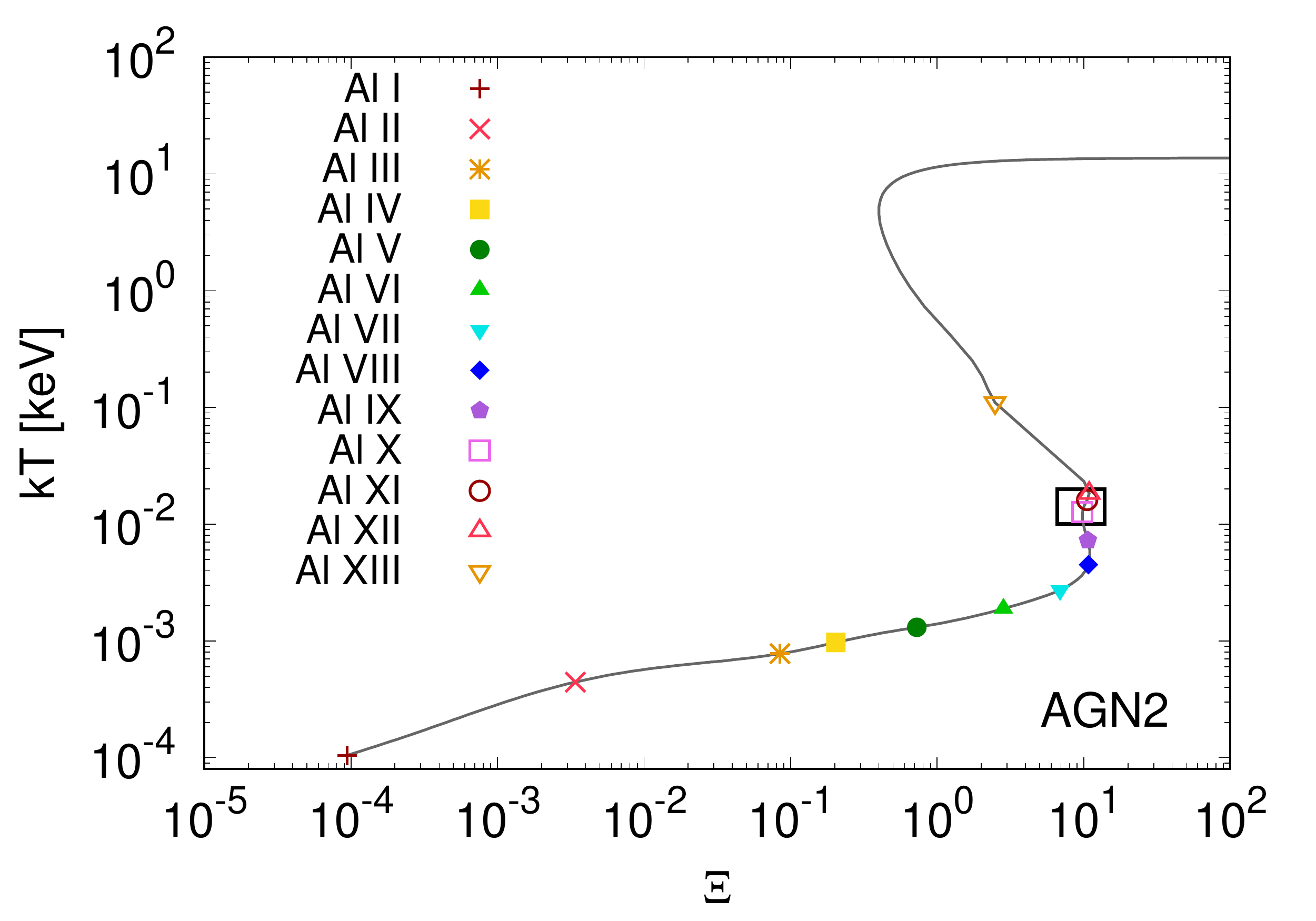}}
	\resizebox{0.85\textwidth}{!}{
		\includegraphics[angle=-90]{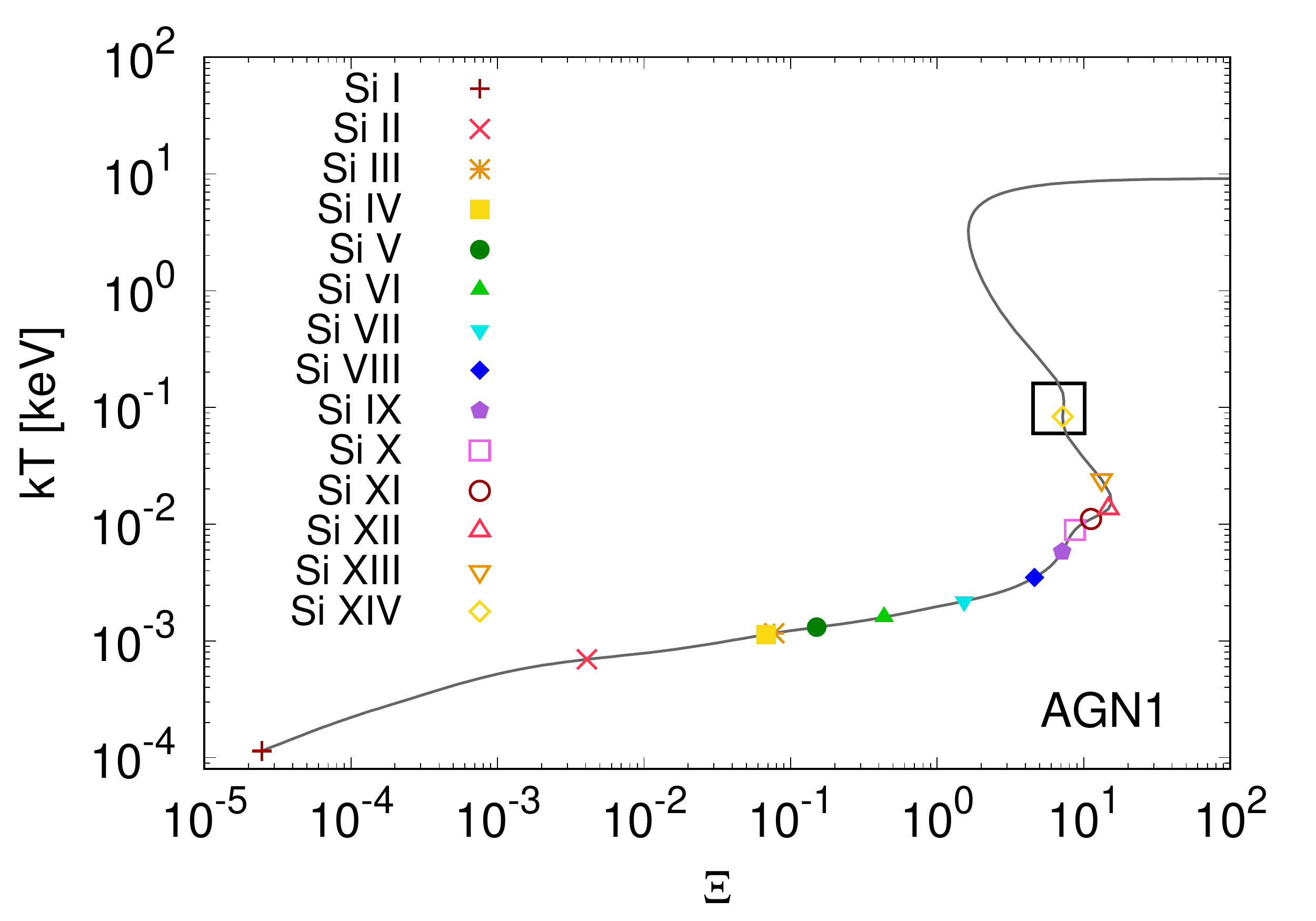}
		\includegraphics[angle=-90]{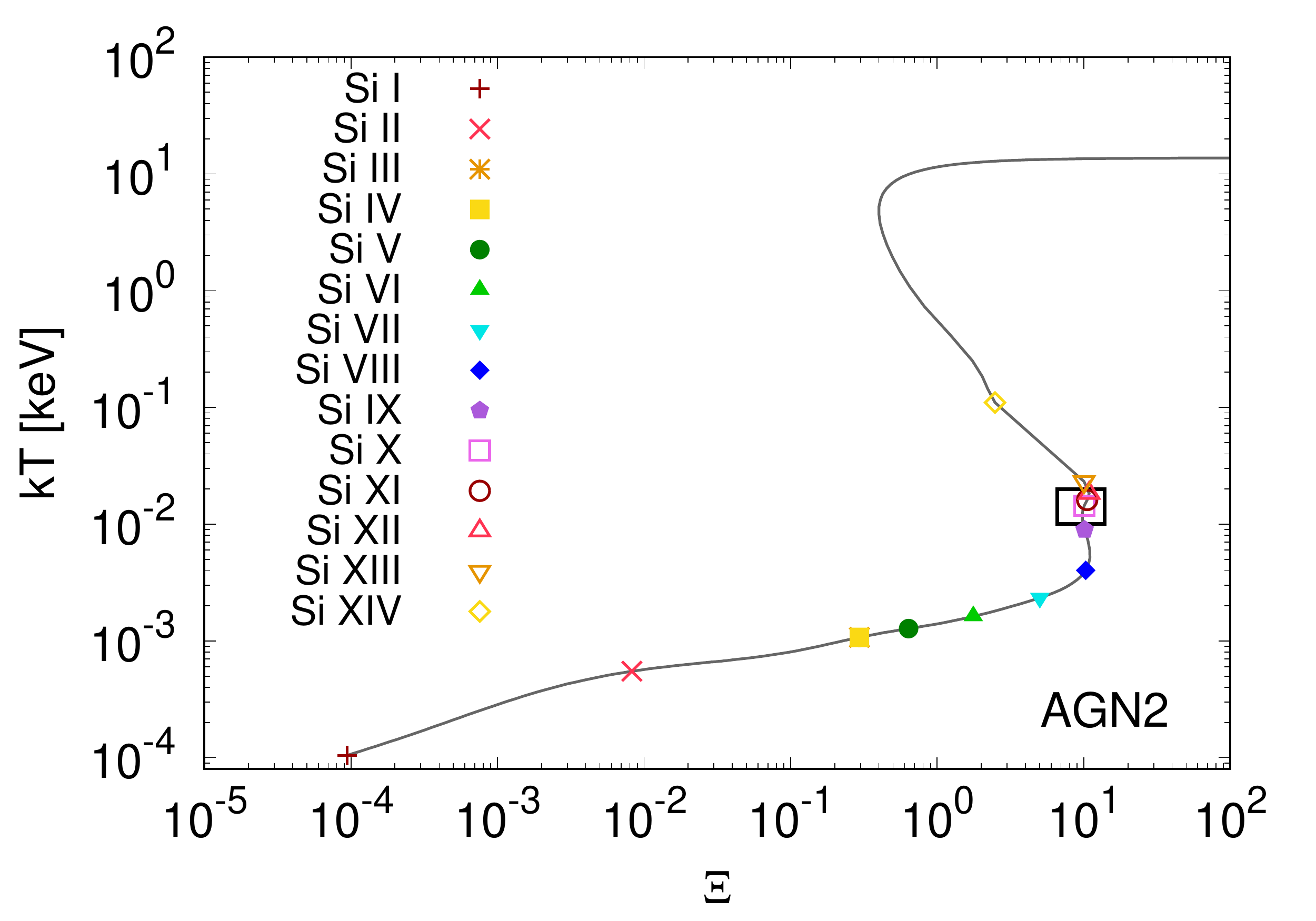}}
	\resizebox{0.85\textwidth}{!}{
		\includegraphics[angle=-90]{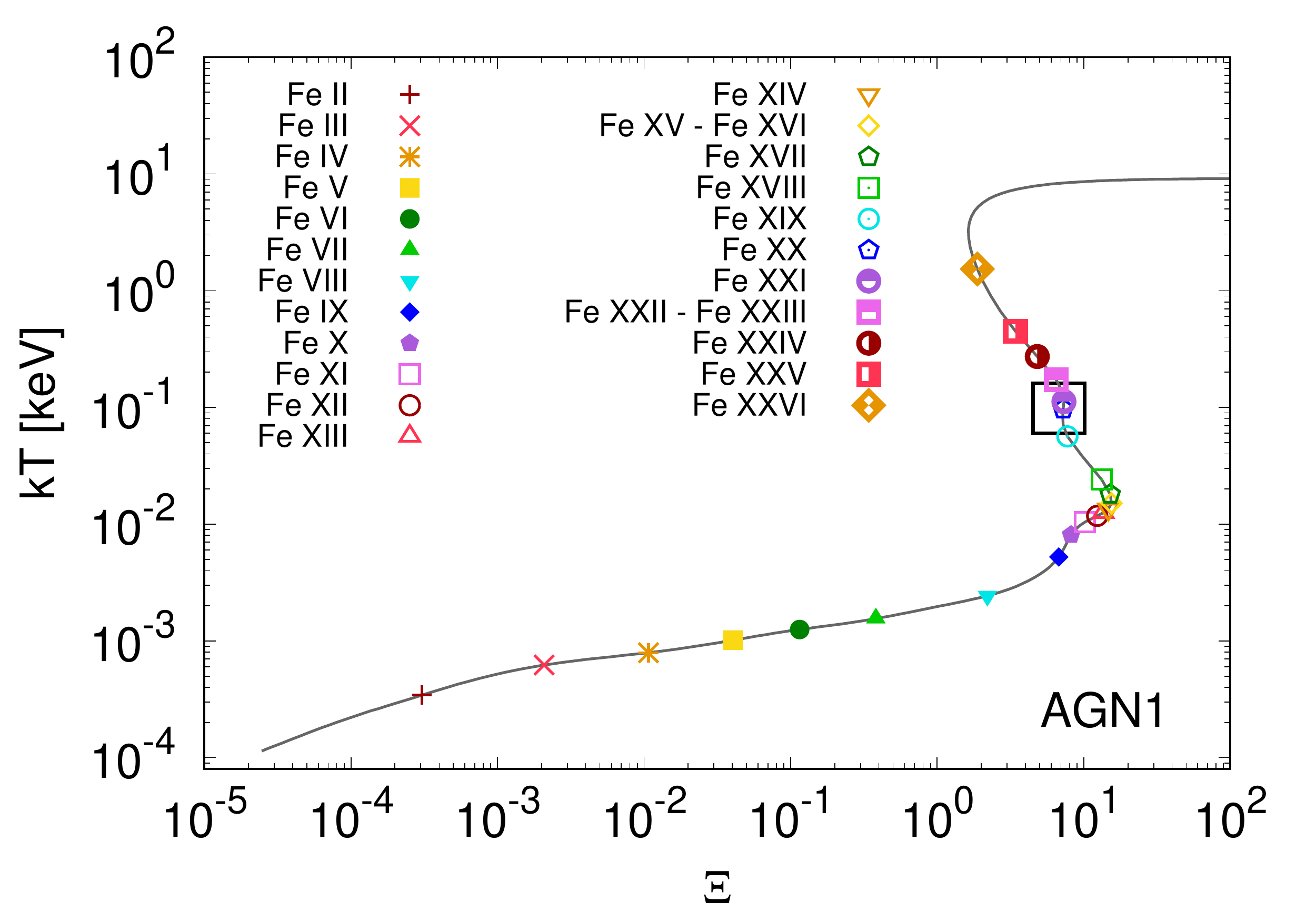}
		\includegraphics[angle=-90]{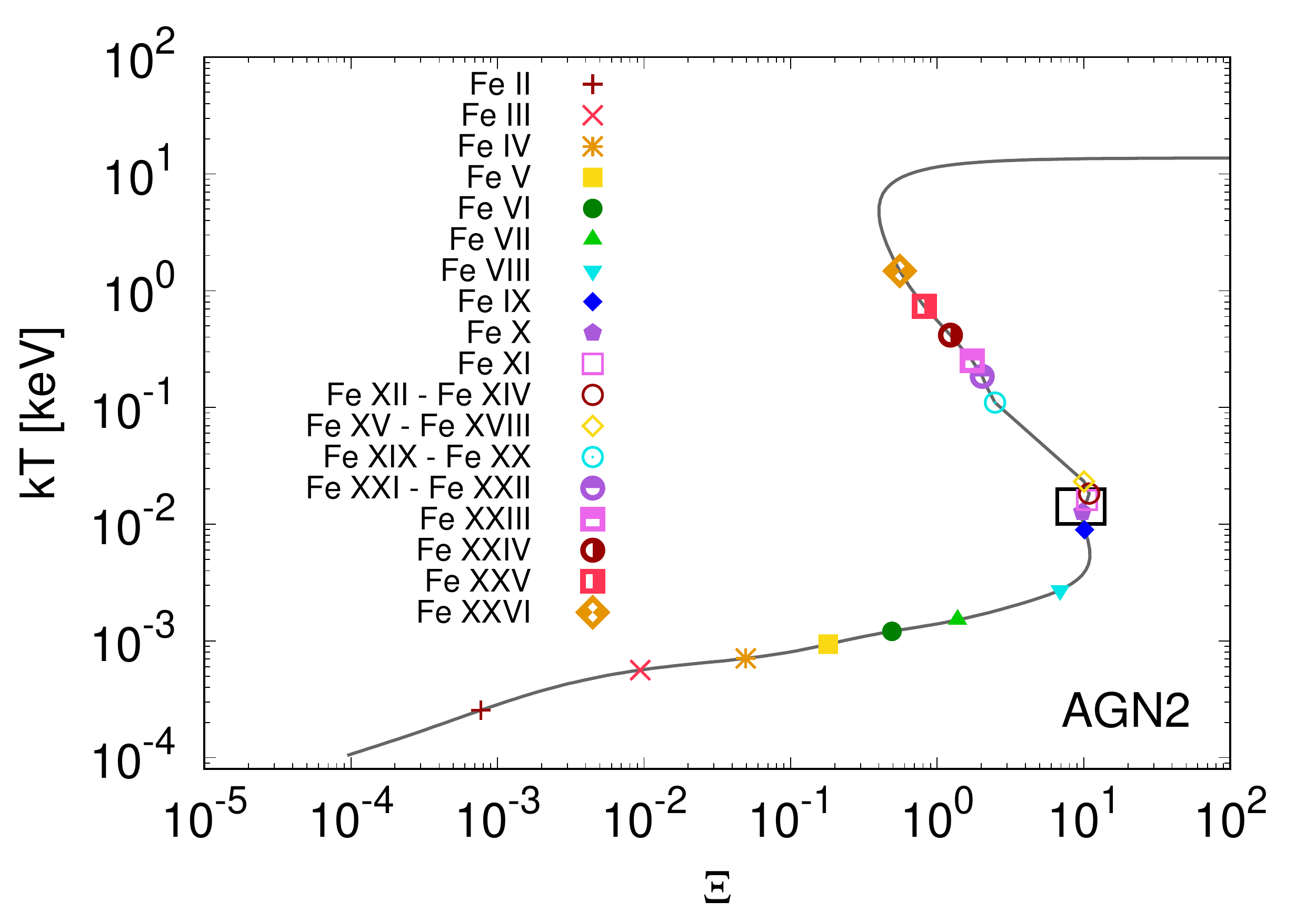}}
	\caption{Same as Fig.\,\ref{Fig:S-curve_ions_1} but for Al, Si, and Fe ions. }
	\label{Fig:S-curve_ions_2}
\end{figure*}

In Fig.\,\ref{Fig:S-curve_ions_1} and Fig.\,\ref{Fig:S-curve_ions_2}, we show where the individual ionic column densities $N_{\rm ion}$ peak on the S-curve in ($kT$, $\Xi$) phase space for AGN1 and AGN2 SEDs, where we find the changes in the S-curve slope (see Section \ref{Sec:S-curve_SPEX_results}). As an example, we show the maximum ionic concentrations for C, O, Ne, Mg, Al, Si, and Fe. For AGN1, we find that \ion{Al}{XIII}, \ion{Si}{XIV}, \ion{Fe}{XX}, and \ion{Fe}{XXI} ions peak at the branch that changed from positive (stable) to almost vertical. In the case of AGN2, the ions which peak on the new stable branch are \ion{O}{VIII}, \ion{Ne}{X}, \ion{Mg}{X}, \ion{Mg}{XI}, \ion{Al}{X}, \ion{Al}{XI}, \ion{Al}{XII}, \ion{Si}{X}, \ion{Si}{XI}, \ion{Si}{XII}, \ion{Fe}{X}, \ion{Fe}{XI}, \ion{Fe}{XII}, \ion{Fe}{XIII}, and \ion{Fe}{XIV}. For the case of AGN2, before the updates these ions were peaking on the unstable branch. After the updated cooling is implemented to pion these ions are found on the new stable branch and could be potentionally observed (based on the column densities) in the photoionised gas with equilibrium temperature around $1$--$2$ $\times 10^{-2}$\,keV.

From Fig.\,\ref{Fig:S-curve_ions_1} and Fig.\,\ref{Fig:S-curve_ions_2}, one might infer, for instance, that there is no highly ionised iron at all (\ion{Fe}{XIX} to \ion{Fe}{XXVI}) because most of it would fall on the unstable branch of the S-curve. This is not the case, because these ions are formed over a broad range of ionisation parameters or temperatures (in these figures, we only show where the ionic column densities peak). For example, \ion{Fe}{XXV} and \ion{Fe}{XXVI} have concentrations larger than $10$\% of their peak concentration over a range spanning a factor of $20$ in temperature. Similarly, for the lower ionised ions of iron, the stable regions can offer sufficient \ion{Fe}{XIX} to \ion{Fe}{XXIV} to be detectable. Obviously, the total measured column density for these ions might be reduced because contributions from the unstable regions are missing.

\section{Discussion}
\label{Sec:discussion}


\subsection{Theoretical calculations}
\label{Sec:Discussion_theoretical_calc}
Because radiative cooling occurs on various scales throughout the Universe, the importance of understanding the details and atomic processes that lead to gas cooling down are central to our knowledge of astrophysical plasmas. Many line complexes and spectral features are still hardly resolved with nowadays instruments and therefore in many cases we need to rely on theoretical calculations of transition probabilities and line energies. 

These theoretical calculations cannot be done without approximations used to solve the Schr$\ddot{\rm o}$dinger equation. These include independent-particle approximation, variational principle, many-body perturbation theory \citep{1955PhRv...97.1353B, 1957RSPSA.239..267G}, Born \citep{1926ZPhy...38..803B}, Born-Oppenheimer \citep{1927AnP...389..457B}, the Coulomb-Born-Oppenheimer or widely used distorted wave \citep{1933tac..book.....M}, and R-matrix approximations \citep{Burke1993-R-matrix}. 

Theoretical line energies can also differ from measured line energies. For example, for the calculations made with CHIANTI (see the ions in Fig.\,\ref{Fig:all_used_ions}), and depending on the ion and the complexity of the system, we find differences as low as  $< 1$\% but also as high as $10$\%. For most ions, the difference between the radiative loss rate (per ion) using theoretical or observational line energies is less than a few percent, with only a few exceptions when the difference can be up to $200$\% (for instance, \ion{Cl}{II}, \ion{Ar}{III}, \ion{Ar}{V}, \ion{Ca}{V}). On the other hand, the wavelengths of the dominant lines in the SPEX code are in general very well benchmarked with observational data, and the energy uncertainties affect the corresponding ionic cooling rates by a much smaller amount than the uncertainties in the collision strengths.

All differences in the above-mentioned approximations then contribute to discrepancies between plasma codes and atomic databases. On top of that, the calculation of spectra depends on physical processes that are very complex and often treated in a simplified manner or are neglected completely (for example, before our updates presented in this paper, the radiative loss by collisional excitation was calculated in the pion model using MEKAL, and the potassium-to-zinc iso-electronic sequences were missing). This only shows the importance of constant updates of atomic data and supports independent development of the plasma codes that can then be compared to measurements.

\subsection{Comparison of SPEX to other plasma codes}
Such a comparison of differences between plasma codes is also shown in our results in Section \ref{Sec:cooling_curves_results}. First, we considered the radiative loss curves for collisional excitation in SPEX and compared them with the radiative loss curves obtained from its precursor MEKAL, which is still being used by the community to analyse the X-ray spectra. The shapes and normalisations of these curves vary from ion to ion. As seen from Fig.\,\ref{Fig:tot_cooling_lines}, the significant difference between MEKAL and SPEX lies in the updates of the collisional excitation rates for \ion{C}{III}, \ion{O}{V}, and \ion{Ne}{VII} ions. The updates resulted in a decrease of the total radiative loss, especially for temperatures between $2 \times 10^{-3}$--$10^{-1}$\,keV. This is also valid for the total cooling curve, as we showed in the comparison to \citet{2009A&A...508..751S}, where the total cooling in the same temperature range decreased by almost $70$\% (for the temperatures around $0.01$\,keV). For this temperature range, the updates of the collisional excitation resulted in a closer agreement of the total cooling curve of SPEX with other codes such as Cloudy or APEC.


As already discussed in Section \ref{Sec:total_cooling_curve}, we found that due to incorrect collision strengths for neutral hydrogen in SPEX up to version $3.06.00$, the SPEX cooling rates were significantly overestimated for temperatures around  $\sim 2.5$\,eV, which is the temperature at which neutral hydrogen is a dominant coolant. After detailed examination, we decided to adopt the collision strength values for neutral hydrogen from the CHIANTI database and used this atomic data in SPEX. This lead to a decrease in SPEX cooling at this temperature and resulted in a closer agreement between SPEX, Cloudy, and APEC.


\newpage 

\subsection{Metastable levels}
With Be-like oxygen, we demonstrated the importance of the radiative loss from metastable levels, which can be even higher than the radiative loss from the ground level for temperatures such as those below $2 \times 10^{-3}$\,keV and above $3$\,keV. However, the radiative loss from metastable levels is not yet included in SPEX, because to have a fully self-consistent model the ionisation and recombination rates from and to these metastable levels also need to be taken into account in the ionisation balance.  Although we show only the metastable levels in the case of collisional ionisation equilibrium, many works point out how these metastable levels are also important for the plasma in photoionisation equilibrium (see for example, \citealt{1974agn..book.....O,1993ApJS...88..253S,1998PASP..110..761F} or one of the recent works of \citealt{2013MNRAS.430.3292B}, where in general the cooling gas is studied via cosmological hydrodynamical simulations in which the cooling rates become an essential ingredient for understanding the Universe on large scales. The processes that become important at temperatures below $10^6$\,K ($0.09$\,keV) are mainly the radiative recombination and two-photon decays of the metastable levels).


\subsection{Update of the pion model in SPEX}
In Section \ref{Sec:implementation_to_SPEX}, we presented an updated model for the cooling due to collisional excitation only that was applied to pion model. The updated radiative loss due to collisional excitation lead to a different contribution from specific processes included in the overall heating/cooling balance of the photoionised plasmas. For three different SEDs, we plotted the stability curve for the updated PIE model and noticed two main differences: (a) at low values of $\Xi$, the slope of the S-curve became steeper in comparison with old calculations; and (b) for $\Xi \in (1,20)$, where the behaviour of the S-curves changed (mainly for AGN1 and AGN2 SEDs): in the case of AGN1, the S-curve changed its slope to almost vertical, while for AGN2 we found a new stable branch. By calculating where the ionic column densities peak, we showed which ions lie in the regions of new stability branches. These ions are then expected to exist under such conditions in the photoionised plasmas (AGN1: \ion{Al}{XIII}, \ion{Si}{XIV}, \ion{Fe}{XX}, \ion{Fe}{XXI}; AGN2: \ion{O}{VIII}, \ion{Ne}{X}, \ion{Mg}{X}, \ion{Mg}{XI}, \ion{Al}{X}, \ion{Al}{XI}, \ion{Al}{XII}, \ion{Si}{X}, \ion{Si}{XI}, \ion{Si}{XII}, \ion{Fe}{X}, \ion{Fe}{XI}, \ion{Fe}{XII}, \ion{Fe}{XIII}, \ion{Fe}{XIV}).  

The S-curves for the same sample of SEDs were also studied in \citet{2016A&A...596A..65M}, where the authors showed the comparison of SPEX (version $3.02.00$) to other photoionisation codes such as Cloudy (version $13.01$) and XSTAR (version $2.3$). In this older version of SPEX, no stable branch for AGN2 was found for $\Xi \sim 10$, whereas Cloudy and XSTAR showed a different behaviour (slope of the S-curves is positive). This shows that by updating the cooling via collisional excitation, which also affects the total cooling in the photoionisation model, the new version of the pion model in SPEX predicts S-curves that are in better agreement with the predictions from Cloudy and XSTAR, mainly for the AGN2 case where Cloudy and XSTAR already predicted the stable branch for $\Xi \sim 10$. This stable branch was found in SPEX after the updates of the radiative loss curve presented in this paper (see Fig.\,\ref{Fig:S-curve}).


In general, the S-curve can be more complicated and can have multiple stable or unstable branches that depend on the source of ionisation and its SED as well as chemical composition and density of the absorbing gas along the line of sight (for instance, see \citealt{1997ApJ...478...94H,2008A&A...487..895R,2009MNRAS.393...83C, 2013MNRAS.430.2650L}). For example, as \citet{2012MNRAS.422..637C} showed, this can be caused by the strength of the soft excess as well as changes in the temperature of the accretion disc. Among the fairly recent observational papers showing such S-curves are \citet{2015A&A...575A..22M}, and \citet{2016A&A...596A..65M}, which shows a very unusual S-curve found for ultra-fast outflow in the quasar PDS\,$456$ \citep{2019ApJ...873...29B}.

Understanding these stable or unstable branches of S-curves can be important for studying the surroundings of AGN. \citet{2015A&A...575A..22M} show how the obscurer in NGC 5548 shields the ionising radiation and creates a significant unstable branch on the S-curve. They studied the source in multiple epochs, and by calculating the S-curves and the peaking ionic column densities they found that highly ionised states of iron (\ion{Fe}{XXIII} and \ion{Fe}{XXVI}) fell on the unstable branches for all five epochs, which lead to the conclusion that these ions might be lacking in the photoionised gas regardless of the obscured or unobscured scenario. 

Updating photoionisation models is also important for the estimation of the so-called absorption measure distribution (AMD), which describes the ionisation structure of the AGN wind (the density radial profile of the AGN wind is constrained by the slope of AMD \citealp{2009ApJ...703.1346B}) and depends on the ionisation parameter $\xi$ and the hydrogen column density $N_{\rm H}$. AMD is defined as \citep{2007ApJ...663..799H}
\begin{equation}
\textnormal{AMD} = 	\dfrac{d N_{\rm H}}{d \log \xi}	\;.
\end{equation}
Such AMD often shows a deep minimum in the column density profile, corresponding most often to $\log{\xi}$ roughly between $1$ and $2$ (but this can differ from source to source). This is interpreted as the thermally unstable gas (for instance, see \citealp{2009ApJ...703.1346B,2011A&A...534A..38D,2014MNRAS.445.3011S,2015ApJ...815...83A} and references therein). If different timescales (for example, recombination, cooling or dynamical timescales) are obtained, the limits to the hydrogen column density and to the distance of the absorber from the ionising source can be obtained (for example, see \citet{2018A&A...615A..72M} for more details).

\section{Conclusions}
\label{Sec:conclusions}


In this work, we highlight the differences of excitation rates among different plasma codes (MEKAL, APEC, Cloudy) and other atomic databases (CHIANTI, ADAS). We show how the updates, mainly of collisional excitation, impact the radiative loss rates as well as the cooling rates in SPEX. Our main findings are as follows:
\begin{itemize}
	\item The comparison between MEKAL and the newest version of SPEX (presented in this paper, version $3.06.01$) shows that the updates of collisional excitation affect the total radiative loss curve mainly in the temperature range between $2 \times 10^{-3}$--$10^{-1}$\,keV, where a significant decrease by $70$\% is found (Fig.\, \ref{Fig:tot_cooling_lines}).
	\item We calculate the total cooling curve (Fig.\,\ref{Fig:tot_cool_elem}) considering the most updated version of SPEX (version $3.06.01$), which includes the updated contribution to the cooling by the collisional excitation and dielectronic recombination. Additionally, we describe an update of the collision strengths of neutral hydrogen, which mainly affects the cooling for temperatures around $2.5$\,eV. All considered updates result in a better agreement of the SPEX cooling curve and the cooling curves obtained by other plasma codes such as Cloudy and APEC (Fig.\,\ref{Fig:tot_cool_Schure_APEC_Cloudy}).
	\item We implement the new cooling to the SPEX photoionisation model pion and demonstrate the impact of the updates on the stability curve of photoionised plasmas. By using the updated cooling tables we find new stable branches on the stability curves (Fig.\,\ref{Fig:S-curve}) and associate them with an example set of ions for which the ionic column densities peak on these stability branches (Fig.\,\ref{Fig:S-curve_ions_1} and \ref{Fig:S-curve_ions_2}).
\end{itemize}
With this work we show the importance of constant updates of atomic databases and plasma codes in order to understand more precisely the high-resolution spectroscopic data that will be obtained by upcoming missions, such as XRISM and Athena. These updates are also crucial for the calculation of the cooling curve, which is an important ingredient for the cooling processes in various astrophysical objects (in this paper we used AGN as an example).



\subsection*{\textbf{Dataset availability}}
The dataset generated and analysed during this study is available in the ZENODO repository 
\citet{lydia_stofanova_2021_5497662}.

\subsubsection*{Acknowledgements}
The authors acknowledge the financial support from NWO, the Netherlands Organisation for Scientific Research and NOVA, the Netherlands Research School for Astronomy.

We thank Ton Raassen, Junjie Mao and Liyi Gu for very useful and detailed discussions about atomic data and various aspects of the calculation of the collision strengths. We thank CHIANTI, PyAtomDB and Cloudy teams for the support they provide to the users of their atomic databases and plasma codes. 

We thank the astrophysics group at SRON and Joop Schaye's group at the Leiden University for the support throughout the various stages of this project. 

Last but not least, we thank the referee of this paper for the detailed referee report which helped us to improve this paper.

\bibliography{bibliography}{}
\bibliographystyle{aa}

\end{document}